\documentclass[a4paper,11pt]{article}
\usepackage{amsmath}
\numberwithin{equation}{section}
\numberwithin{equation}{subsection}

\usepackage{amssymb}
\usepackage{amsfonts}
\usepackage{color}
\usepackage{graphicx}
\usepackage{wrapfig}
\usepackage{slashed}

\def\circlet{\rlap{\raise0.20truecm\hbox{\small$\circ$}}}
\def\à{\`a}
\def\ò{\`o}
\def\ì{\`\i}
\def\ù{\`u}
\def\à{\`a}
\def\è{\`e}
\def\é{\'e}
\def\È{\`E}

\begin{document}

\title{{THE CONFORMAL UNIVERSE III: \\
The Mechanism of Matter Generation}}

\author{{\large \em FINAL VERSION}\\
     \\
     Renato Nobili\\
    {\normalsize    E-mail: renato.nobili@unipd.it}\\}
\date{Padova, 27 March 2016}
\maketitle
\pagestyle{myheadings} \markright{R.Nobili, The Conformal Universe III}

\begin{abstract}
\noindent
This is the last of three papers on Conformal General Relativity (CGR), which ascribes inflation to a spontaneous breakdown
of conformal symmetry, followed by a sudden energy transfer from geometry to matter identified as big bang. This process is 
driven by a conformal--invariant, unitarity--preserving interaction of two Nambu--Goldstone fields: a ghost scalar field 
$\sigma$, invested with geometric meaning, and a physical scalar field $\varphi$ behaving like a Higgs field of varying mass. 
The big bang generates a bulk of Higgs bosons at temperature $T_B\simeq 141$\,GeV, after which the universe evolves adiabatically 
while the Higgs bosons decay into Standard--Model particles and the magnitude of the gravitational coupling constant decreases. 
The process ends when the $\sigma$--$\varphi$ interaction potential vanishes, the amplitudes of these fields converge to their 
expectation values in a final stable vacuum and the Higgs--boson mass converges to about 126 GeV.  The main aspects of this
phenomenology are qualitatively described and accurately exemplified by numerical simulations. The combination of CGR gravitational 
equation at time zero with entropy conservation equation results in striking predictions. The best 
fit to astronomic data is obtained from only standard Higgs boson parameters and a universe age of 19.5 Gyr. The cosmological constant
$\Lambda\simeq 1.35\times 10^{-35}$s$^{-2}$, the scale factor across inflation $Z\simeq 4.54\times 10^{27}$, and the lower bound
of the power spectrum of cosmic background anisotropies $W_{\hbox{\tiny min}}\!\simeq 37.5\,\mu$K$^2$ are thus predicted.
\end{abstract}

\tableofcontents

\section{Basic principles of Conformal General Relativity (CGR)}
\label{introduction}
In two previous papers, here called Part I \cite{PART1} and Part II \cite{PART2}, I impute the
origin of the universe to a local spontaneous breakdown of conformal symmetry occurring in the vacuum state of a
conformal--invariant quantum field system. This primordial event opened up the future cone of a negatively
curved spacetime, first promoting its scale expansion, and then priming in it a huge temporary transfer
of energy from geometry to matter.

A process of this sort is impossible in General Relativity (GR), as here the energy--momentum (EM)
tensors of matter, $\Theta^M_{\mu\nu}(x)$, and of geometry, $\Theta^G_{\mu\nu}(x) \equiv
- G_{\mu\nu}(x)/\kappa$, where $\kappa\simeq 1.6861\times 10^{-37}$ GeV$^{-2}$ is
the gravitational coupling constant, are separately conserved. $G_{\mu\nu}(x)= R_{\mu\nu}(x) -
\frac{1}{2}g_{\mu\nu}(x)R(x)$ is Einstein's gravitational tensor as a function of spacetime parameters
$x = \{x^0, x^1, x^2, x^3\}$, and $R_{\mu\nu}(x)$ and $R(x)$ are respectively the Ricci tensor and scalar of
a (pseudo--)Riemannian manifold, whose metric tensor $g_{\mu\nu}(x)$ has signature $\{+- -\, -\}$.
Instead, the process is possible in a conformal extension of GR, which I call {\em Conformal General Relativity} (CGR),
because -- as explained in \S\,2 of Part I -- in this case, the separate conservation of $\Theta^M_{\mu\nu}$
and $\Theta^G_{\mu\nu}$ does not generally hold.

CGR differs from GR in that the invariance of the total action integral of matter and geometry
under metric diffeomorphisms $x^\mu \rightarrow \bar x^\mu(x)$ -- which change $g_{\mu\nu}(\bar x)$ into
$\bar g_{\mu\nu}(\bar x) = g_{\rho\sigma}[x(\bar x)](d x^\rho\!/\!d\bar x^\mu) (d x^\sigma\!/\!d\bar x^\nu)$
-- is extended to invariance under {\em conformal diffeomorphisms} by the inclusion of Weyl transformations.
These simply consist in the multiplication of each local quantity of dimension $n$ by $e^{n\lambda(x)}$,
where $e^{\lambda(x)}$ is a real scale factor.

Three properties of CGR are worth noting: {\em i}) the group  of conformal diffeomorphisms is the largest
group of coordinate transformations which preserve the causal order of physical events in a
generally curved manifold (\S\,1 of Part II); {\em ii}) CGR is only possible in negatively curved (1+3)D
spacetime manifolds (\S\,3 of Part I); {\em iii}) since the history of the universe is confined to a future cone,
spacetime can profitably be parameterized by {\em hyperbolic polar coordinates}. Time parameter $x^0$ of these
coordinates -- here called {\em kinematic time} $\tau$ -- is the length of the polar--geodesic
segment joining the origin of the future--cone to any given point within that cone.
Since each polar geodesic is one--to--one with its direction $\vec \rho\,$ at the cone origin,
we can write the spacetime parameters as $x = \{\tau, \vec\rho\, \}$, which
implies metric--tensor properties $g_{00}(\tau, \vec\rho\,)=1$, $g_{0i}(\tau, \vec\rho\,)=0$,
$i=1,2,3$. As explained in \S\,5 of  Part II, there are three equivalent ways of implementing CGR.

I. The first way is to ground the theory in a Riemann manifold $H^+$ equipped with
hyperbolic--polar metric $g_{\mu\nu}(\tau, \vec\rho\,)$, and introduce into it a massless
scalar field $\sigma(x)$, called {\em dilation field}, so that the total action integral of
matter and geometry be conformal--invariant. This requires each constant of
dimension $n$ appearing in the Lagrangian density to be multiplied by $\sigma(x)^{n}$, up to a suitable
constant factor. This implementation is called the {\em kinematic--time picture}.
As proven in \S\,3.4 of Part I, in order for CGR to evolve in time toward GR, the following
properties must be satisfied: ({\em i}) The kinetic--energy term of $\sigma(x)$ is negative, which
implies that $\sigma(x)$ is a ghost scalar field; ({\em ii}) the vacuum expectation value (VEV) of $\sigma(x)$
is a monotonic function of $\tau$ for any $\vec\rho\,$; ({\em iii}) gravitational coupling constant $\kappa$ is replaced by
$6/\sigma(x)^2$; ({\em iv}) $\sigma(x)$ converges asymptotically to $\sqrt{6/\kappa}$ in the course of time.
Since in this way $\kappa^{-1}G_{\mu\nu}(x)$ is replaced by $\sigma(x)^2 G_{\mu\nu}(x)/6$, while gravitational
equation $\Theta^M_{\mu\nu}(x) + \Theta^G_{\mu\nu}(x) =0$ still holds, the separate conservation of
$\Theta^M_{\mu\nu}(x)$ and $\Theta^G_{\mu\nu}(x)$ does not hold. As discussed in \S\,7.1 of Part I and
\S\,2 of Part II, these properties allow us to regard $\sigma(x)$ as a Nambu--Goldstone (NG)
boson field, which is created together with a physical NG--boson field $\varphi(x)$, by the spontaneous breakdown
of conformal symmetry. In I and II, is also proven that a huge energy transfer from geometry to matter
may occur provided that $\sigma(x)$ interacts with $\varphi(x)$ in a conformal--invariant, unitarity--preserving
way~\cite{ILHAN}; which makes $\varphi(x)$ a Higgs field of squared--mass term proportional to $\sigma^2(x)$.

II. The second way consists of replacing the metric tensor $g_{\mu\nu}(x)$ of $H^+$ with
{\em fundamental tensor} $\hat g_{\mu\nu}(x)=e^{2\alpha(x)}g_{\mu\nu}(x)$,
with $e^{2\alpha(x)}= \sigma(x)\sqrt{\kappa/6}$ as Weyl scale--factor, of a spacetime
manifold $\widehat{H}^+$ equipped with a (path--dependent) conformal connection of Cartan \cite{CARTAN}, here
called the {\em Cartan manifold}. The connection of $H^+$ is called ``metric'' because
it preserves the spacetime distance between any close points. Instead, that of $\widehat{H}^+$
is called ``conformal'' because it preserves the angles between any spacetime directions passing through
the same point. Note that the coordinates of $\widehat{H}^+$  are not hyperbolic polar because
$\hat g_{00}(\tau,\vec\rho\,) = e^{2\alpha(\tau\!,\,\,\vec\rho\,)}$ and $\hat g_{0j}(\tau,\,\vec\rho\,)=0$.
Now, at variance with the kinematic--time picture, $e^{\alpha(x)}$ should not be envisaged as the
amplitude of a ghost scalar field divided by a dimensional constant, but rather as an additional
degree of freedom of the conformal spacetime geometry. The properties of this geometry are obtained by replacing
the standard tensor calculus of GR with the {\em conformal tensor calculus} (see Appendix to Part I).
Basically, the Christoffel symbols $\Gamma^\lambda_{\mu\nu}$ of the GR metric are replaced with their
conformal extensions $\hat\Gamma^\lambda_{\mu\nu}=\Gamma^\lambda_{\mu\nu} + \delta^\lambda_\nu\partial_\mu
\alpha+\delta^\lambda_\mu \partial_\nu \alpha -g_{\mu\nu}g^{\lambda\rho}\partial_\rho\alpha$,
where $\delta^\mu_\nu$ is a Kronecker delta and $\partial_\mu$ are partial derivatives with respect to $x^\mu$
\cite{EISENHART}. Accordingly, all local quantities $Q_n(x)$ of GR are replaced by $\hat Q_n(x) = e^{n\alpha(x)}Q_n(x)$.
Thus, in particular, $\sigma(x)$ is replaced by $\sigma_0= \sqrt{6/\kappa}$. Since this implementation of CGR
is the analog of the conformal--time representation of standard inflationary cosmology, it is called
the {\em conformal--time picture}.

III. The third way is a variant of the conformal--time picture, which is obtained from the latter by replacing
kinematic--time parameter $\tau$ with $\tilde \tau = \int_0^{\bar\tau} e^{\alpha(\bar\tau,\,\vec \rho\,)}d\bar\tau$.
This is one--to--one with $\tau$ provided that $e^{\alpha(\bar\tau\!,\,\,\vec \rho\,)}$ is monotonic in $\tau$ for any $\vec \rho\,$;
we can then express $x\equiv \{\tau, \vec\rho\,\}$ as functions of $\tilde x =\{\tilde\tau, \vec\rho\,\}$,
or $\tilde x$ as functions of $x$, by writing $x = x(\tilde x)$, or $\tilde x = \tilde x(\tau)$.
We can therefore express any scalar function $\hat f(x)$ of the conformal--time picture as a function of $\tilde x$, or
vice versa, by writing $\tilde f(\tilde x) \equiv \hat f[x(\tilde x)]$, or $\hat f(x)\equiv \tilde f[\tilde x(x)]$.
Since $d\tilde \tau = e^{\alpha(\tau\!,\,\,\vec \rho\,)}d\tau$, we have $\tilde g_{00}(\tilde x)\, d\tilde\tau^2 =
e^{-2\alpha(x)}\hat g_{00}(x)\, d \tau^2 $, hence, $\tilde g_{00}(\tilde x)=1$, $g_{0i}(\tilde x)=0$ and 
$\tilde g_{ij}(\tilde x)= \hat g_{ij}[x(\tilde x)]$, which thus form the metric tensor $\tilde g_{\mu\nu}(\tilde x)$ 
of a Riemann manifold $\widetilde{H}^+$. This is called the {\em proper--time picture}, because it is the analog of the proper--time representation of standard inflationary cosmology \cite{PEACOCK} \cite{MUKHANOV}. In my view, the importance of the
conformal--time picture lies mainly in that it is the bridge between kinematic--time and
proper--time pictures, both of which provide hyperbolic polar representations of spacetime as a Riemannian manifold.

Although equivalent, these three pictures differ in their physical interpretations. The kinematic--time picture provides a
description of the universe from the point of view of an observer today. Looking back to the past, this interprets all
events occurring during the inflationary epoch as subject to the inflating action of the ghost scalar field.

The conformal--time picture provides a description of the universe as it might have been seen by
ideal sets of synchronized observers co--expanding with the universe (in standard cosmology
they are ambiguously described as comoving with the Hubble flow). Since time and length measurement
units also co--expand, these observers cannot detect any change of spacetime scale,
but only a dramatic change in the gravitational coupling strength.

The proper--time picture allows us to describe the universe as it might have been seen
by coeval observers equipped with co--scaling rulers, but not with co--scaling synchronized clocks.
Since in this picture the length measurement--unit only undergoes the Weyl change of scale,
all bodies appear to preserve their size; but their time--courses appear highly
accelerated and gravitational forces strongly affected by the inflation factor.

\subsection{CGR vs standard model of inflationary cosmology}
\label{CGR&standtheor}
Briefly described are here the fundamental differences between the standard model of inflationary
cosmology and the theory of spontaneous breakdown of conformal symmetry.

When we try to infer the spacetime structure from astronomic observations (universe age, duration of the
inflationary epoch, energy and entropy densities in different epochs, etc), we must pay attention
to whether the metric is cylindrical, i.e., sliced into a set of 3D spaces orthogonal to the time axis,
or conical, i.e., partitioned into a set of expanding 3D hyperboloids. The Robertson--Walker metrics
of standard cosmology, briefly surveyed in \S\,4 of Part II are, for instance, cylindrical.
The main problem with these models is their incompatibility with GR, because the separate conservation
of the EM--tensors of matter and geometry implies that the initial state is so singular to be physically
absurd.

In contrast, CGR requires the metric to have a conical structure. This is because the spontaneous breakdown
of conformal symmetry primes the opening of a future cone spanned by the worldlines stemming from its vertex.
It also requires, for suitable field interactions and initial conditions, matter and geometry EM--tensors,
respectively $\Theta^M_{\mu\nu}$ and $\Theta^G_{\mu\nu}$, not to be separately conserved. This makes it possible
for a huge transfer of energy from geometry to matter to take place during the evolution of the universe.
Moreover, the energy densities of matter and geometry in the running spacelike hyperboloid remain finite when
the hyperboloid approaches the boundary of the future cone. This is consistent with the
view that the universe originated from a point of a primordial instable vacuum.

CGR has four remarkable features, which are totally absent in the standard models of inflationary cosmology:
(1) It is only possible in a curved 4D spacetime; (2) the spontaneous breakdown of conformal
symmetry forces spacetime to acquire a small cosmological constant; (3) during the inflationary epoch,
the gravitational coupling parameter decreases by a factor of $1/\sigma(x)^2$; (4) as $\sigma(x)\rightarrow 
\sigma_0 = \sqrt{6/\kappa}$, the time--dependent mass of Higgs bosons converges to $\mu_H\simeq 126$ GeV.

As proved in \S\S~\ref{cosmconstandscale} and \ref{CMBanysot}, CGR leads very naturally to the following remarkable
predictions: assuming an age of the universe of $\simeq 19.5$ Gyr \cite{COWAN}, the value of the cosmological constant is 
$\Lambda\simeq 1.35\times 10^{-35}$s$^{-2}$ or, as vacuum energy--density, $\rho_{\hbox{\tiny vac}} \simeq 3.46\times 10^{-47}$GeV$^4$; the total
scale--expansion factor across inflation is $Z\simeq 4.54\times 10^{27}$; the lower bound of the power band of cosmic microwave
background (CMB) anisotropies is $W_{\hbox{\tiny min}} \simeq 37.5\,\mu$K$^2$.

\section{The geometric structure of the future cone}
As shown in \S\,2.1 of Part II, in CGR, the universe is confined to the future cone of an open spacetime
because it is imprinted by the symmetry of the main stability subgroup of spontaneously broken
conformal symmetry \cite{FUBINI}; i.e., the anti--deSitter group $O(2,3)$, which characterizes
the class of all anti--deSitter spacetimes (of negative curvature). This leads us to
focus on possible optimal parameterizations of these spacetimes and their extensions.

\subsection{Hyperbolic polar coordinates and Milne--universe spacetime}
\label{polhypcoord}
All worldlines stemming from origin $V$ of a future cone immersed in a smooth spacetime manifold
are called {\em polar geodesics} from $V$. By suitable diffeomorphism of the manifold,
we can parameterize the cone near $V$ in Minkowskian coordinates. The complete set of polar
geodesics stemming from $V$ can then be used to define a system of {\em hyperbolic polar coordinates}.
A more detailed description is available in \S\,3.1 of Part II.

As shown in Fig.\,1, any polar geodesic is one--to--one with its direction $\vec\rho\,$ at $V$; thus, we
can denote it as $\Gamma(\vec\rho\,)$. In particular, a polar geodesic, but in general only one -- suppose
$\Gamma(\vec\rho_0)\equiv \Gamma(0)$ -- can be transformed by a second diffeomorphism of the
manifold into a straight axis, without altering the metric near $V$. We identify {\em kinematic time}
$\tau$ of an event $O\in\Gamma(\vec\rho\,)$ as the length of geodesic segment $VO$; then, {\em hyperbolic angle}
$\varrho\,$, ($-\infty\le\varrho\le +\infty$), as the derivative with respect to $\tau$, at $\tau=0$, of the
length of the hyperboloid arc between $\Gamma(0)$ and $\Gamma(\vec\rho\,)$; lastly, we indicate by $\{\theta, \phi\}$
the Euler angles of projection $\vec r$ of $\Gamma(\vec\rho\,)$ onto the 3D--plane orthogonal to $\Gamma(0)$
at $V$. Since the metric in the neighborhood of $V$ is Minkowskian, we can put $\vec\rho=\{\varrho, \theta, \phi\}$
and $\vec\rho_0=\{0, 0, 0\}$.
\begin{figure}[!ht]
\centering
\mbox{%
\begin{minipage}{0.32\textwidth}
\includegraphics[scale=0.65]{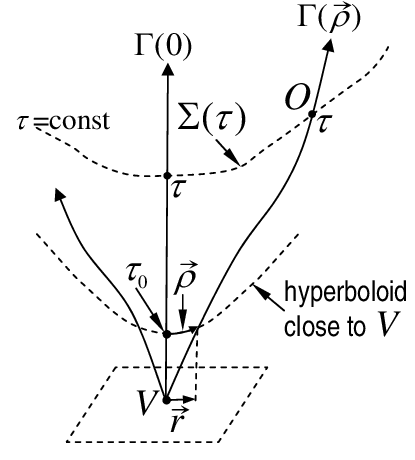}
\end{minipage}%
\quad
\begin{minipage}[c]{0.62\textwidth}
\caption{\small  Geodesics passing through a point $V$ of a spacetime manifold and spanning the interior
of the future cone of origin $V$ can be parameterized by hyperbolic polar coordinates $\{\tau, \vec\rho\,\}$.
This is possible because each geodesic $\Gamma(\vec\rho\,)$ depends uniquely on its direction $\vec\rho\,=
\{\rho, \theta,\phi\}$ at $V$. Kinematic time $\tau$ of an event $O\in\Gamma(\vec\rho\,)$ can then be defined
as the length of geodesic segment $VO$. 3D surface $\Sigma(\tau)$ is the locus of all
comoving observers synchronized at time $\tau$.}
\end{minipage}
}
\end{figure}

Since along each polar geodesic the equations $d\tau/ds =1$ and $\vec\rho= \hbox{constant}$
hold, we can cast the squared line element of the manifold within the future cone in the form
$$
ds^2 = d\tau^2 -\tau^2 \,\gamma_{ij}(\tau, \vec\rho\,)\,d\rho^i\, d\rho^{\,j}\,,\quad (i, j =1,2,3)\,,
$$
where $\rho^1 =\varrho$, $\rho^2 = \theta$ and $\rho^3 =\phi$, and impose the local--flatness conditions at $V$
\begin{eqnarray}
\label{initcond}
\!\!\!\!\lim_{\tau \rightarrow 0}\gamma_{11} = 1;\,\,\, \lim_{\tau \rightarrow 0}\gamma_{22}=(\sinh\rho)^2;\,\,\,
\lim_{\tau \rightarrow 0}\gamma_{33} =(\sinh\rho\,\sin\vartheta)^2;\,\,\,
\lim_{\tau \rightarrow 0}\gamma_{ij}=0\,\, (i\neq j).
\end{eqnarray}

Briefly defining the spacetime parameters as $x=\{\tau, \vec\rho\,\}$, we can write the components of
the metric tensor as $g_{00}(x)=1$, $g_{0i}(x)=0$, $g_{ij}(x) = \gamma_{ij}(\tau, \vec\rho\,)$
and the volume element of the future--cone interior as $\sqrt{-g(x)}\,d^4x \equiv \sqrt{-g(\tau, \vec\rho\,)}
\, d\Omega(\vec\rho\,)d\tau$, where  $d\Omega(\vec\rho\,) =d\rho\,d\theta\,d\phi$.

Denoting the inverse of squared matrix $\big[\gamma_{ij}(\tau, \vec\rho\,)\big]$ by $\big[\gamma^{ij}(\tau, \vec\rho\,)\big]$
and its determinant by $\gamma\equiv \gamma(\tau, \vec\rho\,)$, we can write the squared gradient of  a scalar function
$f(\tau, \vec\rho\,)$ and the Beltrami--d'Alembert operator acting on it, respectively as
\begin{eqnarray}
\label{polarsqrgrad}
&&\hspace{-15mm}(D^\mu f) D_\nu f   = g^{\mu\nu}(x)(\partial_\mu f) \partial_\nu f = (\partial_\tau f)^2 -
\frac{1}{\tau^2}\gamma^{ij}\!\big(\tau,\vec\rho\,\big)(\partial_i f)\partial_j f,\,\,(\mbox{as }\, D_\mu f = \partial_\mu f); \\
\label{polarbeltdalemb}
&&\hspace{-15mm}D^2f  = \frac{1}{\sqrt{-g}}\partial_\mu\big(\sqrt{-g}
\,g^{\mu\nu}\partial_\nu f\big) = \partial_\tau^2 f + \partial_\tau\ln\big(\tau^3\!\sqrt{\gamma\,}\big)\partial_\tau f -
\frac{1}{\tau^2\sqrt{\gamma}}\partial_i\big(\sqrt{\gamma\,}\gamma^{ij}\partial_j f\big).
\end{eqnarray}

The simplest example of a future cone parameterized by hyperbolic polar coordinates is known as
the {\em  Milne universe} \cite{WALKER}, (Mukhanov, 2005, \S\,1.3.5). In this case, the hyperbolic polar coordinates
are related to standard Lorentzian parameters $\{x^0, x^1, x^2, x^3\}$ by equations
$x^0 =\tau\,\cosh\varrho$, $x^1 = \tau\,\sinh\varrho\,\sin\theta\,\cos\phi$, $x^2 = \tau\,\sinh\varrho\,\sin\theta\,\sin\phi$,
and $x^3  = \tau\,\sinh\varrho\,\cos\theta$, from which we obtain $\tau = \sqrt{(x^0)^2 - (x^1)^2 - (x^2)^2 - (x^3)^2}$.

We can check that the squared line--element $ds^2=  (dx^0)^2- (dx^1)^2- (dx^2)^2- (dx^3)^2$
in hyperbolic coordinates is $ds^2 = d\tau^2 - \tau^2\big[d\varrho^2 + (\sinh\varrho)^2 d\theta^2 + (\sinh\varrho\sin\theta)^2\big]$,
the metric--tensor matrix of which is
\begin{equation}
\label{gmunumatrix}
\big[g_{\mu\nu}(\tau, \vec\rho\,)\big] =\hbox{diag}\big[1, - \tau^2, - \tau^2 (\sinh\varrho)^2,
- \tau^2(\sinh\varrho^2 \sin\theta)^2\bigr]\,.
\end{equation}
Accordingly, the 3D and 4D volume elements of the Milne universe are, respectively,
$$
d\Omega(\vec\rho\,)=\big(\!\sinh\varrho\big)^2\!\sin\theta\,d\varrho\,d\theta\,d\phi;\quad
dV(\tau, \vec\rho\,)=\sqrt{-g(\tau, \vec\rho\,)}\,d\varrho\,d\theta\,d\phi\,d\tau \equiv \tau^3\!d\Omega(\vec\rho\,)\,d\tau;
$$
where $d\Omega(\vec\rho\, )$ is the volume element of the hyperbolic--Euler--angle space and $\sqrt{-g(\tau, \vec\rho\,)}= -\tau^6\big(\sinh\varrho\big)^4\!\sin\theta^2$ is the determinant of matrix $\big[g_{\mu\nu}(\tau, \vec\rho\,)\big]$.

The squared gradient of a scalar function $f(\tau, \vec\rho\,)$ is then
\begin{equation}
\label{hyperbsquaregrad} g^{\mu\nu}(\partial_\mu f)\partial_\nu f =
(\partial_\tau f)^2 - \frac{1}{\tau^2}\bigg[(\partial_\rho f)^2 +
\frac{(\partial_\theta f)^2}{(\sinh \rho)^2} + \frac{(\partial_\phi
f)^2}{(\sinh\rho\, \sin \theta)^2}\bigg]\,,
\end{equation}
and the Beltrami--d'Alembert operator acting on a scalar function $f(x)\equiv f(\tau, \vec \rho\,)$ is
\begin{equation}
\label{hyperbdalambert}
D^2 f(x)\equiv\frac{1}{\sqrt{-g(x)}}\,\partial_\mu\Bigl[\sqrt{-g(x)}g^{\mu\nu}(x)\partial_\nu f(x) \Bigr]
= \partial_\tau^2 f(x) + \frac{3}{\tau}\partial_\tau f(x)- \frac{1}{\tau^2}\Delta_\Omega f(x),
\end{equation}
where
\begin{equation}
\label{unitlaplop}
\Delta_\Omega\,f \equiv \frac{1}{(\sinh\varrho)^2} \bigg\{\partial_\varrho
\big[(\sinh\varrho)^2\partial_\rho f\big]+\frac{1}{\sin\theta}\,\partial_\theta
(\sin\theta\, \partial_\theta f) +\frac{1}{(\sin\theta)^2}\,\partial^2_\phi f\bigg\}
\end{equation}
is the 3D Laplacian of $f(x)$ in the hyperbolic--Euler--angle space.

\subsection{The Robertson--Walker metric in hyperbolic polar coordinates}
\label{Friedmod}
Equipping the Milne universe with a metric of Robertson--Walker type \cite{ROBERTSON} \cite{WALKER}, we obtain
a generalized Milne universe, the squared worldline--element and metric--matrix of which are, respectively,
\begin{eqnarray}
\label{curvRWds2}
& & ds^2 = d\tau^2 -  a^2(\tau)\bigl[d\rho^2 + \bigl(\sinh\varrho)^2 d\theta^2 + \bigl(\sinh\varrho\,\sin\theta\bigr)^2d\phi^2
\bigr]\,,\\
\label{curvRWgmunu}
& & \big[g_{\mu\nu}(\tau, \vec\rho\,)\big] =\hbox{diag}\bigl[1, - a(\tau)^2, - a(\tau)^2 (\sinh\varrho)^2,
- a(\tau)^2(\sinh\varrho\sin\theta)^2\bigr],
\end{eqnarray}
where $a(\tau)$ is the Robertson--Walker scale factor of the 3D--hyperboloids. In accordance with Eqs (\ref{initcond}),
we assume $\lim_{\tau\rightarrow 0}a(\tau)/\tau=1$.

The Beltrami--d'Alembert operator derived from metric (\ref{curvRWgmunu}) has the form
\begin{equation}
\label{hyperbdalambert2}
D^2 f\equiv\frac{1}{\sqrt{-g(x)}}\partial_\mu\Bigl[\sqrt{-g(x)}
g^{\mu\nu}(x)\partial_\nu f\,\Bigr]\!= \partial_\tau^2 f +
\frac{3\,\partial_\tau a(\tau)}{a(\tau)}\partial_\tau f-\frac{\Delta_\Omega f}{a(\tau)^2}\,.
\end{equation}

Let us assume that the EM--tensor on the large scale has the form $\Theta_{\mu\nu} = (\rho_E +p)u_\mu u_\nu -
g_{\mu\nu}p + g_{\mu\nu}\rho_{\hbox{\tiny vac}}$, where $\rho_E$ as the energy density of the matter field,
$p$ its pressure, $\rho_{\hbox{\tiny vac}}$ the cosmological constant as energy density of the vacuum, and
$u_\mu=\{1,0,0,0\}$ the 4--velocity of the matter field relative to the comoving reference system. Thus, we have
$\Theta_0^0 = \rho_E + \rho_{\hbox{\tiny vac}}$ and $\Theta^i_i = -p +\rho_{\hbox{\tiny vac}}$, and can
state the Friedmann--Lema\^itre gravitational equations \cite{FRIEDMANN}\cite{LEMAITRE} in the Robertson--Walker
form as $R^0_0 = \frac{1}{2}\kappa\,(\rho_E + 3\,p -2\, \rho_{\hbox{\tiny vac}})\,;\quad R = \rho_E -3\,p + 4\,\rho_{\hbox{\tiny vac}} \,;
\quad R^0_0 - \frac{1}{2}R = \kappa\,(\rho_E + \rho_{\hbox{\tiny vac}})$, where $\kappa \simeq 1.6861\times
10^{-37}$GeV$^{-2}$ is the gravitational coupling constant in natural units, $R_{\mu\nu}$ is the Ricci tensor and
$R$ the Ricci scalar. Therefore, using the Christoffel symbols constructed out of metric (\ref{curvRWgmunu}), which are
listed in Eq (3.4.2) of Part II, we obtain
\begin{eqnarray}
\label{RWRiccitens}
& & R^0_0 = - 3\frac{\ddot a}{a};\quad R^1_1 = R^2_2 = R^3_3 = - \bigg(\frac{\ddot a}{a} +
2\,\frac{\dot a^2-1}{a^2}\bigg);\quad R^\mu_\nu = 0\,\, (\mu\neq \nu)\,;\nonumber\\
& & R^0_0 - \frac{1}{2} R = 3\,\frac{\dot a^2-1}{a^2}\,;
\quad R = -6\,\bigg(\frac{\ddot a}{a}+\frac{\dot a^2-1}{a^2}\bigg)\,;
\end{eqnarray}
showing that we have $R=0=R^\mu_\nu = 0$, if and only if $a(\tau)= \tau-\tau_0$, where $\tau_0$ is an arbitrary constant.
Consequently, the Friedmann--Lema\^itre equations take the form
\begin{equation}
\label{FLRWeqs}
\frac{\ddot a}{a} = - \frac{\kappa}{6}\,\big(\rho_E  + 3\,p -
2\,\rho_{\hbox{\tiny vac}}\big)\,,\quad \frac{\dot a^2-1}{a^2} =
\frac{\kappa}{3}\,\big(\rho_E +\rho_{\hbox{\tiny vac}}\big)\,\Longrightarrow \frac{\dot a}{a}+ \frac{\dot\rho_E}{3\big(\rho_E +p\big)}=0\,,
\end{equation}
which converge to the homologous equations for flat spacetime as $\tau\rightarrow 0$.

\subsection{The accelerated Milne universe in hyperbolic polar coordinates}
\label{Curvedhyperbspacetime}
The EM--tensor of an empty universe with cosmological constant $\Lambda$ has the simple form  $\Theta_{\mu\nu}(x)
= g_{\mu\nu}(x)\,\rho_{\hbox{\tiny vac}} =  g_{\mu\nu}(x)\Lambda/\kappa$. Hence we have,  $R_{\mu\nu} = -\Lambda\,g_{\mu\nu}(x)$
and $R= -4\,\Lambda= -4\,\kappa\,\rho_{\hbox{\tiny vac}}$ (Eisenhart, 1949 p.92). Let $c(\tau)$ be the Robertson--Walker
factor accounting only for the accelerating effect of the cosmological constant. Eqs (\ref{FLRWeqs}) then condense into
\begin{equation}
\label{Rtorhovac}
\frac{\ddot c}{c} = \frac{\dot c^2-1}{c^2} = \frac{\Lambda}{3}\,,
\end{equation}
the solution of which, with the initial condition $\lim_{\tau =0}c(\tau)/\tau= 1$, is
\begin{equation}
\label{atau}
c(\tau) = \tau_\Lambda \sinh (\tau/\tau_\Lambda)=\tau\bigg(1+\frac{\tau^2}{6\,\tau_\Lambda^2}
+\frac{\tau^4}{120\,\tau_\Lambda^4}\dots\bigg)\,,\,\, \hbox{with }\, \tau_\Lambda = \sqrt{\frac{3}{\Lambda}} =
\sqrt{\frac{3}{\kappa\,\rho_{\hbox{\tiny vac}}}}\,,
\end{equation}
$\tau_\Lambda$ is known as the {\em Hubble time} of the empty accelerating universe \cite{BEHAR}.
Recent data indicates that the cosmological constant is very close, if not
exactly equal, to the critical density of the universe, the expansion of which appears
to be in slight acceleration. If the energy density of the universe were exactly equal
to the cosmological constant, the state equation of the universe on the large scale would
be $\rho_E + 3\,p =0$ and the physical state of the universe would be the same as an empty
universe of spacetime curvature $R=- 4\,\kappa\, \rho_{\hbox{\tiny vac}}$.

Astronomic data indicate $\Lambda \simeq 10^{-35}$ s$^{-2}$, which, in terms of vacuum energy density,
is equivalent to $\rho_{\hbox{\tiny vac}}= \Lambda/\kappa \approx  2.56\times 10^{-47}$ GeV$^4$,
in terms of spacetime curvature to $R\approx -1.73\times 10^{-83}$ GeV$^2$, and in terms of Hubble time
to $\tau_\Lambda\approx 5.49\times 10^{17}$\,s. The natural--unit conversion s$^{-1}
\simeq 6.58\times 10^{-25}$\,GeV is used here (see conversion table in \S\,\ref{HiggsEnergyDens}).

\section{The Higgs field in CGR}
\label{HiggsinCGR}
This section introduces the Lagrangian formalism which is necessary to describe in three different
but equivalent ways the basic mechanism of spacetime inflation and matter generation, i.e., the
interaction of a massless ghost scalar field $\sigma$ with a massless physical scalar field $\varphi$,
which thus becomes a Higgs field of dynamically varying mass. These fields originate
as the two NG bosons generated by the spontaneous breakdown of conformal symmetry, respectively
associated to stability subgroups $O(2,3)$ and $O(1, 4)$ of conformal group $O(2, 4)$, as extensively
explained in \S\,2 of Part II. This simple system is capable of representing inflation
as a huge transfer of energy from geometry to matter.

For brevity, we list in advance the symbols and constants used in this section

$\mu_H\simeq 126$ GeV:  Higgs--boson mass;

$\mu=\mu_H/\sqrt{2} \simeq 89.1$ GeV: mass parameter of Higgs--field action integral;

$M_{rP} \simeq 2.4354\times 10^{18}$ GeV: reduced Planck mass;

$\kappa=1/M_{rP}^2\simeq 1.6861\times 10^{-37}$ GeV$^{-2}$: gravitational coupling constant;

$\sigma_0 = \sqrt{6/\kappa} \simeq  5.9654\times 10^{18}$ GeV: limiting amplitude of  $\sigma(x)$ in the post--inflation era;

$G_F \simeq  1.16637\times 10^{-5}$ GeV$^{-2}$: Fermi coupling constant;

$\lambda=mu_H^2 G_F/\sqrt{2} \simeq 0.132$: self--coupling constant of $\varphi$ according to the Standard Model
of elementary particles.

\subsection{The Higgs field in the kinematic--time picture}
\label{higgsOnRiem}
The name kinematic--time picture derives from the fact that it is the analog of the kinematic--time
representation used by Brout {\em et.al.} (1979) in their theory of the causal universe.
The simplest case of conformal--invariant, unitarity--preserving interaction of a ghost
scalar field $\sigma$ with a physical scalar field $\varphi$, is described by action integral
\begin{equation}
\label{Actint}
A \! = \!\!\int_{H^+}\!\!\!\frac{\sqrt{-g}}{2}\bigg[g^{\mu\nu}\bigl(\partial_\mu\varphi\bigr)
\partial_\nu\varphi\!-\!g^{\mu\nu}\bigl(\partial_\mu\sigma\big)\partial_\nu\sigma
\!-\! \frac{\lambda}{2} \bigg(\varphi^2 \!-\!\frac{\mu^2}{\lambda}\frac{\sigma^2}
{\sigma_0^2}\bigg)^2\!\!+\frac{R}{6}\bigl(\varphi^2\!-\sigma^2\bigr)\bigg]d^4x\,,
\end{equation}
which is actually conformal--invariant up to a surface term. Here, $H^+$ is the
Riemann manifold describing the future--cone interior, $x^\mu$ are its spacetime parameters,
$g_{\mu\nu}$ its metric--tensor, $g$ the determinant of matrix $[g_{\mu\nu}]$ and $R$ the Ricci scalar.
The negative signs of the kinetic energy of $\sigma$ and of $\sigma^2$ clearly indicate
that $\sigma$ is not a field provided with physical meaning, but rather a scalar field potentially invested
with geometrical meaning, as discussed in \S\,6.2 of Part I. Because of this peculiar structure of $\varphi$--$\sigma$
interaction, the Hamiltonian of $A$ is bounded from below for almost all initial conditions of $\sigma$, $\varphi$,
$\partial_\mu\sigma$ and $\partial_\mu\varphi$, which prevents the violation of $S$--matrix unitarity on the part of
the scalar ghost, as argued by Ilhan and Kovner (2013).

Expressing the metric tensor in hyperbolic polar coordinates, as described in \S\,\ref{polhypcoord},
we derive from $A$ the motion equations of $\varphi$ and $\sigma$, respectively:
\begin{eqnarray}
\label{varphimoeteq}
&&\hspace{-10mm}
\partial_\tau^2 \varphi + \partial_\tau\!\ln\!\big(\tau^3\sqrt{\,\gamma\,}\,\big)\partial_\tau \varphi -
\frac{\partial_i\big(\sqrt{\,\gamma\,}\,\gamma^{ij}\partial_j \varphi\big)}{\tau^2\sqrt{\,\gamma}}
 + \lambda\bigg(\varphi^2 - \frac{\mu^2}{\lambda}\,
\frac{\sigma^2}{\sigma_0^2}\bigg)\varphi -\frac{R}{6}\,\varphi=0,\\
\label{sigmamoeteq}
&&\hspace{-10mm}
\partial_\tau^2\sigma + \partial_\tau\!\ln\!\big(\tau^3\sqrt{\,\gamma\,}\,\big)\partial_\tau\sigma -
\frac{\partial_i\big(\sqrt{\,\gamma\,}\,\gamma^{ij}\partial_j\sigma\big)}{\tau^2\sqrt{\,\gamma}}+
\frac{\mu^2}{\sigma_0^2}\,\bigg(\varphi^2 -\frac{\mu^2}{\lambda}\,\frac{\sigma^2}{\sigma_0^2}\bigg)\sigma
- \frac{R}{6}\,\sigma=0\,,
\end{eqnarray}
where the expression for $D^2 f$ given by Eq (\ref{polarbeltdalemb}), and the gravitational equation
\begin{eqnarray}
\label{tetamunu}
\hspace{-2mm}\Theta_{\mu\nu} \!\!\!\!\! &\equiv& \!\!\!\!\! \frac{2}{\sqrt{-g}}\bigg[\frac{\delta A}{\delta g^{\mu\nu}}
\!-\! \partial_\lambda\frac{\delta A}{\delta \partial_\lambda g^{\mu\nu}}\bigg] \! =\!
\bigl(\partial_\mu\varphi\bigr)\partial_\nu\varphi\!-\!\bigl(\partial_\mu\sigma\bigr)
\partial_\nu\sigma\!-\!\frac{g_{\mu\nu}}{2}\Bigl[(\partial_\rho\varphi\bigr)\partial^\rho\varphi
\!-\!\bigl(\partial_\rho\sigma\bigr)\partial^\rho\sigma\Bigr]\!+\nonumber\\
& & \hspace{-4mm}\frac{g_{\mu\nu}}{4}\lambda\bigg(\varphi^2 - \frac{\mu^2}{\lambda}\frac{\sigma^2}
{\sigma_0^2}\bigg)^2\!\!+ \frac{1}{6}\bigl(g_{\mu\nu}D^2 \!\!-\!\!D_\mu \partial_\nu\bigr)
\bigl(\varphi^2\!-\!\sigma^2\bigr) \!+\!\frac{\varphi^2\!-\!\sigma^2}{6}\,G_{\mu\nu} =0\,,
\end{eqnarray}
are used. Here, $\Theta_{\mu\nu}(x)$ is the total EM tensor of matter and geometry, $D_\mu$ are the covariant
derivatives in hyperbolic polar coordinates and $G_{\mu\nu}(x)=R_{\mu\nu}(x)- g_{\mu\nu}(x)R(x)/2$.

The potential--energy density term of $A$, that is,
$$
U(x)=\frac{\lambda}{4}\bigg[\varphi^2(x)-\frac{\mu^2}{\lambda}\frac{\sigma^2(x)}
{\sigma_0^2}\bigg]^2-\frac{R(x)}{12}\bigr[\varphi^2(x)-\sigma^2(x)\bigl]\,,
$$
for constant $\sigma(x)^2>0$, has its maximum at $\varphi(x)=0$, and its minimum at
$$
\varphi(x) = \varphi_0(x)\equiv \frac{\mu}{\sqrt{\lambda}}\sqrt{\frac{\sigma^2(x)}
{\sigma_0^2}+\frac{R(x)}{6\,\mu^2}}\,.
$$

Since the energy--densities of $\varphi$ and $\sigma$ are progressively dissipated
by the frictional second terms on the left sides of Eqs (\ref{varphimoeteq}) (\ref{sigmamoeteq}),
we expect that, with $\varphi^2(0) < \varphi^2_0(0)$ as initial condition, $U(x)$ will
evolve towards its minimum.

Note that $\sigma(x)$ is forced to increase in mean because its kinetic--energy is negative, while
$\varphi(x)$ also increases in mean because it is attracted by the minimum of $U(x)$. This may be
interpreted as a process of spacetime expansion and matter creation, which fades away the more and more
as the minimum is approached. The Ilhan--Kovner mechanism which keeps the Hamiltonian bounded
from below is precisely this.

These equations can be considerably simplified if we assume that the universe is homogeneous and isotropic,
implying that $\varphi$, $\sigma$ and $R$ depend only on $\tau$. In which case, the
metric--tensor matrix simplifies to
\begin{equation}
\label{FRWmetmetrictens}
\big[g_{\mu\nu}(\tau, \vec\rho\,)\big] = \mbox{diag}\big[1,  -c(\tau)^2, -c(\tau)^2\bigl(\sinh\varrho)^2,
-c(\tau)^2\bigl(\sinh\varrho\,\sin\theta\bigr)^2\bigr]\,,
\end{equation}
where $c(\tau)$ is a scale factor perhaps accounting for accelerated expansion.
As explained in \S~\ref{Friedmod}, $c(\tau)/\tau $ must converge to 1 as $\tau\rightarrow 0$.
Hence, we have $\sqrt{-g(\tau, \vec\rho\,)\,} = c(\tau)^3\bigl(\sinh\varrho\big)^2\!\sin\theta$.

In this case, the Riemann manifold $H^+$ specializes in that of the accelerated Milne spacetime $M^+$
described in \S\,\ref{Curvedhyperbspacetime} and represented in Fig.\,2.
\begin{figure}[!ht]
\mbox{%
\begin{minipage}{.45\textwidth}
\includegraphics[scale=0.38]{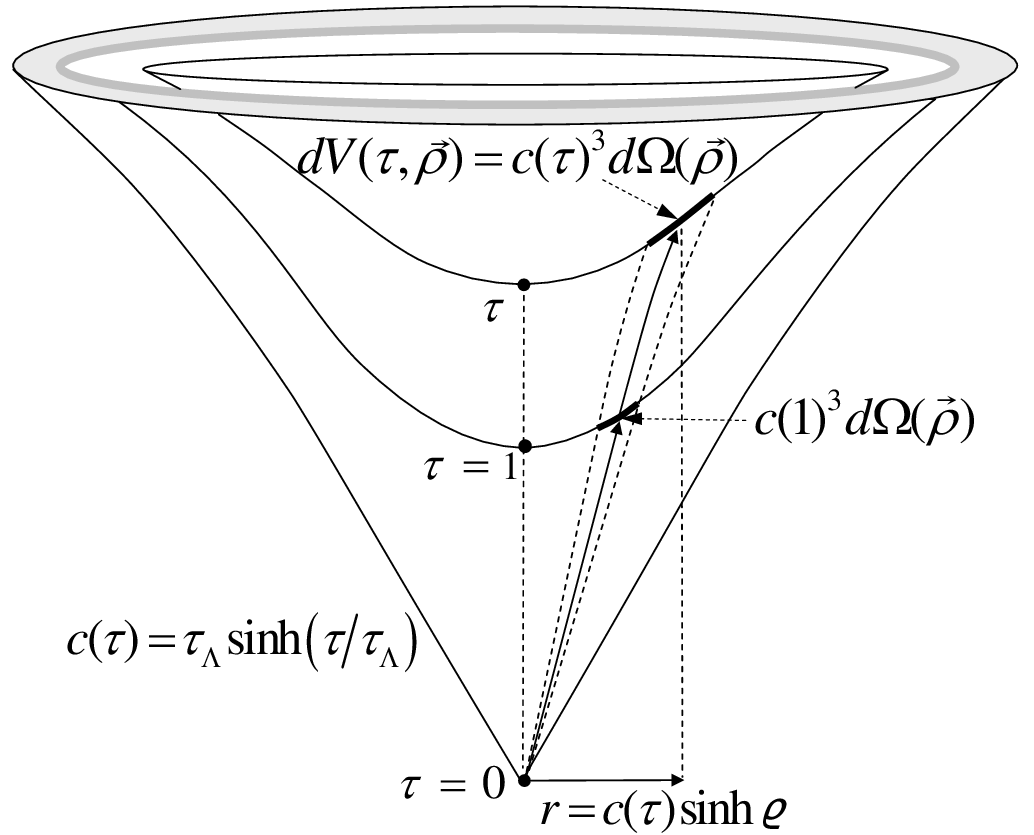}
\end{minipage}%
\quad
\begin{minipage}[c]{.47\textwidth}
\caption{\small Qualitative features of an accelerated Milne spacetime $M^+$ in kinematic--time coordinates:
$\tau$ = kinematic time; $\vec\rho =\{\varrho, \theta, \phi\} = $ hyperbolic--Euler angles;
$d\Omega(\vec\rho\,)$ = volume element of hyperbolic--Euler--angle space; $dV(\tau,\vec\rho\,)
= c(\tau)^3 d\Omega(\vec\rho\,)$ = volume element of hyperboloid at kinematic time $\tau$.}
\end{minipage}%
}
\end{figure}

Correspondingly, Eq (\ref{Actint}) simplifies to
\begin{equation}
\label{SimpleActint}
A =\Omega\int_0^{+\infty}\frac{c^{\,3}}{2}\,\bigg[\bigl(\partial_\tau \varphi \bigr)^2
- \bigl(\partial_\tau\sigma \bigr)^2 - \frac{\lambda}{2} \bigg(\varphi^2 -
\frac{\mu^2}{\lambda}\frac{\sigma^2}{\sigma_0^2}\bigg)^2+\big(\varphi^2\!-\sigma^2\big)\frac{R}{6}\bigg] d\tau\,,
\end{equation}
where $\Omega$ is the infinite volume of the hyperbolic--Euler--angle space, $\partial_\tau$ is the derivative
with respect to $\tau$ and $R$ is negative spacetime curvature. By functional variations of $A$, we can derive
the following motion equations for $\varphi(\tau)$ and scale factor $s(\tau)\equiv \sigma(\tau)/\sigma_0$
\begin{eqnarray}
\label{simpPhiMoteqOnRiem}
& & D^2\varphi \equiv \ddot\varphi + 3\,\frac{\dot c}{c}\,\dot\varphi =
\lambda\,\bigg(\frac{\mu^2}{\lambda}s^2-\varphi^2\bigg)\,\varphi +\frac{R}{6}\varphi\,,\\
\label{simpSMoteqOnRiem}
&& D^2 s \equiv \ddot s + 3\,\frac{\dot c}{c}\,\dot s =\frac{\mu^2}{\sigma_0^2}\,
\bigg(\frac{\mu^2}{\lambda}s^2-\varphi^2\bigg) s +\frac{R}{6}\, s\,,
\end{eqnarray}
where $D^2 f(x)$ is given by Eq (\ref{hyperbdalambert2}) and dot superscripts stand for $\partial_\tau$.

Putting in Eq  (\ref{tetamunu}) $G_{\mu\nu}(x)= - g_{\mu\nu}(x)\,R/4$, as is the case for a constant spacetime
curvature, contracting the indices with $x^\mu x^\nu/\tau^2$, using $x^\mu x^\nu g_{\mu\nu}=\tau^2$, $x^\mu D_\mu=
\tau \partial_\tau$, and denoting spatial indices by $i, j$, we obtain the simplified gravitational equations
\begin{eqnarray}
\label{simpleteta00}
& & \hspace{-16mm}\Theta_{\tau\tau}= \frac{1}{2}\bigg[\dot\varphi^2 -
\dot\sigma^2\!+\!\frac{\lambda}{2}
\bigg(\varphi^2-\frac{\mu^2}{\lambda}\frac{\sigma^2}{\sigma_0^2}\bigg)^2 +
\frac{\dot c}{c}\,\partial_\tau\big(\varphi^2-\sigma^2\big)  -\frac{\varphi^2-\sigma^2}{12}R\bigg] =0\,;\\
\label{simpletetaij}
& & \hspace{-16mm}\Theta_{ij} = \frac{g_{ij}}{2}\bigg[\dot\sigma^2
-\dot\varphi^2 + \frac{\lambda}{2}\bigg(\varphi^2 - \frac{\mu^2}{\lambda}\,
\frac{\sigma^2}{\sigma_0^2}\bigg)^2 +\frac{1}{3}\,\partial_\tau^2\big(\varphi^2-\sigma^2)
+\frac{\dot c}{c}\,\partial_\tau\big(\varphi^2-\sigma^2) -\nonumber\\
& &\frac{\varphi^2-\sigma^2}{12}R\bigg]= -\delta_{ij}\frac{\Theta_{\tau\tau}}{3}=0\,,
\end{eqnarray}
where $\partial_\tau^2 f^2 = 2 f\dot f + 2\dot f^2$, Eqs (\ref{simpPhiMoteqOnRiem})
(\ref{simpSMoteqOnRiem}) and $g_{ij}=-\delta_{ij}$ are used in the last step.
It is then evident that the gravitational equations condense into $\Theta_{\tau\tau}=0$ only.

\subsection{Geometry--to--matter energy transfer}
\label{EnDensCons}
The kinematic--time picture is particularly appropriate for dealing with geometry--to--matter energy transfer.
Let us split action integral (\ref{Actint}) into two parts: (1) the action integral of Higgs field $\varphi(x)$
\vspace{-1mm}
\begin{equation}
\label{AtoM}
A^{(\varphi)} = \int_{H^+}\frac{\sqrt{-g}}{2}\Bigl[g^{\mu\nu}\bigl(\partial_\mu\varphi\bigr)
\partial_\nu\varphi-\frac{\lambda}{2}\,\varphi^4 +\frac{R}{6}\,\varphi^2 +\frac{\mu^2\varphi^2\sigma^2}{\sigma_0^2}\Bigr]d^4x,
\end{equation}
to be viewed as subject to the action of ghost field $\sigma(x)$, now regarded as an external field, and
(2) the action integral of ghost scalar field $\sigma(x)$
\vspace{-1mm}
\begin{equation}
\label{AtoG}
A^{(\sigma)} = -\int_{H^+}\frac{\sqrt{-g}}{2}\Bigl[g^{\mu\nu}\bigl(\partial_\mu\sigma\bigr)\partial_\nu\sigma+
\frac{\mu^4}{2\lambda}\frac{\sigma^4}{\sigma_0^4}+\frac{R}{6}\sigma^2-\frac{\mu^2\varphi^2\sigma^2}{\sigma_0^2}\Bigr]d^4x,
\end{equation}
to be viewed as subject to the action of $\varphi(x)$, now regarded as an external field.

Clearly, these parts are linked to $A$ by equation
$$
A^{(\varphi)} + A^{(\sigma)} = A +\int\sqrt{-g(x)}\,\frac{\mu^2\varphi(x)^2\,\sigma(x)^2}{2\,\sigma_0^2}\,d^4x\,.
$$

Although separately introduced, $A^{(\varphi)}$ and  $A^{(\sigma)}$ provide respectively the same
motion equations as those given by Eqs (\ref{simpPhiMoteqOnRiem}) and (\ref{simpSMoteqOnRiem}),
and two partial EM--tensors
\begin{eqnarray}
\label{ThetaMmunu}
\Theta^{(\varphi)}_{\mu\nu}(x) + g_{\mu\nu}(x)\frac{\mu^2}{2\,\sigma_0^2}\,\sigma(x)^2\,\varphi(x)^2\,,\\
\label{ThetaGmunu}
\Theta^{(\sigma)}_{\mu\nu}(x) + g_{\mu\nu}(x)\frac{\mu^2}{2\,\sigma_0^2}\,\sigma(x)^2\,\varphi(x)^2\,,
\end{eqnarray}
which are clearly non--conservative because they are linked to total EM tensor $\Theta_{\mu\nu}$ by
equation $\Theta^{(\varphi)}_{\mu\nu} + \Theta^{(\sigma)}_{\mu\nu}=
\Theta_{\mu\nu} +g_{\mu\nu}\,\mu^2\varphi^2\,\sigma^2/2\,\sigma_0^2$  and,
since $D^\mu \Theta_{\mu\nu}=0$,
by equation
\begin{equation}
\label{totalnoncors}
D^\mu \Theta^{(\varphi)}_{\mu\nu} + D^\mu \Theta^{(\sigma)}_{\mu\nu} =
\frac{\mu^2}{2\,\sigma_0^2}\,\sigma^2\, \partial_\nu\varphi^2+
\frac{\mu^2}{2\,\sigma_0^2}\,\varphi^2\, \partial_\nu\sigma^2\,.
\end{equation}

Actually, as shown in \S\,5 of Part I, Eq (\ref{totalnoncors}) is the sum of the non--conservative
equations
\begin{equation}
\label{noncons}
D^\mu \Theta^{(\varphi)}_{\mu\nu}(x) = \frac{\mu^2\,\sigma(x)^2}{2\,\sigma_0^2}\,\partial_\nu\varphi(x)^2\,,
\quad D^\mu \Theta^{(\sigma)}_{\mu\nu}(x) = \frac{\mu^2}{2\,\sigma_0^2}\,\varphi(x)^2\,\partial_\nu\sigma(x)^2 \,,
\end{equation}
the second members of which may be interpreted as the power--density delivered by inflationary
forces to produce field--amplitude variations $\partial_\nu\varphi(x)$ and $\partial_\nu\sigma(x)$,
so as to increase the energy densities of matter and geometry, respectively.

The corresponding reduced forms of Eqs (\ref{noncons}) are then
\begin{eqnarray}
\label{nonconserv01}
& & \frac{1}{c(\tau)^3}\partial_\tau \bigl[c(\tau)^3\Theta^{(\varphi)}_{\tau\tau}(\tau)\bigr] =
\frac{\mu^2}{2}\frac{\sigma(\tau)^2}{\sigma_0^2}\, \partial_\tau\varphi(\tau)^2=\frac{\mu^2}{2}
e^{2\alpha(\tau)}\partial_\tau\varphi(\tau)^2\,;\\
\label{nonconserv02}
& & \frac{1}{c(\tau)^3}\partial_\tau \bigl[c(\tau)^3\Theta^{(\sigma)}_{\tau\tau}(\tau)\bigr] =
\frac{\mu^2}{2}\,\varphi(\tau)^2\, \partial_\tau\frac{\sigma(\tau)^2}{\sigma_0^2}= \frac{\mu^2}{2}\,\varphi(\tau)^2\,
\partial_\tau e^{2\alpha(\tau)}.
\end{eqnarray}

Eqs (\ref{noncons}) can easily be generalized to the case in which $A$ is the action integral of the
Higgs field interacting with an unspecified number of other matter fields $\boldsymbol{\Psi}$, all of which may
depend upon general coordinates $x$. In which case, on account of Eq  (\ref{curvRWgmunu})
-- but now with $c(\tau)$ in place of $a(\tau)$ -- the first of Eqs (\ref{noncons}) is replaced by
\begin{equation}
\label{nonconserv}
D^\mu \Theta^{(\varphi, \boldsymbol{\Psi})}_{\mu\nu}(x) \equiv \frac{1}{\sqrt{-g(x)}}
\,\partial_\mu\Bigl[\sqrt{-g(x)}\,g^{\mu\lambda}(x)\,\Theta^{(\varphi,
\boldsymbol{\Psi})}_{\lambda\nu}(x)\Bigr]=\frac{\mu^2 \,\sigma(x)^2}{2\,\sigma_0^2}\,
\partial_\nu \varphi(x)^2\,,
\end{equation}
as explained in detail, but in different notations, in \S\,6 of Part I.

This result can be brought to a form more useful for cosmological computations by averaging both sides of the first
of Eqs (\ref{nonconserv}) over volume $\Omega$ of the hyperbolic--Euler angle space, as formally described by
$$
\langle F(\tau)\rangle = \lim_{\Omega\rightarrow \infty} \frac{1}{\Omega}\int_{\Omega} F(\tau, \vec\rho\,)\,d\Omega(\vec\rho\,)\,.
$$
Thus, for matter energy density in the kinematic--time picture $\rho_M(\tau, \vec\rho)\equiv\Theta^{(\varphi,
\boldsymbol{\Psi})}_{\tau\tau}(\tau, \vec\rho\,)$, and $\sigma(\tau, \vec\rho\,)$ only depending on $\tau$, we obtain
\begin{equation}
\label{averenedens1}
\frac{1}{c(\tau)^3}\,\partial_\tau \bigl[c(\tau)^3 \langle\rho_M(\tau)\rangle\bigr]=
\frac{\mu^2 \sigma(\tau)^2}{2\,\sigma_0^2}\langle \partial_\tau\varphi(\tau)^2\rangle\,,\nonumber
\end{equation}
and consequently, by integration over a kinematic--time interval $[\tau_0, \tau]$,
\begin{equation}
\label{infyield}
\langle\rho_M(\tau)\rangle = \bigg[\frac{c(\tau_0)}{c(\tau)}\bigg]^3 \langle\rho_M(\tau_0)\rangle +
\frac{\mu^2}{2\,\sigma_0^2 c(\tau)^3}\,\int_{\tau_0}^{\tau}\!\!\!c(\tau')^3\sigma(\tau')^2\,
\langle \partial_{\tau'} \varphi(\tau')^2\rangle\, d\tau'\,.
\end{equation}

We can give more physical meaning to this formula by regarding $\langle\rho_M(\tau_0)\rangle$
as the mean energy density of matter at kinematic time $\tau_0$, i.e., the time at which the energy transfer
from geometry to matter begins, and $\langle\rho_M(\tau)\rangle$ as the mean energy density of matter
at a proper time $\tau$ of the post--inflationary era.

Eq  (\ref{infyield}) can be interpreted as the energy--density variation due to the total power
delivered by inflationary expansion during the kinematic--time interval $[\tau_0, \tau]$.
Therefore, this equation represents the first law of thermodynamics applied to the inflationary process.
It shows that the energy transfer  between geometry and matter continues as long as
$\langle \partial_{\tau} \varphi(\tau)^2\rangle\neq 0$.

\subsection{The Higgs field in the conformal--time picture}
\label{higgsOnCart}
The conformal--time version of the kinematic--time picture corresponds to the conformal--time
representation of standard inflationary cosmology. It is obtained by replacing action integral
$A$ of \S\,\ref{higgsOnRiem} with a functionally equivalent action integral $\hat A$ grounded in a
conformally connected Cartan manifold $\widehat H^+$. This means that $\hat A -A$ is a mere
surface term, as demonstrated in \S\,6.2 of Part II.

The squared line--element and metric tensor $\hat g_{\mu\nu}(\tau, \vec\rho\,)$ of $\widehat H^+$
are related to those of $H^+$ by:
\vspace{-4mm}
\begin{eqnarray}
\label{fundtensqline}
& &  \hspace{-5mm}d\hat s^2  =  e^{2\alpha(\tau, \vec\rho\,)}d^2s = e^{2\alpha(\tau, \vec\rho\,)}
\bigl[d\tau^2 -\tau^2 \,\gamma_{ij}(\tau, \vec\rho\,)\,d\rho^i d\rho^j\big],\,\,(i, j =1,2,3);\\
\label{conffundtens}
& &  \hspace{-5mm}\hat g_{00}(\tau, \vec\rho\,) = e^{2\alpha(\tau, \vec\rho\,)};\quad \hat g_{0i}(\tau, \vec\rho\,)=0;
\quad \hat g_{ij}(\tau, \vec\rho\,) = -\tau^2 e^{2\tilde\alpha(\tau, \vec\rho\,)}\gamma_{ij}(\tau, \vec\rho\,);
\end{eqnarray}
where spatial coefficients $\gamma_{ij}(\tau, \vec\rho\,)$ may contain the gravitational field.
Correspondingly: the volume element $\sqrt{-\hat g(\tau, \vec\rho\,)}\,\tau^3 d\Omega(\vec\rho\,)$  is
replaced with $e^{4\alpha(\tau, \vec\rho\,)}\sqrt{-g(x)}\,\tau^3 d\Omega(\vec\rho\,)$,
where $d\Omega(\vec\rho\,)$ is the volume element of the hyperbolic--Euler angles; scalar field
$\varphi(x)$ is replaced by $\hat \varphi(x)=e^{-\alpha(x)}\varphi(x)$;
ghost scalar field $\sigma(x)$ is replaced by scale factor $e^{\alpha(x)}= \sigma(x)/\sigma_0$
of fundamental tensor $\hat g_{\mu\nu}(x)= e^{2\alpha(x)} g_{\mu\nu}(x)$. As shown in \S\,2.2 of Part I,
these changes make gravitational interaction become formally similar to that of standard GR.
Note that the positivity of factor $e^{\alpha(x)}$ is compatible with Eq (\ref{Actint}), because action integral
$A$ is invariant under $\sigma(x)\rightarrow -\sigma(x)$. In more general theories, any local field $\Psi_n(x)$ of
dimension $n$ is replaced by $\hat \Psi_n(x)= e^{n\,\alpha(x)}\,\Psi_n(x)$.
By contrast, the formal structure of Ricci tensors in conformal--time coordinates, $\hat R_{\mu\nu}(x)$ and $\hat R(x)$,
differs substantially from that of their respective homologous $R_{\mu\nu}(x)$ and $R(x)$ of the kinematic--time picture,
which have dimensions 0 and $-2$ , respectively. Instead, they are related to the latter by equations
\begin{eqnarray}
\label{RmunutohatRmunu4}
& &\hspace{-7mm}\hat R_{\mu\nu} =  R_{\mu\nu}+\sigma^{-2}\!\big[4\,(\partial_\mu \sigma)
\,\partial_\nu \sigma - g_{\mu\nu} (\partial^\rho \sigma)\,\partial_\rho \sigma\big]-
\sigma^{-1}\big(2\,D_\mu\partial_\nu\sigma+g_{\mu\nu} D^2\sigma\big)\equiv \nonumber\\
& & \quad\, R_{\mu\nu} + s^{-2}\!\big[4\,(\partial_\mu s)
\,\partial_\nu s - g_{\mu\nu} (\partial^\rho s)\,\partial_\rho s\big]-
s^{-1}\big(2\,D_\mu\partial_\nu s+ g_{\mu\nu} D^2 s\big)\,;\\
\label{RhattoR4}
& & \hspace{-7mm}\hat R =  e^{-2\alpha}\big(R - 6\, \sigma^{-1} D^2 \sigma\big) \equiv s^{-2}\big(R - 6\, s^{-1} D^2 s\big)\,;
\end{eqnarray}
where we have put $\sigma(x)\equiv \sigma_0 e^{\alpha(x)}\equiv \sigma_0 s(x)$ (see details in Appendix to Part I).

From now on, all quantities and symbols pertaining to the conformal--time picture will be marked by the hat superscript.

Carrying out the due replacements, we obtain
\begin{eqnarray}
\label{hatActint}
A \rightarrow \hat A = \int \frac{\sqrt{-\hat g}}{2}\bigg[\hat g^{\mu\nu}(\partial_\mu\hat \varphi)
\partial_\nu\hat \varphi-\frac{\lambda}{2}\bigg(\hat \varphi^2 -\frac{\mu^2}{\lambda}\bigg)^2 -
\frac{\hat R}{6}\bigl(\sigma_0^2-\hat\varphi^2\bigr)\bigg]\,d^4 x\,.
\end{eqnarray}
To prove the functional equivalence of $A$ and $\hat A$, we can first use identity $f \sigma^{-1} D^2 \sigma
= D_\mu(f \sigma^{-1}\partial^\mu\sigma)- \partial_\mu(f \sigma^{-1})\,\partial^\mu\sigma$ in Eq (\ref{RhattoR4}),
and then get rid, by integration, of surface term $\frac{1}{2}\,\partial_\mu \big[\sqrt{-g}\,g^{\mu\nu}(\varphi^2-\sigma^2)
\,e^{-\alpha}\partial_\nu e^\alpha\big]$. Detailed computations are in \S\,6.2 of Part II.

In passing from $A$ to $\hat A$, the conformal--invariance of $A$ ceases to be manifest. The fact is that,
in $\hat A$, it is hidden by the presence of dimensional constants $\sigma^2_0 =\sqrt{6}/\kappa$ and $\mu$.
In compensation, $\hat A$ becomes similar to the action integral of a Higgs field of mass $\mu^2 = \mu^2_H/2$
interacting with gravitation through the terms $-\hat R(x)\,\sigma_0^2/12$ and $\hat R(x)\,\hat\varphi^2(x)/12$.
The first of these terms is formally equal to the Lagrangian density of the standard gravitational action,
as $6/\sigma^2_0$ coincides with gravitational coupling constant $\kappa$. If $\sigma(x)$ were not a ghost
but a physical field, $\kappa$ would be negative and the gravitational forces would therefore be repulsive,
which explains why $\sigma(x)$ must be viewed as a field invested with a geometric meaning.  The second
term represents an additional spacetime--dependent contribution to the mass of $\hat\varphi(x)$, of
negligible magnitude relative to the first.

By functional variation of $\hat A$ with respect to $\tilde\varphi(x)$, we obtain the motion equation
\vspace{-2mm}
\begin{equation}
\label{moteqhatvarphi} \hat D^2\hat\varphi + \lambda\bigg(\hat\varphi^2 -
\frac{\mu^2}{\lambda}\bigg)\hat\varphi - \frac{\hat R}{6}\,\hat\varphi=0\,,
\vspace{-2mm}
\end{equation}
where  $\hat D^2$ is the Beltrami--d'Alembert operator in conformal--time coordinates,
already described in detail in \S~6.2 of Part II. $\hat D^2$ acts on any scalar function
$\hat f(x)$, as follows,
\vspace{-2mm}
\begin{equation}
\label{hatD2fsmall}
 \hat D^2 \hat f \equiv \frac{1}{\sqrt{-\hat g}}\,
\partial_\mu\big(\sqrt{-\hat g}\,\hat g^{\mu\nu}\partial_\nu \hat f\,\big)= e^{-2\alpha}\big[\partial_\mu (g^{\mu\nu}\partial_\nu \hat f) +
\big(2\,\partial_\mu\alpha+ \partial_\mu\ln\sqrt{-g}\,\big)\partial^\mu \hat f\,\big]\,.
\vspace{-2mm}
\end{equation}
\newpage
By functional variation of $\hat A$ with respect to $\tilde g^{\mu\nu}(x)$, we obtain gravitational equation
\vspace{-2mm}
\begin{eqnarray}
\label{hatTetamunu} \hat \Theta_{\mu\nu} & = & \big(\partial_\mu\hat\varphi\big)
\partial_\nu\hat\varphi- \frac{1}{2}\hat g_{\mu\nu}\,
\hat g^{\sigma\rho} \big(\partial_\sigma\hat\varphi\big)\partial_\rho\hat\varphi
+ \hat g_{\mu\nu}\frac{\lambda}{4}
\bigg(\hat \varphi^2- \frac{\mu^2}{\lambda}\bigg)^2 +\nonumber\\
& & \frac{1}{6}\bigl(\hat g_{\mu\nu} \hat D^2 -
\hat D_\mu \hat D_\nu\bigr)\, \hat \varphi^2 -\frac{\sigma_0^2 -
\hat\varphi^2}{6}\bigg(\hat R_{\mu\nu}- \frac{1}{2}\,\hat g_{\mu\nu}\hat R\bigg)=0\,,
\vspace{-3mm}
\end{eqnarray}
where $\hat D_\nu$ are the covariant hyperbolic polar derivatives defined by
\vspace{-2mm}
$$
\hat D_\nu \hat f(x) = \partial_\nu \hat f(x),\quad \hat D_\mu \hat f_\nu(x)  =
\partial_\nu f_\nu(x) - \hat \Gamma_{\mu\nu}^\lambda (x) \hat f_\lambda(x),
\vspace{-2mm}
$$
where $\hat f_\nu(x)$ is any covariant spacetime vector and  $\hat \Gamma_{\mu\nu}^\lambda(x)$
are the Christoffel symbols constructed from $\hat g_{\mu\nu}(x)$.

Contracting Eq (\ref{hatTetamunu}) with $\hat g^{\mu\nu}(x)$, using the identity $\hat D^2\hat\varphi^2
\equiv 2\,\hat g^{\rho\sigma}\bigl(\partial_\rho\hat\varphi\bigr)\,\partial_\sigma\hat\varphi
+ 2\,\hat\varphi\hat D^2\hat\varphi$ and motion equation (\ref{moteqtildevarphi}),  we obtain the
EM--tensor trace equation
\begin{equation}
\label{hattrace}
\hat R = \frac{6\,\mu^2}{\sigma_0^2}\bigg(\hat \varphi^2- \frac{\mu^2}{\lambda}\bigg)\equiv \kappa\,
\mu^2 \bigg(\hat \varphi^2- \frac{\mu^2}{\lambda}\bigg)\,.
\end{equation}
Using Eq (\ref{RhattoR4}) and putting $\hat \varphi(x) = s(x)^{-1} \varphi(x)$, we can rearrange this
equation as
\vspace{-2mm}
$$
D^2 s =\frac{\mu^2}{\sigma_0^2}\,\bigg(\frac{\mu^2}{\lambda}s^2-\varphi^2\bigg) s +\frac{R}{6}\, s\,,
\vspace{-2mm}
$$
which is nothing else but the motion equation for $s$ already given by Eq (\ref{simpSMoteqOnRiem}).

Inserting  Eq (\ref{hattrace}) into Eq (\ref{moteqhatvarphi}), we obtain
\vspace{-2mm}
\begin{equation}
\label{moteqhatvarphi2}
\hat D^2\hat \varphi + \bigg(\lambda -  \frac{\mu^2}{\sigma_0^2}\bigg)
\bigg(\hat\varphi^2 - \frac{\mu^2}{\lambda}\bigg)\,\hat\varphi=0\,,
\vspace{-2mm}
\end{equation}
showing that the dependence of Eq (\ref{moteqhatvarphi}) on $\hat R$ results in the self--coupling constant change
$\lambda\rightarrow \hat\lambda= \lambda-\mu^2/\sigma_0^2$, which is absolutely negligible,
since $\hat\lambda\simeq \lambda(1- 10^{-33})$. Therefore, assuming as initial conditions
$\partial_\tau\hat\varphi=0$ and $\hat\varphi$ very close to zero, Eq (\ref{moteqhatvarphi2})
describes the fall of Higgs field amplitude $\hat\varphi(x)$ into a potential well of depth
$\simeq\mu/\sqrt{\lambda}$.

All equations so far considered can be considerably simplified if we assume that the Higgs field is grounded in the
Cartan manifold of a flat or accelerated Milne universe $\hat M^+$. This means that $\hat\varphi$ and $\hat R$ depend
only on $\tau$, that $R$ in Eq (\ref{RhattoR4}) is zero or a negative constant, and that $R_{\mu\nu}= g_{\mu\nu} R/4$ in Eq
(\ref{RmunutohatRmunu4}). In these conditions, we have $\sqrt{-\hat g(x)} \equiv \sqrt{-\hat g(\tau, \vec\rho\,)}
= e^{4\alpha(t)} c(t)^3(\sinh \rho)^2\sin\theta$ and Eq  (\ref{hatActint}) simplifies to
\vspace{-2mm}
\begin{equation}
\label{simplehatActint}
\hat A = \int_0^{+\infty}\!\!\!\!d\tau \int_{\Omega}\!\!\frac{e^{4\alpha} c^3}{2}
\bigg[\hat g^{\mu\nu}\big(\partial_\mu\hat\varphi\big)\hat\partial_\nu\hat\varphi
-\frac{\lambda}{2}\bigg(\hat\varphi^2 - \frac{\mu^2_H}{2\lambda}\bigg)^2+
\frac{\hat R}{6}\,\big(\hat\varphi^2 - \sigma_0^2\big)\bigg]d\Omega(\vec\rho\,)\,.
\end{equation}

Now, covariant operators $\hat D^2$ and $\hat D_\mu$ act on a scalar function $\hat f(\tau)$, depending
only on $\tau$, as follows: $\hat D_\mu \hat D_\nu \hat f(\tau) = 0\,\,(\mu\neq \nu)$, and
\begin{eqnarray}
\label{simplehatD2}
&&\hspace{-16mm}\hat D^2 \hat f(\tau) = \frac{\partial_\tau\big[s(\tau)^2 c(\tau)^3
\partial_\tau\hat f(\tau)\big]}{s(\tau)^4 c(\tau)^3}=\frac{\partial_\tau^2 \hat f(\tau)}{s(\tau)^2} +
\bigg[\frac{2\,\dot s(\tau)}{s(\tau)^3} +\frac{3\,\dot c(\tau)}{s(\tau)^2c(\tau)}\bigg] \partial_\tau \hat f(\tau)\,;\\
\label{Do2}
&&\hspace{-16mm}\hat D_0\hat D_0 \hat f(\tau) =\partial_\tau^2\hat f(\tau)-
\big[\Gamma^0_{00}(\tau)+ \dot\alpha(\tau)\big]\partial_\tau\hat f(\tau)=\partial_\tau^2\hat f(\tau)- \frac{\dot s(\tau)}{s(\tau)}\partial_\tau\hat f(\tau);\\
\label{D2-Do2}
&&\hspace{-16mm}\big[\hat g_{00}(x)\hat D^2 - \hat D_0 \hat D_0\big]\hat f(\tau) \equiv
\big[s(\tau)^2\hat D^2 - \hat D_\tau \hat D_\tau\big]\hat f(\tau) =
3\bigg[\frac{\dot s(\tau)}{s(\tau)} +\frac{\dot c(\tau)}{c(\tau)}\bigg] \partial_\tau \hat f(\tau)\,;
\end{eqnarray}
since now we have $\hat g_{00}(\tau,\vec\rho\,)=s(\tau)^2$, $\hat g_{0 i}(\tau, \vec\rho\,)=0$,
where $s(\tau)\equiv e^{\alpha(\tau)}$, and $\Gamma^0_{00}(\tau)=0$. Then, as for Eqs (\ref{simpleteta00}) and (\ref{simpletetaij}),
gravitational equations (\ref{hatTetamunu}) condense into the single equation
\begin{equation}
\label{simplehattetamunu}
\hat\Theta_{\tau\tau} = \frac{1}{2}(\partial_\tau\hat \varphi)^2 +\frac{s^2\lambda}{4}\bigg(\hat\varphi^2 -
\frac{\mu^2}{\lambda}\bigg)^2\!+\frac{1}{2}\,\bigg(\frac{\dot s}{s} +\frac{\dot c}{c}\bigg)\partial_\tau\hat\varphi^2
+ \frac{\hat\varphi^2-\sigma_0^2}{6}\,\hat G_{\tau\tau}=0.
\end{equation}

Using in this equation the equalities
\begin{equation}
\label{hatreplacement}
\hat\varphi=\frac{\varphi}{s};\quad \partial_\tau\hat\varphi = \frac{\dot\varphi}{s} - \varphi\frac{\dot s}{s^2};\quad
\partial_\tau\hat\varphi^2 = 2\frac{\varphi\dot\varphi}{s^2} - 2\varphi^2\frac{\dot s}{s^3};\quad
\hat G_{\tau\tau}= -\frac{R}{4} + 6\frac{\dot s\dot c}{sc} +3\frac{\dot s^2}{s^2};
\end{equation}
we find
\vspace{-2mm}
\begin{equation}
\label{hattheta00equiv}
\hat\Theta_{\tau\tau}(\tau) = \frac{\Theta_{\tau\tau}(\tau)}{s(\tau)^2}=0\,;\quad \hat\Theta_\tau^\tau(\tau)
= \hat g^{\tau\tau}(\tau) \,\hat\Theta_{\tau\tau}(\tau) =\frac{\Theta_\tau^\tau(\tau)}{s(\tau)^4}=0\,;
\end{equation}
where $\hat g^{\tau\tau}(\tau)\equiv \hat g^{00}(x)$, showing the expected equivalence of
Eqs (\ref{simpleteta00}) and (\ref{simplehattetamunu}).

\subsection{The Higgs field in the proper--time picture}
\label{proptimepict}
We introduce here a new set of hyperbolic polar coordinates, which correspond to the proper--time representation
of standard inflationary cosmology. Let us define $d\tilde \tau = e^{\alpha(\tau\!,\,\,\vec\rho\,)}d\tau$ as
the {\em proper--time element} corresponding to conformal--time element $d\tau$. We thus have
\begin{equation}
\label{tautotildetau}
\tilde \tau(x)\equiv \tilde \tau(\tau, \vec\rho\,) = \int_0^\tau e^{\alpha(\bar\tau\!,\,\,\vec\rho\,)}d\bar \tau\,.
\end{equation}
This makes sense because $\tau$ is, by definition, the running parameter of the geodesic stemming from
origin $O$ of the future cone at constant hyperbolic--Euler angles $\vec\rho$.

This relationship between $\tau$ and $\tilde \tau$ at constant $\vec\rho\,$ is always one--to--one
if, as we presume and will prove later, $e^{\alpha(\tau\!,\,\, \vec\rho\,)}$ is a monotonic sigmoid--shaped
function of $\tau$ for any $\vec\rho\,$. For a homogeneous and isotropic spacetime, $\tilde\tau(x)$ depends only on $\tau$.
The set of parameters $\tilde x= \{\tilde \tau, \vec\rho\,\}$ are called {\em proper--time coordinates}
corresponding to kinematic-- or conformal--time coordinates $x= \{\tau, \vec\rho\,\}$. We thus have
$\tilde\partial_\mu =\{\partial_{\tilde\tau}, \partial_\rho, \partial_\theta, \partial_\phi\}$.

Vice versa, expressing $\tau$ as a function of $\{\tilde \tau, \vec\rho\,\}$ and defining
$x(\tilde x)\equiv\{\tau(\tilde x), \vec\rho\,\}$, we can write any function
$\hat f$ grounded in the Cartan manifold $\widehat H^+$ in the form $\tilde f(\tilde x)\equiv \hat f[x(\tilde x)]$. In
particular, putting $\tilde\alpha(\tilde x)=\alpha[x(\tilde x)]$ and $\tilde\gamma_{ij}(\tilde x)
\equiv \hat \gamma_{ij}\big[x(\tilde x)\big]$, we can write squared line--element (\ref{fundtensqline}) and
metric tensor (\ref{conffundtens}), respectively, as
\begin{eqnarray}
\label{tildesqlinel2}
& & \hspace{-8mm} d\tilde s^2  = d\tilde\tau^2 -e^{2\tilde\alpha(\tilde \tau\!,\,\,\vec\rho\,)} \tau(\tilde\tau, \vec\rho\,)^2\,
\tilde\gamma_{ij}(\tilde \tau, \vec\rho\,)\,d\rho^i d\rho^j,\,\, (i,j=1,2,3)\,;\\
\label{tildemettens}
& & \hspace{-8mm} \tilde g_{00}(\tilde\tau, \vec\rho\,) = 1;\quad \tilde g_{0i}(\tilde\tau, \vec\rho\,)=0;\quad
\tilde g_{ij}(\tilde\tau, \vec\rho\,) = - e^{2\tilde\alpha(\tau\!,\,\,\vec\rho\,)}\,\tau(\tilde\tau, \vec\rho\,)^2\,
\tilde \gamma_{ij}(\tilde \tau, \vec\rho\,);
\end{eqnarray}
which are manifestly of hyperbolic polar type. Thus, we also have
\begin{equation}
\label{tildedeter}
\sqrt{-\tilde g(\tilde\tau,\vec\rho\,)} =  e^{3\tilde\alpha(\tau\!,\,\, \vec\rho\,)}\,
\tau(\tilde\tau, \vec\rho\,)^3\,\sqrt{\tilde \gamma(\tilde\tau,\vec\rho\,)}\,,
\end{equation}
where $\tilde \gamma(\tilde x)$ is the determinant of $3\times3$--matrix $\big[\tilde \gamma_{ij}(\tilde \tau, \vec\rho\,)\big]$.

With this change, fundamental tensor $\hat g_{\mu\nu}(x)$ of $\widehat H^+$ becomes the metric tensor of a
Riemann manifold $\widetilde H^+\neq H^+$. Correspondingly, conformal Ricci scalar $\hat R(x)$
becomes Riemannian Ricci scalar $\tilde R(\tilde x)\equiv\hat R[x(\tilde x)]$ constructed out of metric
$\tilde g^{\mu\nu}(\tilde x)$, and action integral $\hat A$ becomes the action integral of the proper--time
picture
\begin{equation}
\label{tildeA}
\tilde A =\int_{\tilde H^+}\!\!\!\frac{\sqrt{-\tilde g}}{2} \bigg[\tilde g^{\mu\nu}\bigl(\tilde
\partial_\mu\tilde\varphi\bigr) \tilde \partial_\nu\tilde\varphi
-\frac{\lambda}{2}\bigg(\tilde \varphi^2 - \frac{\mu^2}{\lambda}\bigg)^2
-\frac{\tilde R}{\kappa}\bigg(1-\frac{\tilde\varphi^2}{\sigma_0^2}\bigg)\bigg]d^4\tilde x\,;
\end{equation}
$\kappa \equiv 6/\sigma_0^2$ is the gravitational coupling constant. Since we expect that, for initial conditions
$0< \tilde\varphi(\tilde x)< \mu/\sqrt{\lambda}$ and $\partial_\mu \tilde\varphi(\tilde x)\big|_{\tau=0}=0$,
the amplitude of $\tilde\varphi(\tilde x)$ oscillates in the interval $[0, \mu/\sqrt{\lambda}]$, we see that
$\tilde\varphi(x)^2/\sigma_0^2$ is absolutely negligible relative to 1.

From now on, all quantities and symbols pertaining to the proper--time picture will be superscripted by a tilde.

As in the conformal--time picture, potential--energy density vanishes at $\tilde \varphi= \mu/\sqrt{\lambda}$
and conformal symmetry appears explicitly broken by dimensional constants $\kappa$ and $\mu^2$. Since $\tilde A$
is derived from $\hat A$ by a simple redefinition of time parameter $\tau$, $\tilde A$ equals $\hat A$ in measure.
However, it is only functionally equivalent to $A$ because $\tilde A -A$ inherits the surface term of $\hat A-A$.
The functional variation of $\tilde A$ with respect to $\tilde\varphi(\tilde x)$ gives the motion equation
\begin{equation}
\label{moteqtildevarphi} \tilde D^2\tilde \varphi(\tilde x) +
\lambda\bigg[\tilde \varphi(\tilde x)^2 - \frac{\mu^2}{\lambda}\bigg]\tilde\varphi(\tilde x)
- \frac{\tilde R(\tilde x)}{6}\,\tilde\varphi(\tilde x)=0\,,
\end{equation}
where
\begin{equation}
\hspace{-2mm}
\tilde D^2\tilde f(\tilde x)\!=\!\partial_{\tilde \tau}^2\tilde f(\tilde x)+\frac{\partial_{\tilde \tau}
\big[\tau(\tilde x)^3 e^{3\tilde\alpha(\tilde x)}\sqrt{\tilde\gamma(\tilde x)}\,\,\big]}
{\tau(\tilde x)^3 e^{3\tilde\alpha(\tilde x)}\sqrt{\tilde\gamma(\tilde x)}}\,
\partial_{\tilde \tau}\tilde f(\tilde x)-\frac{\partial_i\big[\tau(\tilde x) e^{2\tilde\alpha(\tilde x)}
\sqrt{\tilde \gamma(\tilde x)}\,\,\tilde\gamma^{ij}(\tilde x)\,\partial_j \tilde f(\tilde x)\big]}
{\tau(\tilde x)^3 e^{4\tilde\alpha(\tilde x)}\sqrt{\tilde\gamma(\tilde x)}},\nonumber
\end{equation}
with $(i,j=1,2,3)$, is the Beltrami--d'Alembert operator constructed from $\tilde g^{\mu\nu}(\tilde x)$.

The functional variation of $\tilde A$ with respect to $\tilde g^{\mu\nu}(\tilde x)$ gives the gravitational equation
\begin{eqnarray}
\label{tildeTetamunu}
&&\hspace{-18mm}\tilde \Theta_{\mu\nu} = \bigl(\tilde
\partial_\mu\tilde\varphi\bigr)\tilde \partial_\nu\tilde\varphi- \frac{\tilde g_{\mu\nu}}{2}\,
\tilde g^{\rho\sigma}\bigl(\tilde \partial_\rho\,\tilde\varphi\bigr)
\tilde \partial_\sigma\,\tilde\varphi + \tilde g_{\mu\nu}\frac{\lambda}{4} \bigg(\tilde \varphi^2- \frac{\mu^2}{\lambda}\bigg)^2 +\nonumber\\
&& \hspace{-6mm}\frac{1}{6}\bigl(\tilde g_{\mu\nu}\tilde D^2 - \tilde
D_\mu \tilde \partial_\nu\bigr)\tilde \varphi^2 -
\frac{\sigma_0^2- \tilde\varphi^2}{6}\,\bigg(\tilde R_{\mu\nu} - \frac{1}{2}\,\tilde
g_{\mu\nu}\tilde R\bigg) =0,\quad \mbox{where} \\
&& \hspace{-18mm} \big(\tilde g_{\mu\nu}\tilde D^2 - \tilde D_\mu\tilde\partial_\nu\big)\tilde \varphi^2 =
\tilde g_{\mu\nu}\bigg\{\partial_{\tilde \tau}^2+\partial_{\tilde \tau}\big[\ln\big(\tau^3 e^{3\tilde\alpha}
\sqrt{\tilde\gamma}\big)\partial_{\tilde \tau}\big]-
\frac{\partial_i\big(\tau e^{2\tilde\alpha}\sqrt{\tilde \gamma}\,\tilde\gamma^{ij}\partial_j\big)}
{\tau^3 e^{4\tilde\alpha}\sqrt{\tilde\gamma}}\bigg\}\tilde \varphi^2 - \nonumber\\
&&\hspace{19mm}\Big[\tilde\partial_\mu\tilde \partial_{\nu} - \tilde \Gamma_{\mu\nu}^\rho\tilde \partial_\rho\Big]\tilde \varphi^2
\end{eqnarray}
Since $\tilde R_{\mu\nu}$ and $\tilde R$ are related to the kinematic time picture via Eqs
(\ref{RmunutohatRmunu4}) (\ref{RhattoR4}), we obtain
\vspace{-2mm}
\begin{equation}
\label{tildeGmunu}
\tilde R_{\mu\nu} =R_{\mu\nu} + \frac{4\,(\partial_\mu s)\,\partial_\nu s - g_{\mu\nu} (\partial^\rho s)\,\partial_\rho s}{s^2}-
\frac{2\,D_\mu\partial_\nu s+ g_{\mu\nu} D^2 s}{s};\quad \tilde R = \tilde g^{\mu\nu} \tilde R_{\mu\nu}\,.
\vspace{-2mm}
\end{equation}

Contracting Eq (\ref{tildeTetamunu}) with $\tilde g^{\mu\nu}$, and then using identity $\tilde D^2\tilde \varphi^2
\equiv 2\,\tilde g^{\rho\sigma}\bigl(\tilde \partial_\rho\tilde\varphi\bigr) \bigl(\tilde \partial_\sigma\tilde\varphi\bigr)
+ 2\,\tilde\varphi\tilde D^2\tilde\varphi$ and motion equation (\ref{moteqtildevarphi}), we find the trace equation
\begin{equation}
\label{tildetraces}
\tilde \Theta(\tilde x) = \mu^2\bigg[\frac{\mu^2}{\lambda}-\tilde \varphi(\tilde x)^2\bigg]+
\frac{6}{\sigma_0^2}\,\tilde R(\tilde x) =0\,;\quad \hbox{then, } \tilde R(\tilde x) =
\frac{6 \mu^2}{\sigma_0^2}\bigg[\tilde \varphi(\tilde x)^2- \frac{\mu^2}{\lambda}\bigg]\,,
\end{equation}
the latter of which is equivalent to the motion equation for $\tilde s$, as was the case for $\hat R(x)$
in Eq (\ref{hattrace}). Inserting the last of these equations into Eq (\ref{moteqtildevarphi}),
we obtain, as we had already for Eq (\ref{moteqhatvarphi2}),
\begin{equation}
\label{moteqtildevarphi2}
\tilde D^2\tilde \varphi(\tilde x) + \bigg(\lambda -  \frac{\mu^2}{\sigma_0^2}\bigg)
\bigg[\tilde\varphi(\tilde x)^2 - \frac{\mu^2}{\lambda}\bigg]\,\hat\varphi(\tilde x)=0\,,
\end{equation}
showing that, due to the frictional term of $\tilde D^2\tilde\varphi(\tilde x)$,
$\tilde\varphi(\tilde x)$ tends to converge to $\mu/\sqrt{\lambda}$.

All equations can be considerably simplified if we assume the homogeneity and isotropy of the universe, which implies
that $\tilde \varphi$ depends only on $\tilde\tau$. In this case, metric tensor (\ref{tildemettens}) specializes into
\begin{equation}
\label{tildemetrictensor}
\tilde g_{\mu\nu}(\tilde \tau,\vec\rho\,) = \hbox{diag}\Bigl[1, - \tilde a(\tilde \tau)^2,
-\tilde a(\tilde \tau)^2\!\sinh\varrho^2,-\tilde a(\tilde \tau)^2\!\sinh\varrho^2 \sin\theta^2\Bigr],
\end{equation}
where $\tilde a(\tilde \tau) = \tilde c(\tilde \tau)\, \tilde s(\tilde \tau)$, with $\tilde c(\tilde \tau)$
and $\tilde s(\tilde \tau)=e^{\tilde\alpha(\tilde \tau)}$, respectively  obtained from $c(x)$
and $s(\tau)\equiv e^{\alpha(\tau)}$, introduced as in \S\,\ref{higgsOnRiem} with substitution rule
$f(\tau) \rightarrow \tilde f(\tilde \tau) \equiv f[\tau(\tilde \tau)]$. Thus, we have $\sqrt{-\tilde g(\tilde \tau,\vec\rho\,)} =
\big[\tilde c(\tilde \tau)\, e^{\tilde\alpha(\tilde \tau)}\big]^3(\sinh\varrho)^2\sin\theta$ and $d^4\tilde x=
\big[\tilde c(\tilde \tau)\,e^{\tilde\alpha(\tilde \tau)}\big]^3 d\Omega(\vec\rho)\,d\tilde\tau$.

\newpage
Since $e^{\tilde\alpha(\tilde \tau)}$ has a sigmoidal profile, the manifold has the form of the
inflated--accelerated Milne universe  $\widetilde M^+$  described in \S\,6.3 of Part II and
represented in Fig.\,3.
\begin{figure}[!h]
\mbox{%
\begin{minipage}{.47\textwidth}
\includegraphics[scale=0.38]{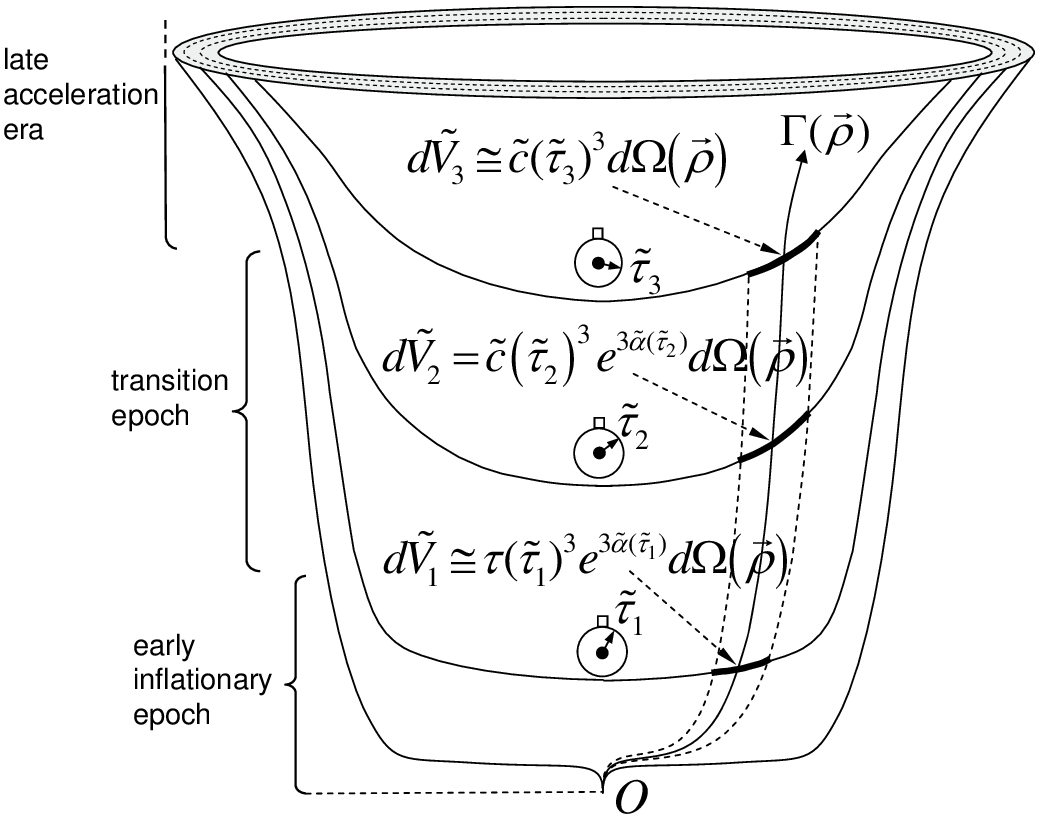}
\end{minipage}%
\begin{minipage}[c]{0.5\textwidth}
\caption{\small Qualitative features of an accelerated--inflated Milne universe $\widetilde M^+$
in proper--time coordinates $\tilde x=\{\tilde\tau, \vec\rho\,\}$; $\tilde c(\tilde\tau)$ = acceleration factor;
$e^{\tilde \alpha(\tilde \tau)}$ = inflationary sigmoid--shaped scale--factor;
$d\tilde V_i=\tilde c(\tilde\tau_i)\,e^{3\tilde\alpha(\tilde \tau_i)} d\Omega(\tilde\rho\,)$, $(i=1,2,3)$
= 3D sections of a worldline tube wrapped around polar geodesic $\Gamma(\vec \rho)$ stemming from
future--cone origin $O$ with direction $\vec\rho$. Note flattening of spacelike surfaces
in early inflationary epoch.}
\end{minipage}%
}
\end{figure}

The motion equations for $\tilde\varphi(\tilde\tau)$ can be directly obtained from Eqs
(\ref{moteqhatvarphi2}) by putting $\partial_{\tilde \tau}\tilde f(\tilde\tau)
= e^{-\tilde\alpha(\tilde\tau)}\big[\partial_\tau f(\tau)\big]_{\tau = \tau(\tilde\tau)}  \equiv
\tilde s(\tilde \tau)^{-1}\big[\partial_\tau f(\tau)\big]_{\tau = \tau(\tilde\tau)} $,
which yields
\begin{equation}
\label{tildevarphiMoteq}
\tilde D^2 \tilde \varphi(\tilde\tau)\equiv
\partial^2_{\tilde\tau}\tilde\varphi(\tilde\tau)\! +\! 3\bigg[\frac{\partial_{\tilde\tau}\tilde c(\tilde\tau)}{\tilde c(\tilde\tau)}\!+\!\frac{\partial_{\tilde\tau}\tilde s(\tilde\tau)}{\tilde s(\tilde\tau)}\bigg]
\partial_{\tilde\tau}\tilde\varphi(\tilde\tau) = \bigg(\lambda\!-\!\frac{\mu^2}{\sigma_0^2}\bigg)
\bigg[\frac{\mu^2}{\lambda}\!-\!\tilde\varphi(\tilde\tau)^2\bigg]\tilde\varphi(\tilde\tau),\\
\end{equation}

In the same way, from Eq (\ref{simplehattetamunu}), and in consideration of Eq (\ref{hattheta00equiv}), we obtain
\begin{equation}
\label{CartEinst}
\tilde\Theta^\tau_\tau= \frac{1}{2}\bigl(\partial_{\tilde\tau}\tilde\varphi\bigr)^2 +\frac{\lambda}{4} \bigg(\tilde \varphi^2
- \frac{\mu^2}{\lambda}\bigg)^2 + \frac{1}{2}\bigg(\frac{\partial_{\tilde\tau}\tilde c}{\tilde c} +
\frac{\partial_{\tilde\tau}\tilde s}{\tilde s}\bigg)\partial_{\tilde \tau}\tilde\varphi^2
-\frac{1}{\kappa}\bigg(1- \frac{\tilde\varphi^2}{\sigma_0^2}\bigg) \tilde G^\tau_\tau =0\,,
\end{equation}
where $\sigma_0^2/6 =1/\kappa$ and, using Eqs (\ref{RmunutohatRmunu4}), (\ref{RhattoR4}) and (\ref{simpSMoteqOnRiem}) with
$s(\tau)=e^{\alpha(\tau)}$,
\begin{equation}
\label{tildeGtautautoRs}
\tilde G^\tau_\tau \equiv \tilde R^{\tilde\tau}_{\tilde \tau} - \frac{1}{2}\tilde R = - \frac{R}{4\,s^2} + 3\,\frac{(\partial_\tau s)^2}{s^4}
+ 6\,\frac{\partial_\tau s}{s^3}\,\frac{\partial_\tau c}{c} =  - \frac{R}{4\,\tilde s^2} +
3\,\bigg(\frac{\partial_{\tilde \tau} \tilde s}{\tilde s}\bigg)^2+
6\,\frac{\partial_{\tilde\tau}\tilde s}{\tilde s}\,\frac{\partial_{\tilde \tau} \tilde c}{\tilde c}.
\end{equation}
where, in the last step, $s(\tau)$ has been replaced by $\tilde s(\tilde \tau)$, $c(\tau)$ by $\tilde c(\tilde\tau)$ and $\partial_\tau$ by $\tilde s\, \partial_{\tilde \tau}$.

\subsection{Conservative quantities}
\label{tildenconserv}
The proper--time picture is particularly appropriate for dealing with conservative properties.
Let $\tilde j_\mu(\tilde x)$ be a conservative vector current in the proper--time picture and  $\tilde j^\mu(\tilde x)$
its contravariant representation. Since $\tilde g_{\mu\nu}(\tilde x)$ is hyperbolic polar and Eqs (\ref{tildemettens}) (\ref{tildedeter}) hold,
the continuity equation of $\tilde j_\mu(\tilde x)$  has the form
\begin{equation}
\label{tildeDmutildejmu}
\tilde D_\mu \tilde j^\mu(\tilde\tau, \vec\rho\,) = \frac{\tilde\partial_\mu\big[\sqrt{-\tilde g(\tilde x)}\,
\tilde j^\mu(\tilde x)\big]}{\sqrt{-\tilde g(\tilde x)}} =
\frac{\tilde\partial_\mu\big\{\big[e^{\tilde\alpha(\tilde \tau, \vec\rho\,)}\tau(\tilde\tau, \vec\rho\,)\big]^3
\!\sqrt{\tilde\gamma(\tilde\tau, \vec\rho\,)}\,\tilde j^\mu(\tilde\tau,\vec\rho\,)\big\}}
{\big[e^{\tilde\alpha(\tilde \tau, \vec\rho\,)}\tau(\tilde\tau, \vec\rho\,)\big]^3\!\sqrt{\tilde\gamma(\tilde\tau, \vec\rho\,)}}=0.
\end{equation}

The spacetime integral of this equation from proper time $\tilde\tau_1$ to proper time $\tilde\tau_2$ is
\begin{eqnarray}
\label{tildeDeltaI)}
\hspace{-10mm}\tilde I(\tilde\tau_2)- \tilde I(\tilde\tau_1)
&= & \int_{\tilde\tau_1}^{\tilde\tau_2}\!\!\!\! d\tilde\tau \!\!\int_\Omega\!\!
\sqrt{-\tilde g(\tilde x)}\,\tilde D_\mu\, \tilde j^\mu(\tilde\tau, \vec\rho\,) d\Omega(\vec\rho\,) \equiv\nonumber \\
\hspace{-10mm}& & \int_{\tilde\tau_1}^{\tilde\tau_2}\!\!\!\!d\tilde\tau \!\!\int_\Omega\!\!\tilde\partial_\mu
\Big\{\big[e^{\tilde\alpha(\tilde \tau, \vec\rho\,)}\tau(\tilde\tau, \vec\rho\,)\big]^3\sqrt{\tilde\gamma(\tilde\tau, \vec\rho\,)}\,\tilde j^\mu(\tilde\tau,\vec\rho\,)\Big\}d\Omega(\vec\rho\,) =0.
\end{eqnarray}

Averaging over $\Omega$ and getting rid of the surface term, we can define the quantities
\begin{eqnarray}
\label{averenedens2}
& & \langle \tilde a(\tilde\tau)\rangle^3\, \tilde j_0(\tilde\tau)\rangle  =
\lim_{\Omega\rightarrow \infty} \frac{1}{\Omega}\!\int_{\Omega}\!\!\big[e^{\tilde\alpha(\tilde \tau, \vec\rho\,)}
\tau(\tilde\tau, \vec\rho\,)\big]^3\sqrt{\tilde\gamma(\tilde\tau, \vec\rho\,)}\,
\tilde j_0(\tilde\tau,\vec\rho\,)\,d\Omega(\vec\rho\,),\nonumber\\
& & \langle\,\tilde j_0(\tilde\tau)\rangle  = \lim_{\Omega\rightarrow \infty}\!\frac{1}{\Omega}\int_{\Omega}
\!\tilde j_0(\tilde\tau,\vec\rho\,)\,(\sinh\varrho)^2\sin\theta\, d\Omega(\vec\rho\,),\nonumber
\nonumber
\end{eqnarray}
where $\tilde{j}_0(\tilde \tau, \vec\rho\,) = \tilde{j}_0(\tilde\tau, \vec\rho\,)$ is the charge density
in the 3D hyperboloid $\widetilde H(\tilde \tau)$ and $\langle \tilde a(\tilde\tau)\rangle$ plays the
role of a mean scale factor. From this obtain and from Eq (\ref{tildeDeltaI)}) we can obtain equality
$\tilde I(\tilde\tau_2)= \tilde I(\tilde\tau_1)$, and hence equation $\langle \tilde a(\tilde\tau_2)\rangle^3
\langle\tilde j_0(\tilde\tau_2)\rangle = \langle \tilde a(\tilde\tau_1)\rangle^3\langle\tilde j_0(\tilde\tau_1)\rangle$.

Since the universe on the large scale approaches inflated--accelerated Milne universe $\widetilde M^+$, in which
$\big[e^{\tilde\alpha(\tilde\tau)}\tau(\tilde\tau, \vec\rho\,)\big]^3\!\sqrt{\tilde\gamma(\tilde\tau, \vec\rho\,)} =
\big[e^{\tilde\alpha(\tilde\tau)}\tilde c(\tilde\tau)\big]^3 (\sinh\varrho)^2\sin\theta$ holds, we expect that
equations $\langle \tilde a(\tilde\tau)\rangle = e^{\tilde\alpha(\tilde\tau)} \tilde c(\tilde\tau)$ and therefore
\begin{equation}
\label{currcons}
\frac{\langle\tilde j_0(\tilde \tau_1)\rangle}{\langle\tilde j_0(\tilde \tau_2)\rangle} =
\frac{\langle\tilde a (\tilde \tau_2)\rangle^3}{\langle\tilde a (\tilde \tau_1)\rangle^3} =
\frac{[e^{\tilde\alpha(\tilde\tau_2)} \,\tilde c(\tilde\tau_2)]^3} {[e^{\tilde\alpha(\tilde\tau_1)}
\,\tilde c(\tilde\tau_1)]^3}=\frac{d\tilde V(\tilde\tau_2, \vec\rho\,)}{d\tilde V(\tilde\tau_1,\vec\rho\,)}\,,
\end{equation}
hold to a very good approximation in the largescale. Here, $d\tilde V(\tilde\tau_i,\vec\rho\,)\equiv \sqrt{-\tilde g(\tilde\tau_i)}\,d\Omega(\vec\rho\,) =
\big[e^{\tilde\alpha(\tilde\tau_i)}\,\tilde c(\tilde\tau_i)\big]^3(\sinh \varrho)^2\,\sin\theta\,
d\Omega(\vec\rho\,)$ are the volume elements of the spacelike hyperboloids at proper--time $\tilde\tau_i$
exemplified in Fig.\,3.

Let $\tilde\tau_0$ be a proper time of the early inflationary epoch, $\tilde \tau$ a proper--time of the
post--inflationary era, and $\tau_0$, $\tau$ their respective kinematic--time counterparts. Then,
on account of Eqs (\ref{atau}), we have $\tilde c(\tilde\tau_0) =\tau_\Lambda\sinh
[\tau(\tilde\tau_0)/\tau_\Lambda]$ and $e^{\tilde\alpha(\tilde\tau_0)}\ll 1$ in the first case,
and $\tilde c(\tilde\tau) =\tau_\Lambda\sinh [\tau(\tilde\tau)/\tau_\Lambda]$ and
$e^{\tilde\alpha(\tilde\tau)} \le 1$ in the second. Eq (\ref{currcons}) therefore becomes
\begin{equation}
\label{longcurrcons}
\bigg[\frac{\langle \tilde j_0(\tilde\tau)\rangle}{\langle \tilde j_0(\tilde \tau_0)\rangle}\bigg]^{1/3}
\simeq \frac{\tilde c(\tilde\tau_0)\,e^{\tilde\alpha(\tilde\tau_0)}}{\tilde c(\tilde\tau)\,e^{\tilde\alpha(\tilde\tau)}} =
\frac{\sinh (\tau_0/\tau_\Lambda)\,e^{\alpha(\tau_0)}}{\sinh (\tau/\tau_\Lambda)\,e^{\alpha(\tau)}}
= \bigg[\frac{dV(\tau_0,\vec\rho\,)}{dV(\tau,\vec\rho\,)}\bigg]^{1/3},
\end{equation}
where, for the sake of simplicity, we passed from the proper--time picture to the
kinematic--time picture in the second step.

This equation will be profitably used in \S\,\ref{caveats} and \S\,\ref{entropycourse} to determine
the time course of mean entropy density because, during the early inflationary epoch, on any scale,
and in all subsequent epochs on the large scale, expansion is adiabatic and adiathermic to a
very good approximation.

\subsection{False--vacuum to true--vacuum decay of the Higgs field}
\label{VEVs}
We have so far regarded $A$, $\hat A$ and $\tilde A$ as classical action integrals of functionally
equivalent representations. In this section, although retaining classical notation,
we move to the quantum--theoretical domain by reinterpreting all classical fields as quantum fields and
their amplitudes as their expectation values in the vacuum state of the universe at the moment
of the spontaneous breakdown of conformal symmetry. The reason for doing this is the following.

In the Heisenberg picture, the physical state of a system remains fixed during the evolution of the system,
whereas the observables evolve in time. Consistent with this view, the vacuum state $\vert\Omega\rangle$ of the
universe at the moment of the spontaneous breakdown of conformal symmetry remains the physical state
of the expanding universe for all subsequent times. This means that any local quantity of physical
interest is the $\vert\Omega\rangle$--VEV of a local operator or a suitably regularized product of local operators. As we shall
see later, the production of matter is a sudden explosion of Higgs--field amplitude, which occurs at a critical
proper time $\tilde\tau_c$ after the instant $\tilde\tau_0=0$ of the spontaneous breaking. During interval
$0<\tilde \tau<\tilde \tau_c$, evolution is governed by a unitary transformation; but, from $\tilde\tau_c$ on,
the evolution can be only described as an irreversible thermodynamic process governed by a time--dependent
Bogoliubov transformation of the observables \cite{UMEZAWA2}, the continuity of the entire process
being thus ensured by the maintenance of $\vert\Omega\rangle$ as the state of the system. However,
the representation of this process is not given in the domain of separable Hilbert spaces,
but in the still scarcely investigated domain of non--separable unitary spaces \cite{KIBBLE}.

Let us now describe the transition from the classical to the quantum--field representation in the framework of
the proper--time picture, which is that of the three pictures which is invested with direct physical meaning.
As is evident from Eq (\ref{moteqtildevarphi2}), which we can rewrite as
$$
\tilde D^2\tilde \varphi(\tilde x) + \tilde\lambda
\bigg[\tilde\varphi(\tilde x)^2 - \frac{\mu^2}{\lambda}\bigg]\,\hat\varphi(\tilde x)=0,\,\,\,\mbox{where } \,
\tilde\lambda = \lambda -  \frac{\mu^2}{\sigma_0^2} \simeq \lambda\,,
$$
$\tilde\varphi(\tilde x)$ eventually tends to converge to its minimum amplitude $\mu/\sqrt{\lambda}$.
Since in quantum field theory particles are quantum excitations of a fundamental state,
it is more appropriate to represent the Higgs--boson field as the difference
$\tilde\eta(\tilde x)=\tilde \varphi(\tilde x)-\mu/\sqrt{\lambda}$ obeying the motion equation
$$
\tilde D^2\tilde\eta(\tilde x) + \mu^2_H\tilde\eta(\tilde x) +\tilde\lambda \sqrt{2}\,
\mu_H\,\tilde\eta^2(\tilde x) + \tilde\lambda\,\tilde\eta^3(\tilde x)= 0\,,
$$
where $\mu_H = \sqrt{2}\mu (\tilde\lambda/\lambda) \simeq \sqrt{2}\mu$ is the mass
of the Higgs boson.

It may seem that, in the proper--time picture, the motion equation of the dilation field has disappeared.
Actually it has not, as it is already contained in the gravitational--trace equation of the proper--time picture.
To prove this, it is sufficient to replace in the second of Eqs (\ref{tildetraces})
the following expression for $\tilde R(\tilde x)$ as a function of $x$
$$
\tilde R[\tilde x(x)] \equiv \hat R(x) = e^{-2\alpha(x)} \big[R(x) - 6\, e^{-\alpha(x)} D^2  e^{\alpha(x)}\big]=
e^{-2\alpha(x)}\frac{6\mu^2}{\sigma_0^2} \bigg[\varphi(x)^2- \frac{\mu^2}{\lambda}\frac{\sigma(x)^2}{\sigma_0^2}\bigg].
$$
This can easily be verified by replacing $s(x)\equiv e^{\alpha(x)}$ with $\sigma(x)/\sigma_0$ in
Eq (\ref{RhattoR4}), which in fact gives exactly the motion equation (\ref{sigmamoeteq}) of the dilation
field in the kinematic--time picture rewritten as
\begin{equation}
\label{D2sofx}
D^2 s(x) +\frac{\mu^2}{\sigma_0^2}\,\bigg[\varphi(x)^2 -\frac{\mu^2}{\lambda}\,s(x)^2\bigg]s(x)
- \frac{R(x)}{6}\,s(x)=0\,.
\end{equation}
Conversely, passing to the proper--time picture, this equation can be rearranged in the form of
trace equation
$$
\tilde R(\tilde x)= \frac{6\,\mu^2}{\sigma_0^2}\,\bigg[\tilde \varphi(\tilde x)^2
-\frac{\mu^2}{\lambda}\bigg]\,,
$$
as given by the second of Eqs (\ref{tildetraces}), since the inequality $\tilde s(\tilde x)>0$ always holds.

After the inflationary epoch, $\langle \Omega |\tilde \eta(\tilde x)|\Omega \rangle$  converges to zero,
$s(x)$ to 1, CGR to GR, the proper--time picture to the kinematic--time picture, and
therefore the asymptotic motion equation for the true Higgs field becomes
$D^2\tilde\eta(x) + 2 \mu^2\tilde\eta(x) +\sqrt{\lambda}\,\mu\,\tilde\eta^2(x) + \lambda\, \tilde\eta^3(x)= 0$.

In the absence of interactions with other fields, this equation describes a gas of  mutually repelling scalar
bosons of mass $\mu_H=\sqrt{2}\,\mu$. For small $\tilde\eta(x)$, the equation simplifies to the free Higgs--boson
field equation
\begin{equation}
\label{etamoteq}
D^2\tilde\eta(x) + \mu^2_H\,\tilde\eta(x)= 0\,.
\end{equation}

Similarly, the asymptotic motion equation for $\sigma(\tau)$ in the kinematic--time picture is
\begin{equation}
\label{ximoteq}
D^2 \sigma(\tau) - \frac{R}{6}\,\sigma(\tau)= 0 \quad (R<0);
\end{equation}
the solution to which is $\sigma(\tau) = \sigma_0- A\, e^{-\tau\sqrt{|R|/6}}$, with $A>0$,
since $\sigma(\tau)$ must converge to $\sigma_0$ from below for $\tau \rightarrow \infty$
(see next section for details).

\section{Higgs--field dynamics in the accelerated Milne universe}
\label{Higgsdyn}
In this section, the motion equations of the Higgs and dilation fields in a Milne universe of constant
curvature $R<0$, already introduced in \S\,\ref{higgsOnRiem} for the kinematic--time picture and in \S\,\ref{higgsOnRiem}
for the proper--time picture, is solved in the semiclassical approximation.

In the kinematic--time picture and in the absence of interactions with other fields, the time courses of $\varphi(\tau, \vec\rho\,)$
and $\sigma(\tau, \vec\rho\,) \equiv \sigma_0 s(\tau, \vec\rho\,)$ are ruled by Eqs (\ref{varphimoeteq}) and (\ref{sigmamoeteq}).
Assuming that the universe is homogeneous and isotropic on each hyperboloid of the Milne spacetime, the motion equations simplify
to Eqs (\ref{simpPhiMoteqOnRiem}) and (\ref{simpSMoteqOnRiem}), which, for convenience, we rewrite in the form
\begin{eqnarray}
\label{varphieq}
&&\hspace{-14mm}\ddot\varphi(\tau) + 3\frac{\dot c(\tau)}{c(\tau)}
\dot\varphi(\tau)=\lambda\bigg[\frac{\mu^2}{\lambda}s(\tau)^2-\varphi(\tau)^2\bigg]\varphi(\tau) +
\frac{R}{6}\varphi(\tau),\quad \varphi(\tau)>0;\\
\label{seq}
&&\hspace{-14mm}\ddot s(\tau) +  3\frac{\dot c(\tau)}{c(\tau)}
\dot s(\tau)=\frac{\mu^2}{\sigma_0^2}\bigg[\frac{\mu^2}{\lambda}s(\tau)^2-
\varphi(\tau)^2\bigg]s(\tau)+ \frac{R}{6} s(\tau),\quad s(\tau)>0\,.
\end{eqnarray}

Here, $\sigma_0$, $\mu$ and $\lambda$ are defined as at the beginning of \S\,\ref{HiggsinCGR},
$c(\tau) = \tau_\Lambda \sinh (\tau/\tau_\Lambda)$, with $\tau_\Lambda$ as Hubble time, is the
scale factor of the accelerated Milne universe $M^+$, and $R\simeq - 1.73\times 10^{-83}$ GeV$^2$
is spacetime curvature, as described in \S\,\ref{Curvedhyperbspacetime}.

As initial conditions of the above equations, we must assume that $\dot\varphi(0)=\dot s(0)=0$, since
otherwise the second terms on the left--hand sides would diverge at $\tau=0$. In the
semiclassical approximation considered here, we cannot assume that $\varphi(0)=0$ and/or $s(0)=0$,
since otherwise the solutions $\varphi(\tau)=0$ and /or $s(\tau)=0$ would follow. In the lack of a
quantum--theoretical refinement of the theory, we assume $\varphi(0)$ to be very close to zero,
so as to approach as nearly as possible the initial state of the absence of matter, but with
$\varphi(0)\ll \mu\,s(0)/\sqrt{\lambda}$, so as to let $s(\tau)$ become sufficiently
large before it starts driving $\varphi(\tau)$.  We therefore expect that $\varphi(\tau)$ will
take a long time to reach the value $\mu\,s(\tau)/\sqrt{\lambda}$.

A precise estimate of the importance of the terms proportional to $R/6$,  in the stated initial conditions,
is provided by the gravitational equation Eq (\ref{simpleteta00}) at $\tau=0$, i.e.,
\begin{equation}
\label{exactR}
\frac{R}{6}= -\lambda\frac{\big[\mu^2s(0)^2/\lambda -\varphi(0)^2\big]^2}{\sigma_0^2 s(0)^2-\varphi(0)^2}\,.
\end{equation}
Since $\varphi(0)^2$ is assumed to be much smaller than $\mu^2s(0)^2/\lambda$, and hence
negligible with respect to $\sigma_0^2s(0)^2$, Eq (\ref{exactR}) does not differ appreciably from
\begin{equation}
\label{approxR}
\frac{R}{6}\simeq -\frac{\mu^4s(0)^2}{\lambda\,\sigma_0^2} + 2\,\frac{\mu^2\varphi(0)^2}{\sigma_0^2}\,,
\end{equation}
which allows us to predict
\begin{equation}
\label{s0fromR}
s(0) \simeq \sqrt{-\frac{2\lambda}{\mu^2}\bigg[\frac{R\,\sigma_0^2}{12\,\mu^2}- \varphi(0)^2\bigg]}
\simeq \frac{\sigma_0}{\mu^2}\sqrt{-\frac{\lambda\, R}{6}} =
 \frac{\sigma_0\sqrt{2\,\lambda}}{\mu^2\tau_\Lambda}
\simeq  4.62\times 10^{-28},
\end{equation}
where $\tau_\Lambda = 2\sqrt{-3/R}= \sqrt{3/\Lambda}$ is Hubble time in kinematic coordinates.
Thus we have,
\begin{equation}
\label{Hubbletime}
\tau_\Lambda =\frac{\sqrt{2\lambda}\,\sigma_0 }{\mu^2 s(0)}\,.
\end{equation}
As shown in \S\,\ref{cosmconstandscale}, computation of $s(0)$ by a totally independent self--consistent method, 
based on the assumption of entropy conservation, ranges in the interval $[2.20,3.56]\times 10^{-28}$.

As long as $\varphi(\tau)\neq \mu s(\tau)/\sqrt{\lambda}$,  Eq (\ref{varphieq})
is not significantly altered by this value of $R/6$, but unfortunately Eq (\ref{seq}) is.
If this were not so, for $\varphi(\tau)\ll \mu\,s(\tau)/\sqrt{\lambda}$,
Eq (\ref{seq}) would be very well approximated by equation
\vspace{-1mm}
\begin{equation}
\label{centralseq}
\ddot s(\tau) +  \frac{3}{\tau} \dot s(\tau)=\frac{\mu^2}{\sigma_0^2}\bigg[\frac{\mu^2}{\lambda}s(\tau)^2-
\varphi(\tau)^2\bigg]s(\tau)\,,
\vspace{-1mm}
\end{equation}
because $c(\tau)=\tau$ for $R=0$. Actually, for $\tau$ near zero or very large, Eq (\ref{seq}) deviates significantly
from Eq (\ref{centralseq}). In fact, replacing Eq (\ref{approxR}) into Eq (\ref{seq}) gives
\vspace{-1mm}
\begin{equation}
\label{seqaboutzero}
\ddot s(\tau) +  3\frac{\coth(\tau/\tau_\Lambda)}{\tau_\Lambda}
\dot s(\tau)\simeq \frac{\mu^2}{\sigma_0^2}\bigg\{\frac{\mu^2}{\lambda}\big[s(\tau)^2-s(0)^2\big]
-\big[\varphi(\tau)^2-\varphi(0)^2\big] +\varphi(0)^2\bigg\}s(\tau)
\vspace{-1mm}
\end{equation}
to a very good approximation. Near $\tau=0$, and as long as $\varphi(\tau)< \mu\,s(\tau)/\sqrt{\lambda}$,
this equation simplifies to $\ddot s(\tau) \simeq s(\tau)\,\mu^4\varphi(0)^2/4\lambda\sigma_0^2$, since
$\dot s(\tau)\simeq \dot s(0) + \ddot s(0)\,\tau +\dots \simeq \ddot s(\tau)\,\tau$;
whereas, for $s(\tau)\gg s(0)\gg \varphi(0)$ and large $\tau$, when  $\varphi(\tau)\simeq
\mu\,s(\tau)/\sqrt{\lambda}$, it is well approximated by $\ddot s(\tau) +3\,\dot s(\tau)/\tau_\Lambda
=  -\mu^4s(0)^2s(\tau)/\lambda\sigma_0^2 = R\,s(\tau)/6 = -\Lambda\,s(\tau)/24$.

In the first case, for $\tau < \tau_0 = 2\sigma_0\sqrt{\lambda}/\mu\,\varphi(0)$, the solution is
$s(\tau) \simeq s(0)\sinh(\tau/\tau_0)$; thus, $\tau_0$ is  about the time after which $R\,s(\tau)/6$
becomes relatively negligible in the right--hand side of Eq (\ref{seq}).
In the second case, as $\varphi(\tau)$ approaches $\mu s(\tau)^2/\sqrt{\lambda}$ and $\tau\rightarrow \infty$,
the solution with $\dot s(0)=0$ is $s(\tau)= A e^{-\tau/\tau_\Lambda}+B$, hence $\varphi(\infty)= B\mu/\sqrt{\lambda}$.
To ensure the convergence of CGR to GR, we must assume $s(\infty) =1$, hence $B=1$ (cf. \S\,4.1 of Part I).
Between these limiting cases, Eqs (\ref{varphieq}) and (\ref{seq}) are approximated very well by
\vspace{-1mm}
\begin{eqnarray}
\label{varphieq3}
& &  \partial_\tau^2 \varphi(\tau) + \frac{3}{\tau}\,\partial_\tau \varphi(\tau)=
\lambda\Bigl[\frac{\mu^2}{\lambda}s(\tau)^2-\varphi(\tau)^2\Bigr]\varphi(\tau);\\
\vspace{-1mm}
\label{seq3}
& & \partial_\tau^2 s(\tau) + \frac{3}{\tau}\,\partial_\tau s(\tau)=
\frac{\mu^2}{\sigma_0^2}\Bigl[\frac{\mu^2}{\lambda}s(\tau)^2-\varphi(\tau)^2\Bigr]s(\tau)\,,
\vspace{-1mm}
\end{eqnarray}
with $s(\infty)=1$ as asymptotic condition.

Numerical simulations show that, even for moderately large values of $\varphi(0)$, $\varphi(\tau)$ first
becomes very small; then, at a certain time $\tau_j$, it suddenly jumps to a certain value $\varphi(\tau_j)$
close from below to $\varphi_{\hbox{\tiny max}}(\tau_j) =\sqrt{2}\mu\, s(\tau_j)/\sqrt{\lambda}$.
It then oscillates with decreasing amplitude, while approaching more and more to $\varphi(+\infty)=\mu/\sqrt{\lambda}$.
This happens because $s(\infty)=1$. Note that, as long as $\varphi^2(\tau)\ll \mu^2 s^2(\tau)/\lambda$,
Eq (\ref{seq}) is approximated very well by
\begin{equation}
\label{seq0}
\partial_\tau^2 s(\tau) + \frac{3}{\tau}\,\partial_\tau s(\tau)=
\frac{\mu^4}{\lambda\,\sigma^2_0}\,s(\tau)^3\,,
\end{equation}
whose general solution is
\begin{equation}
\label{s0oftau}
s(\tau)=\frac{s(0)}{1-\tau^2/\tau^2_c}\,,\quad\mbox{where $s(0)$ is
arbitrary and }\, \tau_c = \frac{\sqrt{8\,\lambda}\,\sigma_0}{\mu^2 s(0)} = 2 \tau_\Lambda\,,
\end{equation}
on account of Eq (\ref{Hubbletime}). Since $\tau_0/\tau_c= [2\sigma_0\sqrt{\lambda}/\mu\,\varphi(0)]/[\sqrt{8\lambda}\sigma_0/\mu^2 s(0)]=
\mu s(0)/\sqrt{2}\varphi(0)$ and we assumed $\varphi(0)\ll \mu s(0)/\sqrt{\lambda}$, we see that the kinematic time required for
the solution to Eq (\ref{seq}) to approach that of Eq (\ref{seq3}) is a negligible fraction of $\tau_c$.

If $\varphi=0,\partial_\tau\varphi=0$ at $\tau = 0$, we have $\varphi(\tau)=0$ for any $\tau$,
as Eq (\ref{varphieq}) clearly shows. In this case, Eq  (\ref{seq0}) is exact and Eq  (\ref{s0oftau})
shows that $s(\tau)$ diverges at {\em critical kinematic time} $\tau =\tau_c=2\tau_\Lambda$, {\em thus
providing an unexpected connection between the critical time of inflation and the cosmological
constant}. In any case, provided that the second member of Eq  (\ref{seq}) remains sufficiently small,
which is indeed the case since $\mu^2/\sigma_0^2 = \mu^2/6M^2_{rP}$ is very small, Eq  (\ref{varphieq})
ensures that this behavior tends to persist for a while. This means that, in the early stage of
inflation, the universe is well approximated by a Milne universe.

However, if $\partial_\tau\varphi=0$ at $\tau = 0$ and $\varphi(0)$ is positive, although
negligible with respect to $\mu\,s(0)/\sqrt{\lambda}$, $\varphi(\tau)$ starts increasing more
and more, as shown by Eq  (\ref{varphieq}), so that, in the long run, the behavior of
$s(\tau)$ departs significantly from that described by Eq  (\ref{s0oftau}).
More in general, as we can easily understand, no matter how close to zero the initial state is,
at some kinematic time $\tau_j$ sufficiently close to $\tau_c$, $\varphi(\tau)$ jumps to
a certain value $\varphi_j$. Hence, there is always a first instant $\tau_j + \varepsilon$,
with $\varepsilon < \tau_c - \tau_j$, at which the jump occurs and the equality
$(\mu^2/\lambda) s^2(\tau_j+\varepsilon)-\varphi^2(\tau_j+\varepsilon)$ becomes negative.
In these conditions, the curvature of the profile of $s(\tau)$ changes sign and an
oscillatory regime starts.

It is important to note that starting from a positive value of $\varphi(0)$,
negligibly small with respect to $\mu\,s(0)/\sqrt{\lambda}$, is critical in determining
the precise value of $\tau_j$ and a very large value of the jump amplitude.

Since the oscillation tends to decrease as $\tau$ approaches infinity, we can interpret the change in the
$\varphi$--amplitude minimum as a transition from the VEV of $\varphi$ in an initial {\em false vacuum}
to its VEV in a final {\em true vacuum}. So, at the end of this process, in the Heisenberg picture the
former can be re--interpreted as the {\em true physical state} of the matter field. At the end of this
process, that is after a kinematic time $\tau_q$, $\varphi(\tau)$ is very close to its asymptotic value
$\mu \,s(\tau)/\sqrt{\lambda}$ and consequently the equation for the final portion $s_f(\tau)$ of $s(\tau)$
is very well approximated by $\partial_\tau^2 s_f(\tau) + 3\,\tau^{-1}\partial_\tau s_f(\tau)=0$, the general
solution of which, for $\tau_q < \tau <\infty$, with $s_f(\tau_q)=s_q$ and $s_f(\infty) = 1$, is
\begin{equation}
\label{squeue}
s_f(\tau)= 1+\bigl(s_q -1\bigr)\frac{\tau^2_q}{\tau^2}\,.
\end{equation}
\begin{figure}[!h]
\centering
\mbox{%
\begin{minipage}{.45\textwidth}
\includegraphics[scale=0.6]{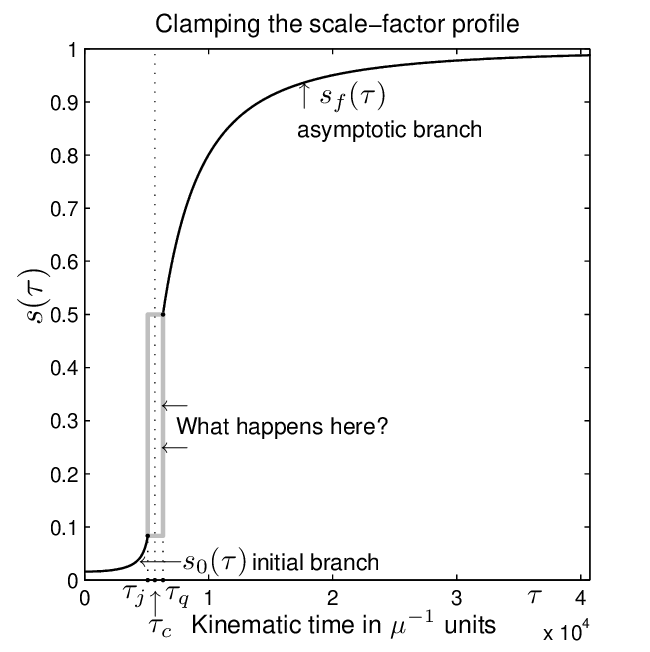}
\end{minipage}%
\begin{minipage}[c]{.45\textwidth}
\caption{\small The initial and final branches of scale factor profile are determined
by Eq  (\ref{seq}) alone, but its intermediate portion is determined by
Eq  (\ref{seq}) combined with Eq  (\ref{varphieq}).}
\end{minipage} \quad
}
\end{figure}
Thus, as shown in Fig.\,4, the entire scale factor profile is clamped by two curvilinear
branches: 1) at the initial end, from $\tau = 0$ to $\tau = \tau_c$, by the profile of
$s_0(\tau)$, characterized by a positive curvature; 2) at the asymptotic end,
from  $\tau = \tau_q$  to $\tau = \infty$, by the profile of $s_f(\tau)$,
characterized by a negative curvature. The behavior of $s(\tau)$ in the
joining region $\tau_c <\tau < \tau_q$, as well as the precise value of $s_q$ and
consequently of $s(0)$, cannot be determined as easily, since it depends  on
the details of the $\varphi(\tau)$--$s(\tau)$ interaction \cite{BROUT}.

Unfortunately, solving Eqs  (\ref{seq}) and (\ref{varphieq}) by numerical methods, even for moderate values
of the parameters, is difficult, because of time--grid problems due to the fact that $s(\tau)$
initially takes a very long time to reach appreciable values whereas, in the intermediate stage, it undergoes
a huge variation in a very small time interval.

These problems appear less acute in the proper--time picture as here the time scale initially is
strongly compressed. Posing $\tilde\varphi= s^{-1} \varphi$, $\partial_{\tau}= s\,\partial_{\tilde \tau}$,
expressing all functions of $\tau$ as functions of proper time $\tilde\tau = \int_0^\tau s(\tau')\,d\tau'$,
writing $\tilde s(\tilde\tau)\equiv s[\tau(\tilde\tau)]$ and $\tilde\varphi(\tilde\tau)
\equiv \tilde\varphi[\tau(\tilde\tau)]$, and using the second of Eq (\ref{tildetraces}), we can put
Eqs (\ref{seq}) and (\ref{varphieq}) in the form
\begin{equation}
\label{eqsystem}\left\{\begin{array}{l}
\partial^2_{\tilde\tau}\tilde\varphi(\tilde\tau)+
\displaystyle{3\Bigl[\frac{1}{\tau(\tilde\tau)}+
\frac{\partial_{\tilde\tau}\tilde s(\tilde\tau)}{\tilde s(\tilde\tau)} \Bigr]
\partial_{\tilde\tau}\tilde\varphi(\tilde\tau)
=\Bigl(\lambda-\frac{\mu^2}{\sigma_0^2}\Bigr)\Bigl[\frac{\mu^2}{\lambda}-
\tilde\varphi^2(\tilde\tau)\Bigr]\tilde\varphi(\tilde\tau)\,;}\\
\vspace{2mm}
\tilde R(\tilde \tau)= \displaystyle{\frac{6\,\mu^2}{\sigma_0^2}\Big[\tilde\varphi^2(\tilde\tau)-
\frac{\mu^2}{\lambda}\Big]},\,\,\mbox{equivalent to Eq (\ref{seq}), as proved in \S\,\ref{VEVs}}\,; \\
\vspace{2mm}
\tau(\tilde\tau) = \displaystyle{\int_0^{\tilde\tau} \frac{d\tilde\tau'}{\tilde
s(\tilde\tau')}\,.}
\end{array}\right.
\end{equation}

Both the system formed of Eqs  (\ref{varphieq}) (\ref{seq}) and system (\ref{eqsystem}) can easily be
integrated for moderate values of the parameters, so as to provide qualitative examples of time courses
of the Higgs field and the scale factor. As already noted in \S\,5.2 of Part II, factor $\lambda-\mu^2/\sigma_0^2$
on the right side of Eq  (\ref{eqsystem}) can safely be replaced by $\lambda$, as $\mu^2/\sigma_0^2 < 10^{-16}$ and
$\lambda > 10^{-3}$. Since in the proper--time picture, the profile of potential energy density $\tilde U(\tilde \varphi) =
\frac{1}{4}\lambda(\tilde\varphi^2 - \mu^2/\lambda)^2$ is fixed, i.e., it does not change in the course of
proper time, the qualitative behavior of $\tilde \varphi(\tilde \tau)$ can be illustrated as in Fig.\,5.
\vspace{-1mm}
\begin{figure}[!h]
\centering
\mbox{%
\begin{minipage}{.46\textwidth}
\includegraphics[scale=0.65]{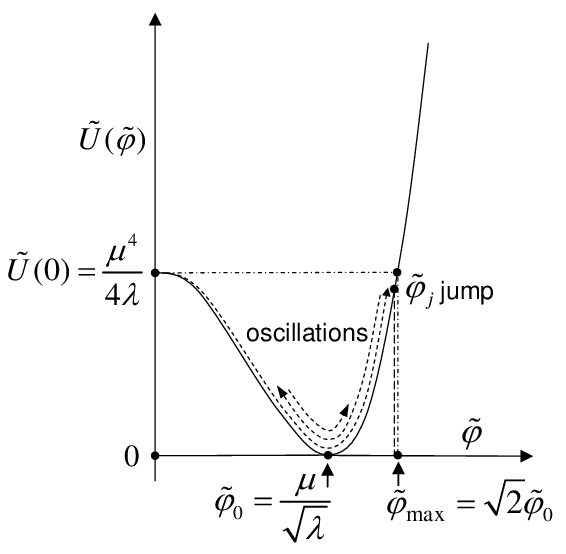}
\end{minipage}%
\begin{minipage}[c]{.5\textwidth}
\caption{\small Behavior of Higgs--field amplitude $\tilde \varphi(\tilde \tau)$
in proper--time picture. At $\tilde\tau =0$, $\partial_{\tilde \tau}\tilde\varphi =0$ and $\tilde\varphi$
is very close to 0 and the potential--energy density is very close to
$\tilde U(0)$. At time $\tilde \tau=\tilde \tau_j$ corresponding to kinematic--time
$\tau_j$, $\tilde\varphi$ jumps to a value $\tilde\varphi_j$ close from below to
$\tilde\varphi_{\hbox{\tiny max}}(\tilde\tau_j)=\sqrt{2} \mu/\sqrt{\lambda}$ with
potential energy close to $\tilde U(0)$. Then it oscillates with
decreasing amplitude closer and closer to the value $\tilde\varphi_0=\mu/\sqrt{\lambda}$
at the bottom of the well.}
\end{minipage}
}
\vspace{-1mm}
\end{figure}

The kinematic--time picture can easily be recovered by performing the inverse transformations $\tau(\tilde\tau)\rightarrow \tau$,
$\tilde s(\tilde\tau)\rightarrow s(\tau)$, $\partial_{\tilde \tau} \rightarrow s^{-1}\,\partial_{\tau}$,
$\tilde \varphi \rightarrow s \,\varphi$.

For $\tau <\tau_c$,  we have $s(\tau)= s_0(\tau)$ and $s_0(0)= s(0)$. Therefore, the third of Eqs (\ref{eqsystem})
can easily be integrated in the interval $0\le \tau < \tau_j$ by using Eq (\ref{s0oftau}), which yields
\begin{equation}
\label{tau2tildetau}
\tilde\tau(\tau)=s(0)\frac{\tau_c}{2}\ln\frac{\tau_c+\tau}{\tau_c-\tau}\,;\quad
\tau(\tilde \tau)=\tau_c\tanh\frac{\tilde\tau}{s(0)\,\tau_c}\,;\quad \tilde
s(\tilde\tau)=s(0)\Big[\cosh\frac{\tilde\tau}{s(0)\,\tau_c}\Big]^2\,.
\end{equation}

The complete integration of Eqs  (\ref{varphieq}) and (\ref{seq}) was carried out numerically by a
Matlab (The MathWorks, 2007) for non--realistic parameter values using the Euler method with a
time--grid of 10$^6$ points. Commented routines are available upon request. The results are
summarized in Figs 6, 7 and 8.
\begin{figure}[!h]
\centerline{\includegraphics[scale=0.72]{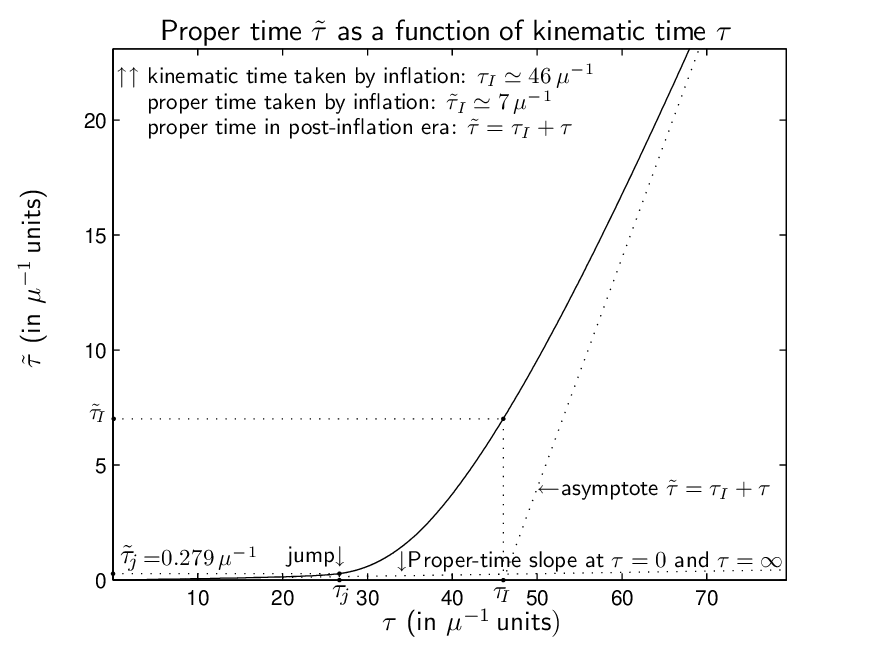}}
\vspace{-4mm}
\caption{\small Proper time $\tilde\tau$ as a function of kinematic time $\tau$ during
acute stage of inflation for moderate values of parameters. Higgs--field jump
occurs at kinematic time $\tau_j\simeq 26.7 \,\mu^{-1}$ corresponding to proper time
$\tilde\tau_j \simeq 0.279\,\mu^{-1}$. Note high scale compression of $\tilde\tau$
relative to $\tau$ for $\tau < \tau_j$. For small $\tau$, we have $\tilde\tau = \tau$
and for $\tau\rightarrow\infty$ $\tilde\tau = \tau_I + \tau$, where $\tau_I$
is amount of kinematic time taken by inflation. A realistic value of $\tilde\tau_I$ is
in fact expected to be much shorter than $\tilde\tau_I \simeq 7\, \mu^{-1}$,
reported in the figure.}
\end{figure}

We recall that, as explained in \S\,5 of Part II, $\tilde\varphi(\tilde \tau)$
represents the Higgs--field amplitude as measured by ideal comoving and
coeval reference frames with the expanding universe (cf. \S\,7.3 of Part I), i.e., at rest
at proper time $\tilde\tau$ on an expanding 3D portion of spacetime. The units of
measure of this reference frame are themselves subjected to scale expansion.
In contrast, $\varphi(\tau)$ represents the same amplitude as described with respect
to the reference frame of observers living in the post--inflation era. In this reference
frame, all dimensional quantities, both geometric and physical, are imagined to undergo
considerable changes of scale. Looking back to the past, these observers interpret the
events which occurred during universe inflation as subjected to the action of the
dilation field.

In Fig.\,7, scale factor $s(\tau)$ and Higgs--field amplitude $\varphi(\tau)$ on the Riemann
manifold are represented as functions of kinematic time $\tau$.

\begin{figure}[!ht]
\centering\includegraphics[scale=0.98]{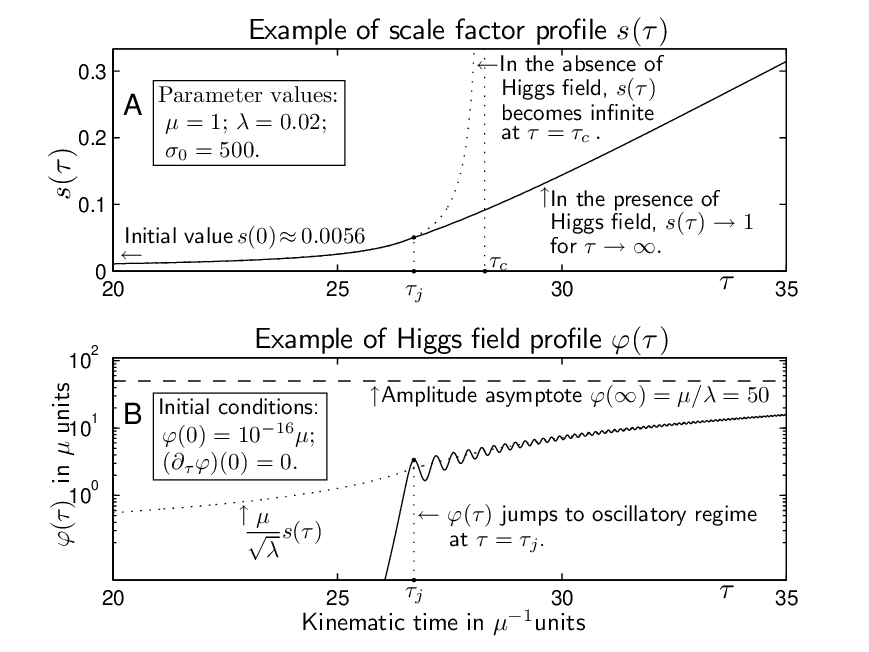}
\vspace{-4mm}
\caption{\small Example of scale factor profile $s(\tau)$ and Higgs field amplitude $\varphi(\tau)$
in a kinematic--time coordinates as functions of kinematic time $\tau$. {\bf A}. Solid line:
scale factor $s(\tau)$. Dotted line: scale factor in absence of Higgs field; it becomes
infinite at critical time $\tau_c$. At $\tau=\tau_j$, accelerated expansion transits
smoothly to decelerated expansion. {\bf B}. Solid line: profile of oscillating Higgs field
amplitude $\varphi(\tau)$, asymptotically adhering to profile of $\mu\,s(\tau)/\sqrt{\lambda}$
(dotted line). Dashed line: scale factor asymptote times $\mu/\sqrt{\lambda}$. All profiles
computed for parameters reported in inset of panel {\bf A}: they render realistic profiles
very poorly, since a realistic value of $s(0)$ is about $10^{-27}$, rather than $0.0056$, as
indicated in the left--bottom corner of panel {\bf A}.}
\end{figure}
Starting from about $\varphi(0) =10^{-16}\mu$, $\varphi(\tau)$ jumps almost suddenly to its maximum at
$\varphi_m =\sqrt{2}\,\mu \,\sigma(\tau_c)/\sqrt{\lambda}\sigma_0$ of about the same initial potential
energy, as shown in Fig.\,5, at a certain time $\tau\simeq \tau_0$, very close to critical time $\tau_c$; it
then oscillates coherently about its mean value of $\mu/\sqrt{\lambda}\times s(\tau)$ and progressively
decreases, while its potential energy is converted to kinetic energy through a sort of rarefaction--condensation
process.

In Fig.\,8A, the scale factor profile shown in Fig.\,7A is plotted as a function of proper
time $\tilde\tau$ for time--scale comparison. In Fig.\,8B, the example of
Higgs--field amplitude $\tilde\varphi$ shown in Fig.\,7B is reported as a function of $\tilde\tau$
as it appears in the Cartan picture. Actually, in this picture, the scale factor is always one
because $\tilde\sigma = \sigma_0$. At $\tilde\tau\simeq \tilde\tau_j$, $\tilde\varphi$ jumps to a
maximum close to $\sqrt{2}\mu/\sqrt{\lambda}$, then oscillates up and down its asymptotic
value $\mu/{\sqrt{\lambda}}$.

\begin{figure}[ht]
\centering\includegraphics[scale=0.98]{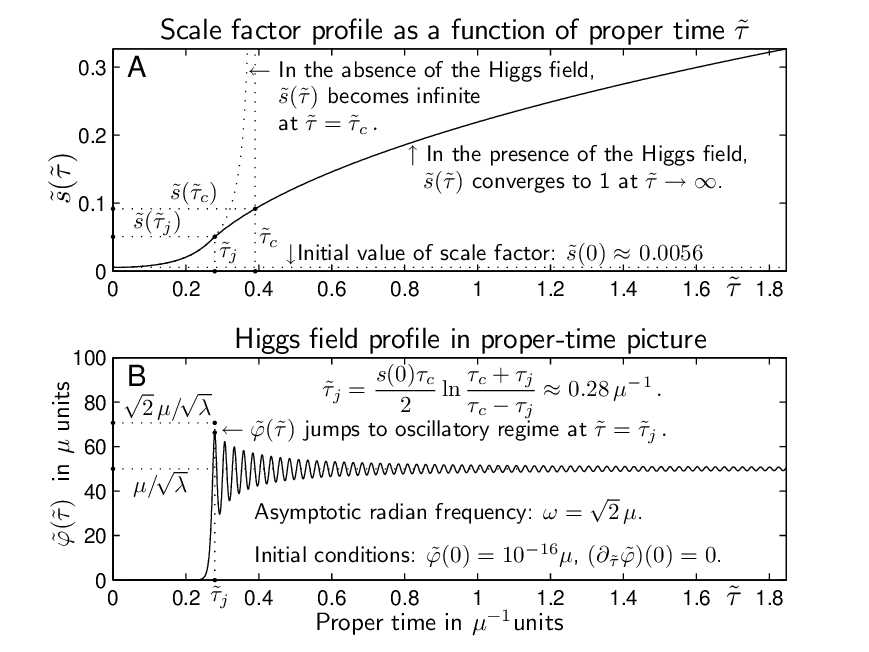}
\vspace{-4mm}
\caption{\small Example of scale factor profile $\tilde s(\tilde\tau)$, for
comparison with profile of $s(\tau)$ shown in Fig.\,7, and Higgs field
amplitude  $\tilde\varphi(\tilde\tau)$ in proper--time coordinates as functions of proper time
$\tilde\tau$. {\bf A}. Solid line: scale factor profile during inflation. Dotted line: scale factor in
absence of Higgs field; it coincides with that of accelerated Milne spacetime,
which becomes infinite at critical time $\tilde \tau_c$. At proper time $\tilde \tau_j$,
accelerated expansion of inflated Milne spacetime stops abruptly and transits to a decelerated
regime, which lasts until $\tilde s(\tilde\tau)$ becomes 1.
{\bf B}. Solid line: Higgs field amplitude $\tilde\varphi(\tilde\tau)$ converging
to final VEV $\mu/\sqrt{\lambda}$ at $\tilde\tau =\infty$.}
\end{figure}
Of note, the reason why $\tilde\varphi(\tilde\tau)$ exhibits pronounced oscillations whereas
$\tilde s(\tilde\tau)$ does not, is that factor $\mu^2/c_0^2$ on the right hand side
of the first of Eqs (\ref{eqsystem}) is enormously smaller than factor
$\lambda-\mu^2/\sigma_0^2$ on the same side of the second equation.

\subsection{The time course of inflation}
\label{predictions} Numerical simulations of the $\tilde\varphi(\tilde\tau)$ and
$\varphi (\tau)$ profiles showed that the time interval of appreciable oscillation amplitude
shrinks more and more as $\tilde\tau_j$ and $\tau_j$ get closer and closer to $\tilde\tau_c$
and $\tau_c$, respectively. Correspondingly, the sigmoidal profiles
shown Figs.\,9A and 9B, which are formed by the direct smooth joining of the initial branch with the
asymptotic branch exemplified in Fig.\,4, approach closer and closer to the true scale factor
profiles $s(\tau)$ and $\tilde s(\tilde \tau)$, exemplified in Figs.\,7A and Fig.\,8A, respectively.
\begin{figure}[!ht]
\includegraphics[scale=0.63]{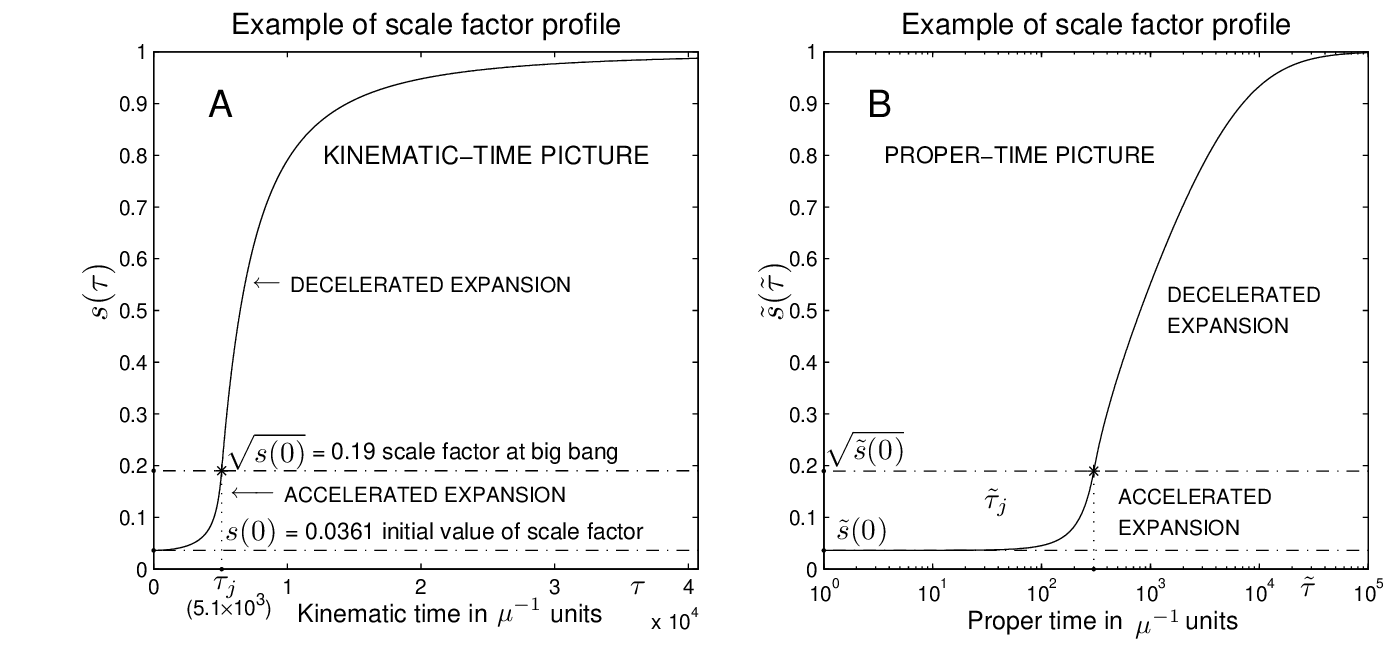}
\vspace{-4mm}
\caption{\small {\bf A}: Example of scale factor profile as a function of kinematic time $\tau$.
{\bf B}: Example of scale factor profile as function of proper time $\tilde\tau$. Note
scale compression of $\tilde\tau$ relative to $\tau$. Since profile slope at $\tau_j$
is expected to be enormously larger than shown here, time taken by false--vacuum
to true--vacuum transition of Higgs field state is actually much shorter.
The entire profile is thus well approximated by joining smoothly accelerated--expansion
branch and decelerated--expansion branches (Brout {\em et al}, 1978), which occurs when
$s(\tau_j)= \sqrt{s(0)}$.}
\end{figure}

Since the order of magnitude of scale expansion across inflation is estimated to be very large, we
conclude that the jump from geometry potential--energy to Higgs field potential--energy
is equivalent to a virtually instantaneous proliferation of a huge amount per unit volume of
Higgs bosons at rest in the comoving and co--expanding reference frame at kinematic time $\tau_j$,
which may therefore be described as a sort of cold big bang.

The smooth junction between branches $s_0(\tau)$ and $s_f(\tau)$ at $\tau = \tau_j$ is obtained
from joining conditions $s_0(\tau_j)=s_f(\tau_j)$ and $\dot s_0(\tau_j)=\dot s_f(\tau_j)$ (with $\dot s\equiv \partial_\tau s$).
From these conditions, and first using Eq  (\ref{squeue}) and then Eq  (\ref{s0oftau}), we obtain
\begin{eqnarray}
\label{s0tosj}
\hspace{-17mm}&& s(\tau)\!=\!s_0(\tau)\!=\!\frac{(1\!-\!\tau_j^2/\tau_c^2)^2}{1\!-\!\tau^2/\tau_c^2}
\,\, \mbox{for } 0 \leq \tau \leq\tau_j;\quad s(\tau)\!=\!s_f(\tau)\!=\! 1\!-\!\frac{\tau^4_j}{\tau_c^2\tau^2}
\,\,\mbox{for }\tau_j\!\leq\!\tau\!<\!\infty; \\
\label{sdot}
\hspace{-17mm}&&\dot s(\tau)\!=\!\dot s_0(\tau)\!=\!\frac{2\tau (1\!-\!\tau_j^2/\tau_c^2)^2}
{\tau_c^2(1-\tau^2/\tau_c^2)^2}\,\,\mbox{for } 0 \leq \tau \leq\tau_j;\quad \dot s(\tau)\!=\!\dot s_f(\tau)\!=\!
\frac{2\tau^4_j}{\tau_c^2\tau^3}\,\,\mbox{for }\tau_j\!\leq\!\tau\!<\!\infty;\\
\label{s0stauj}
\hspace{-17mm}&& s_0(\tau_j)\!=\!s_f(\tau_j)\!=\!\sqrt{s(0)}\!=\!
1\!-\!\frac{\tau_j^2}{\tau_c^2}\,\,\mbox{and}\,\,\dot s_0(\tau_j) \!=\!\dot s_f(\tau_j)\!=\!\frac{2\tau_j}{\tau_c^2}\!\simeq\!
\frac{2}{\tau_j}\,\mbox{in both cases}.
\end{eqnarray}

Although $\tau_j$ is very close to $\tau_c$, the scale expansion during
interval $[\tau_j, \tau_c]$ is not so small, as is evident from the levels reached by
$s(\tau)$ at $\tau=\tau_j$ and $\tau=\tau_c$ in Fig.\,7A. In fact,
as $\tau_j$ approaches $\tau_c$ from above we have
\begin{equation}
\label{sc2sj}
s_f(\tau_c)-s_f(\tau_j)\!=\!\frac{\tau^2_j}{\tau_c^2}\Big(1\!-\!\frac{\tau^2_j}{\tau_c^2}\Big)
\!=\!  \frac{\tau^2_j}{\tau_c^2}s_f(\tau_j),\,\,\mbox{then}\, s(\tau_c)\!\simeq\!2s_f(\tau_j)\,
\mbox{and}\,\,\tau_c -\tau_j\!\simeq\! \frac{\tau_j}{2} \sqrt{s(0)}.
\end{equation}
Similar relations also hold for $\tilde s(\tilde\tau_c)$ and $\tilde s(\tilde\tau_j)$.

Unfortunately, due to the discontinuity of $\ddot s(\tau)$ at $\tau = \tau_j$,
the scale factor constructed in this way is somewhat imprecise. In effect, from
$$
\ddot s_0(\tau) = \frac{2\,s(0)}{\tau_c^2(1-\tau^2/\tau_c^2)^2} +
\frac{8\,\tau^2 s(0)}{\tau_c^4 (1-\tau^2/\tau_c^2)^3}\quad
\mbox{and}\quad\ddot s_f(\tau_j) = -6\frac{\tau_j^4}{\tau_c^2 \tau^4}\,, $$
we derive
$$
\ddot s_0(\tau_j) =
\frac{2}{\tau_c^2}\Big[1+ \frac{4\,\tau_j^2}{\tau_c^2 \sqrt{s(0)}}\Big]
\simeq \frac{8}{\tau_c^2\sqrt{s(0)}}\,,\quad \ddot s_f(\tau_j)  = -\frac{6}{\tau_c^2}\,,
$$
as $\tau_j/\tau_c \simeq 1$ and $\sqrt{s(0)} \ll 1$, while we wish to find $\ddot s_0(\tau_j)
= \ddot s_f(\tau_j)=0$ instead.

This contrasts with the expected flatness of true scale factor $s(\tau)$ at the moment of the
accelerated--to--decelerated transition. Of the two second derivatives, the more deceptive is
clearly $\ddot s_0(\tau_j)$, as it is greater than $\ddot s_f(\tau_j)$ by about $8/(\tau_c^2 \sqrt{s(0)}$.
This means that the true $s(\tau_j)$ and $\dot s(\tau_j)$ must be somewhat smaller than $\dot s_0(\tau_j)$.

However the discrepancy is negligible. In fact, as can be evinced from the coefficients of
$\big(s^2 - \lambda\,\varphi^2/\mu^2\big)$ -- in Eqs (\ref{varphieq}) and (\ref{seq})
-- the ratio between the rising times of $\varphi(\tau)$ and $s(\tau)$ at
$\tau=\tau_j$ is $\sqrt{\lambda}\sigma_0/\mu \simeq 2.423\times 10^{16}$. This means that
the time taken by $\sqrt{\lambda}\,\varphi(\tau)/\mu$ to pass from a very small value to $s(\tau)$, as $\tau$
approaches $\tau_j$, is in the order of magnitude of $\tau_j\times 10^{-16}$. Correspondingly,
the time taken by $\ddot s(\tau)$ to deviate from $\ddot s(\tau) \simeq \ddot s_0(\tau) > 0$
to $\ddot s(\tau)\simeq \ddot s_f(\tau)<0$ across $\tau=\tau_j$ is negligibly small.
We can therefore regard $s(\tau_j)$ and $\dot s(\tau_j)$ as virtually equal to
$s_0(\tau_j)=s_f(\tau_j)$ and $\dot s_0(\tau_j)=\dot s_f(\tau_j)$.

\subsection{Higgs field energy density at big bang}
\label{HiggsEnergyDens}
Performing the measurement--unit conversions
\smallskip

\centerline{
\begin{tabular}{l r @{.} l}
1 eV  as mass ($\times \,c^{-2}$)  & $\rightarrow \quad $  1&78$\times 10^{-36}$ Kg\,,\\
1 eV$^{-1}$ as length ($\times \,\hslash\,c$)  & $\rightarrow \quad $ 1&97$\times 10^{-7}$ m\,,\\
1 eV$^{-1}$ as time ($\times \,\hslash$)  & $\rightarrow\quad$ 6&58$\times 10^{-16}$ s\,,
\end{tabular}}
\smallskip

\noindent where $c$ is the speed of light and $\hslash$ the Planck constant divided by $2\pi$,
we derive
\begin{eqnarray}
& & 1 \,\mbox{Kg} \simeq 5.62 \times 10^{26}\, \mbox{GeV} \,;\quad 1\, \mbox{GeV}\simeq
1.78\times 10^{-27}\,\mbox{Kg}\simeq 1.52\times 10^{24}\mbox{s}^{-1}\,; \nonumber\\
& & 1 \,\mbox{GeV} \simeq  5.076\times 10^{15}\,\mbox{m}^{-1} \,;\quad 1 \,\mbox{GeV}^{-1}
\simeq 1.97 \times 10^{-16}\,\mbox{m} \simeq 6.58\times 10^{-25}\,\mbox{s} \,; \nonumber\\
& & 1 \, \mbox{m}^{-1} = 1.97\times 10^{-16}\,\mbox{GeV}\,;\quad 1\,\mbox{s}^{-1} = 6.58\times
10^{-25} \,\mbox{GeV}\,; \nonumber \\
& & 1 \,\mbox{Kg/m}^{3}\simeq 4.297\times 10^{-21}\, \mbox{GeV}^4 \,;\quad 1\, \mbox{GeV}^4\simeq
 2.327\times 10^{20}\,\mbox{Kg/m}^{3}\,. \nonumber
\end{eqnarray}

Using the first of Eqs (\ref{tau2tildetau}) together with the first of Eqs (\ref{s0tosj}),
we obtain $\mu_H\simeq  2.243 \times 10^{-25}$ Kg and, from Eq (\ref{sc2sj}),
$\tau_j \simeq \tau_c/[1+ \sqrt{s(0)}/2]\simeq \tau_c $. Hence we have
\begin{equation}
\label{Sotauj}
\tau_j \simeq  \tau_c = \frac{\sqrt{8\,\lambda}\,\sigma_0}{s_0(0)\,\mu^2} \equiv
\frac{8\sqrt{3\lambda}\,M_{rP}}{s(0)\,\mu^2_H}\simeq \frac{7.691\times 10^{14}}{s(0)} \mbox{GeV}^{-1}
\simeq \frac{5.061\times 10^{-10}}{s(0)}\, \mbox{s}\,.
\end{equation}
Here $M_{rP}= \sigma_0/\sqrt{6} = 2.435\times 10^{18}$ GeV is the reduced Planck mass and $\mu_H\simeq 126$ GeV
the Higgs boson mass, as stated at the end of \S\,\ref{VEVs}. The self--coupling constant
$\lambda \simeq 0.131$ is provided by the Standard Model, where it appears related to the Fermi coupling constant
$G_F =  1.16637\times 10^{-5}$ GeV$^{-2}$ by equation $\lambda =\sqrt{2}\,\mu_H^2 G_F$.

Using Eq (\ref{Sotauj}) and the last of Eqs (\ref{s0stauj}), we obtain the Hubble constant as a function
of proper time at $\tilde \tau=\tilde\tau_j$ (see \S\,3.2 of Part II)
\begin{equation}
\label{Hubble_j}
\tilde H_j\!\equiv\!\tilde H(\tilde \tau_j)\!=\!\frac{\partial_{\tilde\tau} \tilde s(\tilde \tau_j)}{\tilde s(\tilde\tau_j)}
\! =\! \frac{\dot s(\tau_j)}{s(\tau_j)^2}\!\simeq\! \frac{2}{\tau_j s(0)}\!=\! 2.6\times 10^{-15}\,\,  \hbox{GeV}\! \simeq\!
3.95\times 10^9\,\,\hbox{s}^{-1}.
\end{equation}

As shown in Fig.\,9B, in the Cartan representation, $\tilde\varphi(\tilde\tau)$ remains
virtually zero for $\tilde\tau <\tilde\tau_j$, then jumps almost abruptly to $\tilde \varphi_{\hbox{\tiny max}}=
\mu\,\sqrt{2/\lambda}\equiv \mu_H/\sqrt{\lambda}$ at $\tilde\tau=\tilde\tau_j$ and, for $\tilde\tau >
\tilde\tau_j$, oscillates with decreasing amplitude of limiting radian frequency $\mu/\sqrt{\lambda}
\equiv\mu_H /{\sqrt{2\lambda}}$. A number of cycles after $\tilde\tau_j$, as $\tilde \varphi(\tilde\tau)$
approaches $\mu/\sqrt{\lambda}$, the damped oscillation tends to become harmonic
with proper--time period
\begin{equation}
\label{ptimeperiod}
\Delta\tilde\tau_H = 2^{3/2}\pi/\mu_H \simeq 3.28\times 10^{-26}\,\hbox{sec}\,,
\end{equation}
while $[\partial_{\tilde \tau}\tilde \varphi(\tilde\tau)]^2$ fades away as $1/\tilde\tau^3$,
so that the total energy of the oscillator is conserved.

As is evident from Fig.\,6, the maximum energy density above the minimum potential density
is attained immediately after $\tilde\tau_j$ and, the closer $\tilde\tau_j$ to $\tilde\tau_c$, the closer
it is to
\begin{equation}
\label{HiggsEdens}
\tilde U_{\hbox{\tiny max}} (\tilde\tau_j)=\frac{\mu^4}{4\lambda}\equiv \frac{\mu_H^4}{16\lambda} =
\frac{(126)^4}{16\times 0.131} \simeq 1.20\times 10^8 \,\mbox{GeV$^4$} \simeq  2.80\times 10^{28}\,\mbox{Kg/m}^{3}\,,
\end{equation}
which corresponds to $U_{\hbox{\tiny max}}(\tau_j)= \mu_H^4 s(\tau_j)^4/16\lambda=
\mu_H^4 s(0)^2/16\lambda$ in the Riemann picture.

This figure may be compared with the presently estimated energy density of universe
$\rho_{\hbox{\tiny Univ}} = 6 \times 10^{-27}$ Kg/m$^3 = 2.58\times 10^{-47}\, \mbox{GeV}^4$.

As discussed in \S\,5.1 of Part II and in \S\,\ref{EnDensCons}, an important aspect of universe
dynamics during inflation is that the continuous energy transfer from dilation field
$\sigma$ to Higgs field $\varphi$ and its decay products, through the work done by the
negative pressure of the former, all contribute to preserving the initial energy density, i.e.,
the original energy density of the vacuum. We may say that the total energy density of matter
and geometry is the same before and after the event of conformal--symmetry breakdown. This is consistent
with assuming that the universe originated from a local phase transition of an empty world
and explains why the present energy density of the universe is equal to the critical energy density
(thus solving the coincidence problem).

To include in our theory the decay products of the Higgs field, we can extend the action
integral $A= A^M[\varphi, \sigma]+A^G[\sigma]$ defined by Eq  (\ref{Actint}) of
\S\,\ref{higgsOnRiem}, with $A^M[\varphi, \sigma]$ and $A^G[\sigma]$  defined by Eqs
(\ref{AtoM}) and (\ref{AtoG}) of \S\,\ref{EnDensCons}, to the total action integral $A^T$,
as follows
\begin{equation}
\label{AtoT}
A^T = A^M[\varphi,\sigma] + A^G[\sigma] + \int_{H^+}\!\!\!\sqrt{-g}\,L^{(\varphi, {\boldsymbol \Psi})}(x)d^4x\,,
\end{equation}
where $H^+$ is the interior of the future cone and $L^{(\varphi, {\boldsymbol \Psi})}(x)$ is the Lagrangian density
of all SM fields ${\boldsymbol \Psi}$ with the exception of $\varphi$, which receives mass parameters from the VAV
of $\varphi$. Hence, motion equations (\ref{varphimoeteq}) and (\ref{sigmamoeteq}) extend to
\begin{equation}
\label{extmoteq} D^2\varphi + \lambda\Bigl(\varphi^2 - \frac{\mu^2}{\lambda}\,\frac{\sigma^2}{\sigma_0^2} \Bigr)\varphi -
\frac{R}{6}\,\varphi = \frac{\delta L^{(\varphi, {\boldsymbol \Psi})}}
{\delta \varphi}\,,\quad D^2\sigma + \frac{\mu^2}{\sigma_0^2}\,\Bigl(\varphi^2 -
\frac{\mu^2}{\lambda}\,\frac{\sigma^2}{\sigma_0^2} \Bigr)\sigma - \frac{R}{6}\,\sigma=0\,,
\end{equation}
while the total EM tensor of matter and geometry and the gravitational
equation given by Eq  (\ref{tetamunu}) extend respectively to
\begin{equation}
\label{exttetamunu}
\Theta^T_{\mu\nu}(x) = \Theta_{\mu\nu}(x) + \Theta^{(\varphi,  {\boldsymbol \Psi})}_{\mu\nu}(x)\,;\quad
\Theta^T_{\mu\nu}(x)=0\,.
\end{equation}
However, for the purposes of our investigation, we do not need to solve Eqs  (\ref{extmoteq}) and (\ref{exttetamunu}).

\subsection{The geometry and timing of the CGR universe}
\label{caveats}
The proper--time age of the universe provided by various authors in the last 15 years ranges
from 13.4 to 20.2 Gyr, for instance, $13.8\pm 0.04$ Gyr in Ref \cite{PLANCK}, $14.2\pm 1.7$ in \cite{RIESS},
$14.9\pm 1.3$ in \cite{PERLMUTTER} and $15.6 \pm 4.6$ in Cowan {\em et al.} (1999). The first three of these
are standard--model dependent, but the last is not, as it is inferred from the abundances of long
half--life radioactive elements in extremely old metal--poor stars of the galactic halo.

Thus, a problem arises about which of these values should be imported into the proper--time picture of CGR.
Considering the enormous flattening of the spacelike hyperboloids of the accelerated--inflated Milne universe
along directions of small parallax (Fig.\,3 of \S~\ref{proptimepict}) and the co--expansion of the length
measurement unit, we are inclined to assume that, in the proper--time picture of CGR, the age of the universe
from big bang time $\tilde\tau_c$, should not differ much from $\tilde \tau_U\simeq 15$ Gyr
$\simeq 4.73\times 10^{17}$s. In the next computations, we adopt this figure. To be meticulous,
we should add to $\tilde \tau_U$ the proper time taken by pure geometric inflation to reach critical time:
\vspace{-1mm}
$$
\tilde\tau_c \simeq  \tilde\tau_j=\int_0^{\tau_j}\!\! s_i(\tau)\, d\tau
= s(0)\frac{\tau_c}{2}\ln\frac{\tau_c+\tau_j}{\tau_c-\tau_j}\simeq  s(0)\frac{\tau_c}{2}\ln\frac{4}{s(0)} \approx
10^{-8}\,\hbox{s }\ll \tilde\tau_u\,,
$$
but it is negligibly small. Here the first of Eq (\ref{tau2tildetau}), the last of Eqs (\ref{sc2sj}) and
$s(0)\approx 3.17\times 10^{-28}$ are used, the latter of which matches the age universe age of the universe
self--consistently determined in \S\S\,\ref{entropycourse} and \ref{cosmconstandscale}.

To determine the kinematic time $\tau$ that corresponds to the proper time $\tilde\tau$ of the deceleration era,
we must first carry out the integration:
\vspace{-1mm}
\begin{equation}
\tilde \tau - \tilde\tau_c \!=\!\!\int_{\tau_c}^{\tau}\!\!\!\!s_f(\bar\tau)\,d\bar\tau
\! =\!\!\int_{\tau_c}^{\tau}\!\!\!\bigg(1 -\frac{\tau_j^4}{\tau_c^2\bar\tau^2}\bigg)d\bar\tau = \tau -\tau_c
+\frac{\tau_j^4}{\tau_c^2} \bigg(\frac{1}{\tau} - \frac{1}{\tau_c}\bigg) \simeq
\tau + \frac{\tau_c^2}{\tau} - 2\tau_c, \nonumber
\vspace{-1mm}
\end{equation}
where we have put $\tau_j/\tau_c\simeq 1$, then solve this equation for $\tau$, which gives
$$
\tau \simeq \tau_c + \frac{\tilde\tau-\tilde\tau_c}{2} \Bigg(1+\sqrt{1+ \frac{2\,\tau_c}{\tilde\tau-\tilde\tau_c}}\,\,\Bigg).
$$
Thus, in particular, the age of the universe in kinematic time units is
\begin{equation}
\label{tildetau2tau}
\tau_U = \tau_c +  \frac{\tilde\tau_U}{2} \Bigg(1+\sqrt{1+ \frac{2\,\tau_c}{\tilde\tau_U-\tilde\tau_c}}\,\,\Bigg)
\simeq \tau_c +  \frac{\tilde\tau_U}{2} \Bigg(1+\sqrt{1+ \frac{2\,\tau_c}{\tilde\tau_U}}\,\Bigg),\,\, \hbox{as }\,
\frac{\tilde\tau_c}{\tilde\tau_U} \approx 10^{-25}.
\end{equation}
Using Eq (\ref{Sotauj}), with the value of $S(0)$ given above, we obtain $s(0)\,\tau_c \approx 3.17\times 10^{-28}\,\tau_c
\approx 5.06\times 10^{-10}$s, then $\tau_c\approx 2.00\times 10^{18}$s. Thus, Eq (\ref{tildetau2tau}) yields
$\tau_c\approx 1.60\times 10^{18}$ s.

In the same way, from the proper time of plasma recombination $\tilde\tau_r\simeq 0.38$
Myr $= 1.19\times 10^{15}$s $\gg \tilde\tau_c$, we obtain the corresponding kinematic time:
\vspace{-2mm}
\begin{equation}
\label{kinrecombtime}
\tau_r =\tau_c +  \frac{\tilde\tau_r}{2} \Bigg(1+\sqrt{1+ \frac{2\tau_c}{\tilde\tau_r}}\,\,
\Bigg)\simeq 1.21\times 10^{18}\hbox{s}.
\vspace{-2mm}
\end{equation}

In Fig.\,10, the geometric proportions and timing, as well as the light--ray profiles to the observer,
of the accelerated Milne universe $M^+$ are faithfully represented.
\begin{figure}[!h]
\vspace{-4mm}
\centering
\includegraphics[scale=0.65]{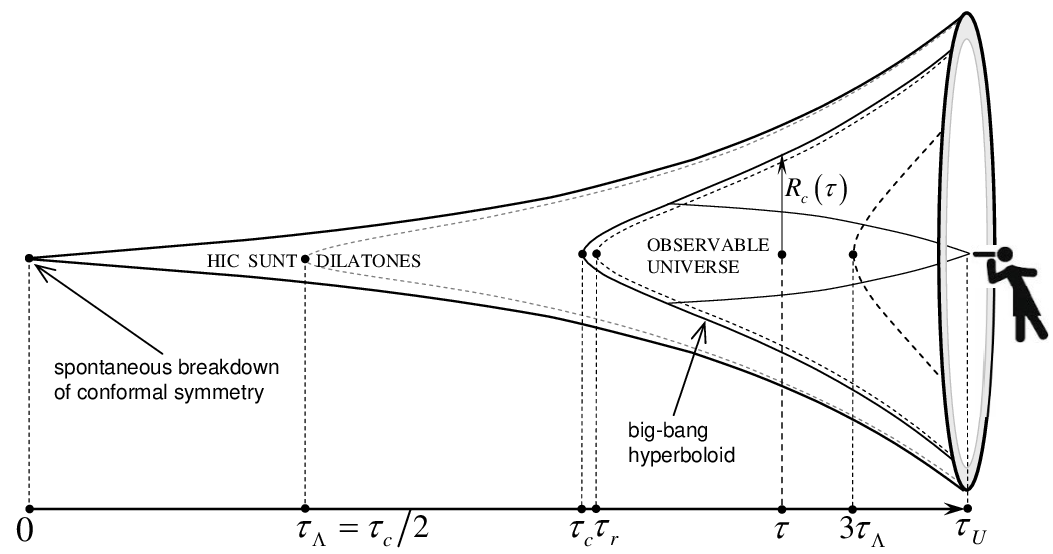}
\caption{\small {\em Kinematic--time picture of accelerated Milne universe $M^+$}. $\tau_U$: age of universe
in kinematic time units; $\tau_\Lambda = \sqrt{3/\Lambda}$: Hubble time (see \S\, \ref{Curvedhyperbspacetime});
$\Lambda$: cosmological constant; $\tau_j\simeq\tau_c = 2 \tau_\Lambda$: kinematic time of big--bang
hyperboloid $H^+_{\tau_c}$, as given by Eq (\ref{s0oftau}). $H^+_{\tau_c}$ divides $M^+$ in two parts: domains of
dilation and matter fields. Visible part of the latter is confined to past light--cone
of today's observer; $R_c(\tau)$: running radial distance of time axis from big--bang hyperboloid, as
a function of time--axis parameter $\tau> \tau_c$ (see text for explanation); $\tau_r$: kinematic time
of plasma--recombination (dotted hyperboloid).}
\vspace{-2mm}
\end{figure}

The running radial distance $R_c(\tau)$ of the future--cone axis of $M^+$ from the big--bang hyperboloid
$H^+_{\tau_c}$ with apex at $\tau_c$, is given by the formula:
\vspace{-2mm}
\begin{equation}
\label{Rctau}
R_c(\tau) = \int_{\tau_c}^\tau \frac{c(\bar\tau)\,\bar\tau}{\sqrt{\tau^2- \bar\tau^2}}\, d\bar\tau\,,
\vspace{-2mm}
\end{equation}
where $c(\bar\tau)$ is the scale factor of metric (\ref{FRWmetmetrictens}).

To realize this, let us first consider that the analogous distance of the future--cone axis of a flat Milne universe
$M^+_0$, from a hyperboloid $H^+_{\bar\tau}$ with apex at $\bar\tau$, is $\bar R_{\bar\tau}(\tau)=
\sqrt{\tau^2- \bar\tau^2}$. Since, for increasing time $\tau\ge \bar\tau$, $\bar R_{\bar\tau}(\tau)$ expands
with radial velocity $\bar v_{\bar\tau}(\tau) = -\partial_{\bar\tau}\bar R_c(\tau)=\bar\tau/\sqrt{\tau^2- \bar\tau^2}$,
we can write $\bar R_c(\tau) = \int_{\tau_c}^\tau \bar v_{\bar\tau}(\tau)\,d\bar\tau$. It is therefore evident
that Eq (\ref{Rctau}) is obtained by replacing $\bar v_{\bar\tau}(\tau)$ with $v_{\bar\tau}(\tau) =
c(\bar\tau)\,\bar v_{\bar\tau}(\tau)$, i.e., running radial velocity $\bar v_{\bar\tau}(\tau)$ enhanced
by scale factor $c(\bar\tau)$. Note that $R_c(\tau)$ is the integral of radius variations $dR_{\bar\tau}(\tau)
= v_{\bar\tau}(\tau)\,d\bar\tau$ from all the hyperboloids of the family $\{ H^+_{\bar\tau};
\tau_c\le \bar\tau\le\tau\}$. The same formula applies to any other hyperboloid of $M^+$, including
the future--cone boundary.

Passing from the kinematic--time to the proper--time picture, the accelerated Milne universe
becomes inflated--accelerated Milne universe $\widetilde M^+$, the geometric proportions and
timing of which, as well as light--ray profiles, are faithfully represented in Fig.\,11.
\begin{figure}[!h]
\vspace{-3mm}
\centering
\includegraphics[scale=0.65]{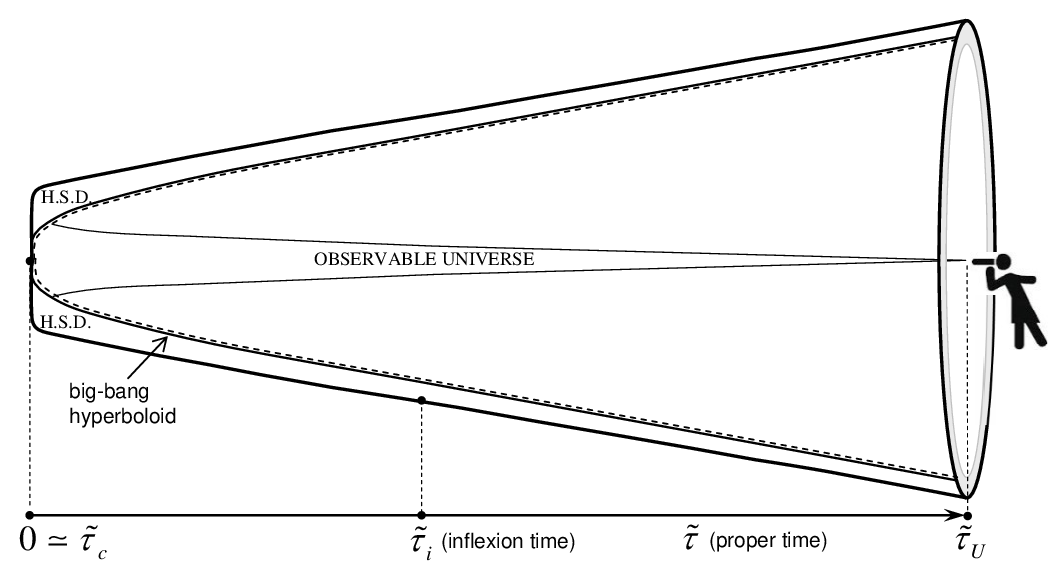}
\caption{\small  {\em Proper--time picture of inflated--accelerated Milne universe $\widetilde M^+$},
from Fig.\,10 with substitution $\tau\rightarrow\tilde\tau =\int_0^\tau s(\bar\tau)\,d\bar\tau$, where $s(\tau)$
is well--approximated scale factor of Eqs (\ref{s0tosj}). This substitution results in enormous compression
of the time interval $0\le \tilde\tau<\tilde\tau_c$ and consequent flattening of all the hyperboloids of apex less than
or close to $\tilde\tau_c$. Flattening of future cone is so pronounced as to resemble a truncation, so that
the portion of universe seen by an observer today is about the same as for a cylindrical universe. As
side--effect of proper--time parameterization, the calyx--shaped form of $M^+$ in Fig.\,3 of \S~\ref{proptimepict}
is, quite surprisingly, greatly depressed, which makes inflexion of the future--cone profile at
$\tilde\tau = \tilde\tau_i$ poorly visible.}
\vspace{-2mm}
\end{figure}

In kinematic--time units, volume element $dV(\tau, \vec \rho\,)$ running along polar geodesic
$\Gamma(\vec\rho\,)$ stemming from the future--cone origin, expands by a factor of $c(\tau)^3 =
[\tau_\Lambda \sinh(\tau/\tau_\Lambda)]^3$, where $\tau_\Lambda = \tau_c/2 \simeq 8\times 10^{17}$s,
matching Eq (\ref{s0oftau}).

For future reference, we derive from this the cubic root of the ratio of the volume--element
at kinematic time $\tau_c$ to the volume--element at kinematic time $\tau \ge \tau_c$:
\begin{equation}
\label {volumeratio}
\bigg[\frac{dV(\tau_c, \vec\rho\,)}{dV(\tau,\vec\rho\,)}\bigg]^{1/3}\!\!\!\! =
\frac{e^{\alpha(\tau_c)}\sinh(2\tau_c/\tau_c)}{e^{\alpha(\tau)}\sinh(2\tau/\tau_c)}=
\frac{e^{\alpha(\tau_c)}\big(1- e^{-4}\big)\,e^{-2\tau/\tau_c}}{e^{\alpha(\tau)}
\big(1- e^{-4-4\tau_u/\tau_c}\big)}.
\end{equation}

\section{Higgs--field dynamics in quantum field theory}
\label{zpirrel}
Eqs (\ref{varphieq}) and (\ref{seq}) of the previous section were introduced as semi--classical approximations of
quantum field theory (QFT) equations, which raises the problem of how much they may be altered, or even destroyed, by quantization.
As amply discussed in \S~2 of Part II, $\sigma(\tau)$ and $\varphi(\tau)$ are the kinematic--time dependent VEVs of soft
scalar bosons, which originate from the Nambu--Goldstone (NG) bosons, respectively associated to anti--deSitter stability--subgroup
$O(2,3)$ and deSitter stability--subgroup $O(1,4)$ of conformal group $O(2,4)$. Unlike possible NG bosons associated to
the Poincar\'e stability subgroup $O(1,3)$ of $O(2,4)$, their VEVs are not constants of motion but evolve as classical solutions
to conformal--invariant equations, $\sigma(\tau)$ behaving as a scalar ghost and $\varphi$ as a scalar particle field. Fields of
this sort were introduced in 1976 by Fubini, who deferred the study of quantization effects to another more extensive
paper, which unfortunately never appeared. Let us try to fill this gap in the light of {\em effective--action} methods 
\cite{JONA} \cite{COLEMAN1} \cite{JACKIW}.

As anticipated in \S~\ref{VEVs}, the quantization of $\varphi(\tau)$ and $\sigma(\tau)$ can be carried out
by introducing quantum fields $\varphi_Q(x)=\varphi(\tau) + \eta(x)$ and $\sigma_Q(x)= \sigma(\tau)+\xi(x)$,
where $\eta(x)$ and $\xi(x)$ are respectively quantum fluctuations of vanishing VEV about $\varphi$ and $\sigma$.
Hence, we have $\varphi(\tau) =\langle \Omega| \varphi_Q(x) |\Omega\rangle$ and $\sigma(\tau) =
\langle \Omega| \sigma_Q(x)|\Omega\rangle$.

Action integral $A^T$ introduced in Eq.(\ref{AtoT}) must therefore be regarded as the classical part of a
renormalizable quantum--field action--integral $$A^T_{\hbox{\tiny Q}}[\varphi_Q, \sigma_Q]= A^T_{\hbox{\tiny cl}}
[\varphi, \sigma]+ \Delta A^T[\varphi, \sigma; \eta, \xi; \bar\Lambda]\,,$$ the last term of which depends
explicitly on $\eta$ and $\xi$ and implicitly on renormalization subtraction terms depending on an overall
momentum cutoff $\bar\Lambda$. The transition amplitude from the initial state to the final state of
the system is then given by the path integral
$$
e^{(i/\hbar)A^T_{\hbox{\tiny eff}}[\varphi, \sigma;\bar \Lambda]} =
e^{(i/\hbar)\,A^T_{\hbox{\tiny cl}}[\varphi, \sigma]}
\int \langle\Omega|e^{(i/\hbar)\,\Delta A^T[\varphi, \sigma;\,\eta,\xi;
\bar\Lambda]}|\Omega\rangle\prod_x {\cal D}\eta(x)\,{\cal D}\xi(x)\,,
$$
which defines $A^T_{\hbox{\tiny eff}}[\varphi, \sigma;\bar \Lambda]$ as the effective action of the system. The 
dependence on Planck constant $\hbar$ is shown here for reasons which will soon become apparent.

Defining
$$
\Delta A^T_{\hbox{\tiny Q}}[\varphi, \sigma; \bar \Lambda] =-i \hbar \ln \Big[\int \langle\Omega|e^{(i/\hbar)\,\Delta
A^T[\varphi, \sigma;\,\eta,\xi; \bar \Lambda]}|\Omega\rangle\prod_x {\cal D}\eta(x)\,{\cal D}\xi(x)\Big]
$$
as the quantum correction to the classical action, we obtain the effective action in the form $A^T_{\hbox{\tiny eff}}[\varphi, \sigma]=
A^T_{\hbox{\tiny cl}}[\varphi, \sigma] +\Delta A^T_{\hbox{\tiny Q}}[\varphi, \sigma;\bar\Lambda]$.

$A^T_{\hbox{\tiny eff}}[\varphi, \sigma]$ can also be regarded as the spacetime integral of effective
total Lagrangian density $L^T_{\hbox{\tiny eff}}(\varphi, \partial_\tau\varphi,  \sigma,\partial_\tau\sigma; \Lambda)
= L^T_{\hbox{\tiny cl}}(\varphi,\partial_\tau\varphi,  \sigma,  \partial_\tau\sigma)- \rho_{\hbox{\tiny Q}}
(\varphi, \sigma;\bar\Lambda)$, where $\rho_{\hbox{\tiny Q}}$ is the quantum correction to the potential--energy
density term of the classical Lagrangian density or, to tell it differently, the zero--point energy density
due to the quantum fluctuations about classical fields $\varphi(\tau)$ and $\sigma(\tau)$, as described in
the fundamental papers on effective action referenced above. As we learn from those papers,
$\rho_{\hbox{\tiny Q}}$ can be expanded in powers of $\hbar$
\begin{equation}
\label{rhoQ}
\rho_{\hbox{\tiny Q}}\big[\varphi(\tau), \sigma(\tau);\bar\Lambda\big] = \sum_{n=1}^\infty \,\frac{1}{\hbar^n\, n!}\,\,
\rho^{(n)}\!\big[\varphi(\tau), \sigma(\tau);\bar\Lambda\big]\,,
\end{equation}
where $\rho^{(n)}$ is the contribution to the zero--point energy--density coming from
the $n$--loop 1PI (one particle irreducible) diagrams of all fields interacting with $\varphi_q(\tau)$ and
$\sigma_q(\tau)$.

The most important consequence of this result is that any nonzero $\rho_{\hbox{\tiny Q}}$ destroys the conformal invariance
of $L^T_{\hbox{\tiny cl}}$, thus heavily altering spacetime geometry. This is evident when we realize that it
changes classical gravitational equation $\Theta^T_{\hbox{\tiny cl}\,\mu\nu}(x)=0$ to the effective gravitational
equation $\Theta^T_{\hbox{\tiny eff}\,\mu\nu}(x)= \rho_{\hbox{\tiny Q}}\big[\varphi(\tau), \sigma(\tau);\,\bar\Lambda\big]
\,g_{\mu\nu}(x)$, which therefore violates the zero--trace property of conformal invariance and imparts an
additional time--dependent contribution $-4\,\rho_{\hbox{\tiny Q}}\big[\varphi(\tau), \sigma(\tau);\,\bar\Lambda\big]$ to
the geometric curvature of the universe.

It is thus clear that the only way to save the mechanism of the spontaneous breakdown of conformal symmetry is
to require the structure of true total Lagrangian density to yield exactly $\rho_{\hbox{\tiny Q}}\big[\varphi(\tau),
\sigma(\tau);\,\bar\Lambda\big]=0$, which is in principle possible in the one--loop approximation because boson
loops and fermion loops contribute with opposite signs.

Actually, this requirement is very strong, as it entails the separate vanishing of all $n$--loop terms $\rho^{(n)}$
appearing on the right side of Eq.(\ref{rhoQ}), which may in turn imply that the total Lagrangian density of CGR
has a wide strongly symmetric structure, or that, for still unknown reasons, the conformal invariance of
the quantum action integral acts as a custodial symmetry. There are three important reasons for advancing this conjecture:
\begin{itemize}
\item[1.] The observed value of the cosmological constant is in the order of magnitude of $10^{-47}$ GeV$^4$,
which is unlikely compatible with the hypothesis that it is totally or partially due to the zero--point
energy--density of the vacuum state.

\item[2.] As we prove in \S~\ref{cosmconstandscale}, if we assume equation $\rho_{\hbox{\tiny Q}}(\varphi,
\sigma; \Lambda)=0$, the value of the observed cosmological constant can be predicted with astonishing
precision uniquely as a property of the classical EM-tensor of conformal invariant Lagrangian density.

\item[3.] The vanishing of $\rho_{\hbox{\tiny Q}}(\varphi, \sigma; \Lambda)$ implies that Bohr's correspondence
principle, which states the convergence of quantum mechanics to classical mechanics as $\hbar\rightarrow
0$, should also hold in QFT as a {\em General Principle of Correspondence}.
\end{itemize}

We shall not attempt to prove the validity of this conjecture, mainly because, at
the present state of knowledge, the repertoire of Higgs field interactions is still incomplete.
In particular, the nature of dark matter is still a mystery and the validity of supersymmetry is not
yet experimentally confirmed, etc. It is perhaps preferable to reverse this attitude and assume the observed property as a
new fundamental principle. 

\subsection{The cosmological--constant problem in standard cosmology}
\label{cosmconstprob}
Assuming that the main contribution to cosmological constant $\rho_{\hbox{\tiny vac}}$ derives from
the zero--point energy density of quantum--field fluctuations and that the Planck mass is the natural
wave--number cutoff of quantum fluctuations, S.~Weinberg \cite{WEINBERG1} came to the striking conclusion
that $\rho_{\hbox{\tiny vac}} = 2\times 10^{71}$ $\simeq 10^{120}$ GeV$^4$, and others after him produced
smaller but still exaggerated values \cite{MARTIN}. In fact, the value provided by astronomers is about
$10^{-47}$ GeV$^4$, which is very close, if not equal, to the critical density of the universe.

Since all attempts to determine the observed value of $\rho_{\hbox{\tiny vac}}$ by zero--point energy
arguments have so far failed, we are faced with the conjecture that quantum fluctuations play a minor
role, if at all. Let us prove that this conjecture is not so fanciful when we reconsider the following
argument of Weinberg.

Summing the zero-point energies of all normal modes of a free scalar field $\psi(x)$ of mass $m$ up
to the Planckian momentum cutoff $\bar\Lambda=M_{rP}$ yields a vacuum--energy density
\begin{eqnarray}
\label{rhoetavac}
\rho^{(\psi)}_{\hbox{\tiny vac}} \!\!\!& = &\!\!\! \frac{4\pi}{(2\pi)^2} \int_0^{M_{rP}}\!\frac{1}{2} \sqrt{\bar k^2+m^2}\,\bar k^2 \bar dk = \frac{M_{rP}^4}{16\,\pi^2} + \frac{m^2 M_{rP}^2}{16\,\pi^2} - \nonumber\\
& & \!\!\!\frac{m^4\ln M_{rP}}{32\,\pi^2} + O( M_{rP}^{-1}) \simeq \frac{M_{rP}^4}{16\,\pi^2} \simeq 2\times 10^{71}\, \hbox{GeV}^4\,.
\end{eqnarray}

From our point of view, what appears to be totally ignored in this approach are the negative contributions
from the quantum fluctuation of ghost scalar field $\sigma(x)$. Since the action integral of a free
ghost scalar field $\xi(x)$ of mass $m_{\xi}$ is negative, so is its zero--point energy--density.
Let us indicate by $\eta(x)$ and $\xi(x)$ the quantum--amplitude deviations of $\sigma(x)$ and $\varphi(x)$
from their respective VEVs $\langle\sigma\rangle= \sigma_0$ and $\langle\varphi\rangle = \mu/\lambda$
in the post--inflationary era, and in the absence of interactions with other fields. Eqs~(\ref{ximoteq})
(\ref{etamoteq}) give the motion equations of $\eta$ and $\xi$ in the linear approximation
$$
D^2\eta + \big(\mu^2_H+\epsilon^2\big)\,\eta= 0\,,\quad D^2\xi + \epsilon^2\,\xi= 0\,,
$$
where $\epsilon^2 = -R/6$. Thus, for a general momentum cutoff $\bar\Lambda$, the total zero--point energy density
$\rho^{(\varphi + \sigma)}_{\hbox{\tiny vac}}\equiv \rho^{(\sigma)}_{\hbox{\tiny vac}}+
\rho^{(\varphi)}_{\hbox{\tiny vac}}$ is
\begin{eqnarray}
\rho^{(\varphi + \sigma)}_{\hbox{\tiny vac}}(\bar\Lambda)& = & \frac{\mu^2_H}{16\,\pi^2}\,\bar\Lambda^2 -
\frac{(\mu^4_H+2\,\mu^2_H\epsilon^2)}{32\,\pi^2}\ln\bar\Lambda + O(\bar\Lambda^{-1}) \,.\nonumber
\end{eqnarray}

Although the $\bar\Lambda^4$--term has now disappeared, the $\bar\Lambda^2$ and $\ln\bar\Lambda$ terms still remain.
Hence, the question naturally arises about whether these terms can be canceled by zero--point energy contributions
from other fields. For the $\bar\Lambda^2$ term, the answer may be positive because, as first
suggested by Veltman in 1981 \cite{VELTMAN}, the condition for this term to vanish provides a precise relation
between the masses of Standard Model particles. Unfortunately, we are not either in a position to verify
this relation or to solve the residual problem of logarithmic divergence, since the Standard Model is still
incomplete and we do not know what obscure matter really is. However, rather than insisting on this sort of
speculation to infer that the solutions to motion equations (\ref{extmoteq}) of \S~\ref{HiggsEnergyDens}
are totally divergence--free, it would perhaps be preferable ``to take the bull by the horns'' by invoking
the general principle of correspondence stated at the end of the previous subsection.

Regarding problems of gravitational--field quantization, we must consider that the inclusion in
the Lagrangian density of a conformal--invariant term $-\frac{1}{2}\beta^2 C^2(x)$,
introduced by Eq (A-29) of Part I, where $C^2(x)$ is the square of Weyl tensor $C_{\mu\nu\rho\sigma}(x)$,
i.e., the totally antisymmetric part of the Riemann tensor, ensures quantum--gravity renormalizability
\cite{STELLE} and asymptotic freedom \cite{TOMBOULIS1}, although at the price of introducing
gravitational ghosts of mass $M_G=M_{rP}/\beta$. This point deserves further study, because it
is still unclear whether in CGR these gravitational ghosts violate unitarity or not. Regarding the contribution
of the gravitational field to zero--point energy, we merely note that, in the linear one--loop
approximation and harmonic gauge, the contribution vanishes, because the trace of the graviton
propagator is zero.

\newpage

\subsection{A brief digression on thermal vacua}
\label{onthermalvacua}
The connection between thermodynamics and quantum field theory (QFT) has been investigated by several authors
since the early 1960s  \cite{ARAKI} \cite{KUBO} \cite{HAAG}. The entire subject is rooted in the theory of
infinite direct products and unitarily inequivalent quantum--field representations over a continuum of vacuum states
\cite{VNEUMANN} \cite{BRATTELI}. In fact, what are usually called quantum fields are only unitarily inequivalent
representations of algebraic entities called fundamental fields \cite{UMEZAWA1}, which may differ from each other
in the VEVs of one or more scalar fields, particle density in momentum space, or input--output coherent swarms
of infrared photons (Kibble, 1968). The most familiar kind of vacuum state is that of the Fock representation,
which is characterized by zero densities of particles, local currents and energy, as if the temperature of the
vacuum state were the unphysical absolute zero.

In this general view, physical particles must be regarded as quantum excitations of a specific vacuum
state. Thus, all the states of a particular unitary space can be viewed as a finite superposition
of quantum excitations and weighted statistics of infrared swarms. Since these unitarily inequivalent
representations form a continuum of mutually orthogonal spaces, each of which has its own fundamental state,
they are suitable for describing the classical limit of the macroscopic world, as well as its continuous
irreversible evolution.

Although unitarily inequivalent, these representations may be mutually related by algebraic maps ${\cal U}_B$,
called Bogoliubov transformations \cite{BOGOLIUBOV} -- generally depending on one, several or even infinite
parameters $\theta$ -- which preserve the canonical commutation relations of all fundamental fields, so that
they can be formally manipulated as unitary operators.

Any ${\cal U}_B$ can be viewed in two equivalent ways: either {\em \`a la} Heisenberg, as
an invertible transformation of all bounded operators $X$, constructed algebraically out of
fundamental fields and represented in a given unitary space $\cal H$, onto bounded operators $X'$,
non--equivalently represented in the same Hilbert space, i.e., by operations of the form
$X\rightarrow X'= {\cal U}_B X\,{\cal U}_B^{-1}$; or {\em \`a la} Schr\"odinger, i.e., by replacement of
the vacuum state $|\Omega\rangle$ of $\cal H$ with the vacuum state $|\Omega' \rangle$ of
${\cal H}'$. In this case, we write $|\Omega' \rangle = {\cal U}_B^{-1}|\Omega\rangle$.
The two modes are equivalent as $\langle \Omega| X'|\Omega\rangle = \langle \Omega| {\cal U}_B\, X\,
{\cal U}_B^{-1} |\Omega\rangle =\langle \Omega'| X |\Omega'\rangle$.

The simplest example of Bogoliubov transformations is that formally defined by
\begin{equation}
\label{lingen}{\cal U}(\theta) = e^{iG(\theta)}\,,\quad \mbox{where }\, G(\theta)=
- i\sum_k\theta[a(k) - a^\dag(k)]\,.
\end{equation}
It maps the annihilation--creation operators $a(k), a^\dag(k)$ of a fundamental scalar field,
represented in a Fock space of vacuum $|\Omega\rangle$, onto the representation
\begin{equation}
a'(k) = {\cal U}(\theta)\,a(k)\,{\cal U}^\dag(\theta) = a(k) + \theta\,, \quad
a'^{\,\dag}(k) = {\cal U}(\theta)\,a^\dag(k)\,{\cal U}^\dag(\theta) = a^\dag(k) + \theta\,,\nonumber
\end{equation}
of the same fundamental field in a second Fock space of vacuum $|\Omega'\rangle =
{\cal U}(\theta)|\Omega\rangle$, showing that ${\cal U}(\theta)$ performs a simple translation
of the boson field amplitude.

Denoting by $N(k) = a(k)\,a^\dag(k)$ and $N'(k) = a'(k)\,a'^{\,\dag}(k)$ the particle--number
operators, respectively in the first and second representations, we can easily verify equations
$\langle \Omega| N(k) |\Omega\rangle = \langle \Omega'| N'(k)|\Omega'\rangle =0$ and $\langle \Omega| N'(k)
|\Omega\rangle = |\theta|^2$.  Since ${\cal U}(\theta)$ changes the particle--number 0 into $|\theta|^2$
without modifying the spectrum of the Hamiltonian, it may be interpreted as an adiabatic transformation
at zero temperature. Therefore, the thermal properties of $|\Omega\rangle$ and $|\Omega'\rangle$ are trivial.

Vacuum states with non--trivial thermal properties are called thermal vacua. These are characterized
by the unboundedness from below of the number of possible quantum annihilations. Thus, in order for a thermal
vacuum to be a cyclic state, the fundamental--field representation needs a twofold number of degrees of freedom
(Araki \& Woods, 1963): one representing "positive" thermal excitations -- say {\em particles} --
the other representing "negative" thermal excitations -- say {\em particle holes}. For instance,
the state of an empty box immersed in a thermal reservoir of temperature $T$ is of this sort (Fig.\,12).
\begin{figure}[!h]
\centering
\includegraphics[scale=0.8]{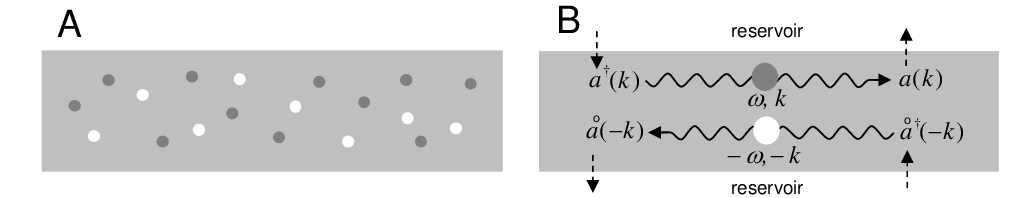}
\caption{\small {\bf A}: Thermal vacuum as an incoherent superposition of particles (dark spots)
and holes (white spots). {\bf B}: Exchange of thermal quanta with reservoir occurs in two modes:
1) by creation and annihilation of particles of energy--momentum $\{\omega, k\}$,
respectively represented by operators $a^\dag(k)$ and $a(k)$; 2) by annihilation and creation
of holes of energy--momentum $\{-\omega, -k\}$, respectively represented by operators $\mathring{a}(-k)$
and $\mathring{a}^\dag(-k)$. Both modes result in same amount of energy--momentum exchanged with
reservoir. Since particles and holes are independent degrees of freedom, all $\mathring{a}(-k)$
and $\mathring{a}^\dag(-k)$ commute with all $a(k)$ and $a^\dag(k)$. Simultaneous creations or annihilations
of particles and holes of opposite energy--momentum represent thermal fluctuations.}
\end{figure}

If a number of particles of energy--momentum  $(\omega, k)$ and an equal number of holes of
energy--momentum $(-\omega, -k)$ are simultaneously created or annihilated, the energy--momentum
exchanged between system and reservoir is zero. We can regard these zero--sum processes as
internal fluctuations of the thermal vacuum \cite{MANN}. Since these are unobservable,
the Heisenberg indetermination relations of the matter fields appear to be affected by an
additional entropic indetermination representing {\em thermal noise} with Gaussian standard
deviation of both field amplitudes and their time--derivatives (Umezawa, 1993).

On this basis, the thermal vacuum of an infinite system can be ideally obtained by expanding the volume
of the box to infinity. Since, at this limit, the reservoir disappears, the vacuum itself must be
regarded as its own reservoir. In this case, the thermal fluctuations are more appropriately described
as quantum fluctuations of a mixture of virtual particles and holes. In the following, we only refer to
infinite systems.

It is intuitive that the ratio between hole density and particle density  varies with temperature
and approach zero as $T\rightarrow 0$. If this limit could be reached, all holes would disappear,
which is impossible, in accordance with the third principle of thermodynamics.

Let $a^\dag(k)$, $a(k)$ respectively be the creation and annihilation operators of a boson of energy--momentum
$\omega, k$, and $\mathring{a}^\dag(-k)$, $\mathring{a}(-k)$ respectively be the creation and annihilation
operators of a boson--hole of energy--momentum $-\omega, -k$. Since particles and holes are independent
degrees of freedom, all $\mathring{a}(-k)$, $\mathring{a}^\dag(-k)$ commute with all $a(k)$, $a^\dag(k)$.
Therefore, as far as energy--momentum balance is concerned, the actions of $a(k)$ and $\mathring{a}(-k)$
produce the same effects. It is therefore natural to introduce, as creation and annihilation operators of
thermal fluctuations, linear combinations
\begin{equation}
\label{parthole}
a(k, T)\!=\! c(k, T) a(k)\! +\! d(k,T)\,\mathring{a}^\dag(-k),\quad a^\dag(k, T)\! =\! c^*(k, T) a^\dag(k)\! +\!
d^*(k,T)\,\mathring{a}(-k),
\end{equation}
where $c(k, T), d(k, T)$ are suitable coefficients. Since the phases of these coefficients can be
absorbed by a redefinition of $a(k)$ and  $\mathring{a}^\dag(-k)$, there is no loss of generality in
assuming $c(k, T)$ and $d(k,T)$ to be real and positive. The requirement that $a(k, T),
a^\dag(k, T)$ should satisfy the canonical commutation relations (c.c.r.) $[a(k, T), a^\dag(k', T)] = \delta^3(k-k')$
leads to equations $c(k, T)^2 - d(k, T)^2 = 1$. Eqs  (\ref{parthole})
can be written as $a(k, T) = {\cal U}_T\,a(k)\,{\cal U}_T^{-1}$, $a^\dag(k, T) =
{\cal U}_T\,a^\dag(k)\,{\cal U}_T^{-1}$, by the formal action of the Bogoliubov operator
\begin{equation}
\label{quadgen} {\cal U}_T= e^{i\,G_T}\quad \mbox{with }\, G_T=
- i\sum_k d(k, T) \big[\mathring{a}(-k)\,a(k)  - \mathring{a}^\dag(-k)\,a^\dag(k)\big]\,.
\end{equation}
Denoting by $N(k)=a(k)\,a^\dag(k)$ and $\mathring{N}(-k)=\mathring{a}(-k)\,\mathring{a}^\dag(k)$,
respectively, the number operators of particles and holes of momentum $k$ in the
Fock space representation, we find $[G_T, N(k)] =[G_T, \mathring{N}(-k)]$, showing that
$N(k) - \mathring{N}(-k)$ are the invariants of ${\cal U}_T$.

Let us denote by $|\Omega_F \rangle$ the Fock vacuum state of $a(k), a^\dag(k),
\mathring{a}(-k), \mathring{a}^\dag(-k)$, by $|\Omega_T \rangle\! =\!{\cal U}_T^{-1}|\Omega_F\rangle$
the thermal vacuum and by $N(k,T)\!=\!a(k,T)a^\dag(k, T)$ the number of thermal excitations of momentum $k$
in the Fock representation. We thus have $a(k)|\Omega_F \rangle\! =\!\mathring{a}(-k)|\Omega_F \rangle\!=\!a(k, T)|\Omega_T \rangle\!=\!\mathring{a}(-k,T)|\Omega_T \rangle\!=0$, hence $N(k)|\Omega_F \rangle\!=\!
\mathring{N}(-k)|\Omega_F \rangle\!=\!N(k,T)|\Omega_T \rangle=0$.

By developing ${\cal U}_T$ in series of powers of $G_T$ and rearranging the terms by repeated commutations
\cite{MANN} \cite{UMEZAWA2}, we can prove the equation
$$
|\Omega_T \rangle = {\cal U}_T^{-1}|\Omega_F\rangle =\sum_{n,k}\frac{d(k,T)^n}{n!\,\exp[\,\ln\cosh d(k,T)]}
\big[\mathring a^\dag(-k)\,a^\dag(k)\big]^n |\Omega_F \rangle\,,
$$
showing that, in the Fock--space representation, the thermal vacuum is a quantum--entan\-gled superposition
of particle--hole pairs of zero energy and zero momentum fluctuations.

For any operator $X$ in $\{a(k), a^\dag(k)$, $\mathring{a}(-k), \mathring{a}^\dag(-k)\}$ algebra, there is
a corresponding operator $X(T)= {\cal U}[\theta]\,X\,{\cal U}[\theta]^{-1}$  in $\{a(k,T), a^\dag(k,T)$,
$\mathring{a}(-k,T), \mathring{a}^\dag(-k,T)\}$ algebra, satisfying equation $\langle \Omega_T| X |\Omega_T \rangle
= \langle \Omega_F|X(T)|\Omega_F \rangle$. In particular, we have $\langle \Omega_T|N(k)|\Omega_T \rangle =
\langle \Omega_F|N(k,T)|\Omega_F \rangle = d^2(k,T)V$, where $V = (2\pi)^3\delta^3(0)=\int\! e^{ikx}|_{k=0}d^3x$
is the space volume.

Since, in accordance with Bose--Einstein statistics, particle density $n(k)=N(k)/V$ at thermal equilibrium is
$\langle \Omega_T |n(k)|\Omega_T \rangle =[e^{\omega(k)/T} -1]^{-1}$,
we find for Eqs (\ref{parthole})
$$
d(k, T) = \frac{1}{\sqrt{e^{\omega(k)/T} -1}}\,,\quad  c(k, T) =\sqrt{1+d^2(k,T)}=
\frac{e^{\omega(k)/2\,T}}{\sqrt{e^{\omega(k)/T} -1}}\,.
$$

Similar results are obtained for fermion particles and holes, in which case the coefficients
are $d(k, T)= 1/\sqrt{e^{\omega(k)/T} +1}, c(k, T)= e^{\omega(k)/2T}/\sqrt{e^{\omega(k)/T} +1}$.

We thus realize that ${\cal U}_T$ makes a boson field at temperature $T=0$, in the
Fock--space representation, jump to a boson gas at temperature $T>0$ in the same representation.

Equivalently, ${\cal U}_T^{-1}$ causes the zero--temperature vacuum of the Fock representation
to jump to a thermal vacuum of temperature $T$, leaving formally unvaried the algebra of fundamental
fields. We can also build a thermal Bogoliubov operator which depends on time. This would then allow us
to represent a continuous thermal evolution of the vacuum state. When applied to the fundamental state
of an initially empty system, generating a boson gas which remains in thermodynamic equilibrium at
a continuously varying temperature.

\subsection{The big bang as a thermodynamic process}
\label{thermpro}
As noted in \S\S~\ref{zpirrel} and \ref{cosmconstprob}, in order for the effective potential to be zero,
$\varphi$ must interact with other fields besides $\sigma$. In the semi--classical approximation, it must obey
the first of Eqs (\ref{extmoteq}), with $\delta L^{(\varphi, {\boldsymbol \Psi})}/\delta \varphi$ averaged
over the unit hyperboloid at $x=\{\tau, \hat x\}$ but, in the quantum field representation, this classical
interaction term must be replaced by the VEV of its quantum--theoretical counterpart, i.e., equation
\begin{equation}
\label{extmoteq2} D^2\varphi(\tau) + \lambda\Bigl[\varphi^2(\tau) - \frac{\mu^2}{\lambda}\,\frac{\sigma^2(\tau)}{\sigma_0^2} \Bigr]
\varphi(\tau) -\frac{R}{6}\,\varphi(\tau) = \lim_{V_1\rightarrow \infty} \frac{1}{V_1}\int_{V_1}\!\!
\langle\Omega|\frac{\delta L^{(\varphi, {\boldsymbol \Psi})}(\tau, \hat x)}
{\delta \varphi(\tau, \hat x)}|\Omega \rangle dV_1(\hat x).\nonumber
\end{equation}

Since the effective potential is zero and the quanta of fields ${\boldsymbol \Psi}$ are created
by the Higgs field only decay after $\tau > \tau_j$, the right side of this equation is zero
over the whole kinematic--time interval $0\le \tau \le \tau_j$. During this interval, the motion equations
of $\varphi(\tau)$ and $\sigma(\tau)$ in hyperbolic coordinates are the same
as equations (\ref{simpPhiMoteqOnRiem}) and (\ref{simpSMoteqOnRiem}).

Looking at Figs. 7 and 8 of \S~\ref{Higgsdyn}, one may wonder that the two fields behave so differently.
Before the occurrence of the big bang jump, the evolutions of $\varphi(\tau)$ and $\sigma(\tau)$
proceeded smoothly and differently, as adiabatic processes at virtually zero temperature,
without interacting with other fields.  Then, in the classical approximation, $\varphi(\tau)$
jumped abruptly to a damped oscillation regime, while $\sigma(\tau)$ appeared to
evolve smoothly, as it were insensitive to the behavior of $\varphi$. This depended on the fact
that the kinetic--energy variation of $\sigma$ during the interaction was much smaller than that
of $\varphi$ by the factor
$$
K=\frac{\sigma^{-1} D^2\sigma}{\varphi^{-1}D^2\varphi}= \frac{\mu^2}{\lambda\,\sigma_0^2} \simeq 1.70\times 10^{-33} \,,
$$
as shown in Eqs (\ref{extmoteq}). In fact, after the amplitude jump, the evolution did not have the
properties of the coherent oscillatory regime exemplified in the above--mentioned figures. Rather, from
the moment of the jump to the beginning of gravitational collapse, due to the violent expansion of
space volume \cite{TAKAHASHI}, an incoherent crowd of Higgs bosons was created, which started to interact
with each other and with all the particles created by their decay. Then, evolution proceeded
as an approximately adiabatic thermodynamic process. After the end of the inflationary epoch,
matter continued to expand freely, the temperature decreased, and the process tended to become
inhomogeneous and anisotropic as a consequence of gravitational forces.

From a quantum theoretical standpoint, the sudden creation of the Higgs--boson bulk
at proper time $\tilde\tau_j$ cannot be represented as a unitary transformation of a quantum state, but
rather as a thermal Bogoliubov transformation ${\cal B}(T)$, which can be manipulated as a
unitary operator. As explained in the previous section, ${\cal B}(T)$ dynamically maps Higgs--boson
creation--annihilation operators $a_-^\dag(k), a_-(k)$, defined in Hilbert space ${\cal H}_-$ of
the Higgs field representation at $\tau_j^- = \tau_j-d\tau$, onto isomorphic operators
$a_+^\dag(k), a_+(k)$, defined in a different Hilbert space ${H}_+$ at $\tau_j^+ = \tau_j+d\tau$, and
the cyclic state $|\Omega_-\rangle$ of ${\cal H}_-$ onto the cyclic state $|\Omega_+\rangle$ of ${\cal H}_+$.

Indicating by $N_-(k) = a_-(k)\,a_-^\dag(k)$ and $N_+(k)= a_+(k)\,a_+^\dag(k)$ the number--operators
of the quanta of momentum $k$, respectively at $\tau_-$ and $\tau_+$, we have, by definition
of vacuum, $a_-(k)|\Omega_-\rangle = a_+(k)|\Omega_+\rangle=0$, then $N_-(k)|\Omega_-\rangle=0$,
but the nonzero density of bosons of momentum $k$ at temperature $T$
\begin{equation}
n_+(k)\equiv V^{-1}N_+(k)|\Omega_+\rangle  =  \frac{1}{e^{E(k)/T}-1}|\Omega_+\rangle\,,\nonumber
\end{equation}
where $V$ is the space volume defined in the previous section.

If we want the description of the universe to be always relative to a comoving and co--expanding reference
frame, we must solve the equations in the conformal--coordinate representation, i.e., for $\tilde \varphi(\tilde \tau)$,
$\partial_{\tilde x}\tilde \varphi(\tilde x)$, $\tilde{\boldsymbol\Psi}(\tilde \tau)$, etc. In this case,
the motion equation of $\tilde\varphi(\tilde \tau)$ obeys Eqs~(\ref{varphieq}) over the whole proper time
interval $0\le \tilde\tau \le  \tilde\tau_j$. Correspondingly, at the amplitude jump, the maximum energy density
of the Higgs field in the co--expanding reference frame is the same as the classical one given by Eq
(\ref{HiggsEdens}); i.e.,
\begin{equation}
\label{HiggsEdensQ}
\tilde U_{\hbox{\tiny max}} (\tilde\tau_j)=\frac{\mu_H^4}{16\,\lambda}
\simeq   1.186\times 10^8 \,\mbox{GeV$^4$}\,.
\end{equation}

The Higgs--field bulk created at proper time $\tilde\tau_j$  is expected to behave as a fluid at rest
in the reference frame of comoving observers. If it were regarded as a gas of classical particles,
all of them would be forced to be at rest in the comoving reference frame, due to the high viscosity
of the dilation field in expansion, which would be consistent with the argument discussed in \S\,7.3
of Part I. In this case, the profile of $\varphi(\tau)$ as a coherent oscillation of decreasing amplitude
would be justified. Actually, the story is quite different, because, during the inflationary epoch and beyond,
the Higgs field interacts, directly or indirectly, with all its decay products, i.e., presumably, with
all known or still unknown particles of the Standard Model, in manners and ways which must be consistent with
renormalizability, absence of triangle anomalies and matter--antimatter asymmetry.

\section{Entropy conservation from big bang to now}
\label{entropycourse}
Since the hot bulk of Higgs bosons created by the big bang is in thermodynamic equilibrium,
no heat can flow, nor work can be done among comoving volume elements of the expanding hyperboloids.
This holds exactly until the beginning of gravitational collapse and, to a good approximation, on a
large scale during the subsequent era. Therefore, the entropy of the universe is exactly or
almost exactly conserved.

In the proper--time picture (in natural units and with Boltzmann constant $k_B=1$), the energy density
$\epsilon$, pressure $p$, entropy density $s$, temperature $T$ and particle density $n$, of a
gas of particles of resting mass $m$ and degeneracy factor $g$, are given by the integrals
\begin{eqnarray}
\label{energydens}
\epsilon(T) &\equiv & g\,a_{\epsilon}(m/T)\,T^4 = \frac{g}{2\pi^2}\int_0^\infty \frac{E(m,p)\, p^2}{e^{(E(m,p)-\mu_p)/T} \pm 1} dp\,;\\
\label{pressure}
p(T) &\equiv& g\,a_p(m/T)\,T^4 = \frac{g}{6\pi^2}\int_0^\infty \frac{p^4}{E(m,p)\big[e^{(E(m,p)-\mu_p)/T} \pm 1\big]}dp\,;\\
\label{entropydens}
s(T) &\equiv& g\,a_s(m/T)\,T^3 = \frac{g}{T}(a_{\epsilon} + a_p)\,;\\
\label{particledens}
n(T) &\equiv &  g\,a_n(m/T)\,T^3 = \frac{g}{2\pi^2}\int \frac{p^2}{e^{(E(m,p)-\mu_p)/T} \pm 1}dp\,;
\end{eqnarray}
where, $p$ is the momentum of the particles, $E(m, p) = \sqrt{m^2 + p^2}$ their energy, and $\mu_p$ their
chemical potential \cite{WEINBERG4}. The latter is zero for massless particles and can be neglected
for large $T$. The signs + or $-$ in the denominator correspond to the case of fermions or
bosons, respectively. For $m=0$ and $m\ll T$, we have $a_\epsilon = \pi^2/30$ for bosons,
$a_\epsilon = (7/8)(\pi^2/30)$ for fermions and $a_p= a_\epsilon/3$; hence, $a_s = (4/3)\,a_\epsilon\,T^3$,
and $a_n= (3/4)\zeta(3)\,a_\epsilon$, where $\zeta(3)=1.20206\dots$ is the Riemann zeta function of 3.
In the presence of several species of particles, we must write the sum of similar expressions over all species
\cite{WALD}, i.e.:
\begin{equation}
\label{stareqs}
\epsilon_*(T) = T^4 \sum_i g_i a_\epsilon\Big(\frac{m_i}{T}\Big);\,\, p_*(T) =
T^4 \sum_i g_i a_p\Big(\frac{m_i}{T}\Big);\,\,
s_*(T) = T^3 \sum_i g_i a_s\Big(\frac{m_i}{T}\Big).
\end{equation}

To simplify the notation and make it uniform with the massless case, let us introduce
the following {\em effective degeneracy factors} for energy--density, pressure and entropy--density
$$
g_{\epsilon*}(T) = \frac{30}{\pi^2}\sum_{i} g_i a_\epsilon\Big(\frac{m_i}{T}\Big),\quad
g_{p*}(T) = \frac{30}{\pi^2}\sum_{i} g_i a_p\Big(\frac{m_i}{T}\Big),\quad  g_{s*}(T) =
\frac{45}{2\pi^2} \sum_{i} g_i a_s\Big(\frac{m_i}{T}\Big)\,.
$$

We can therefore rewrite Eqs (\ref{stareqs}) in the general form
\begin{equation}
\label{effectivefactors}
\epsilon_*(T)\!\equiv\!\frac{\pi^2g_{\epsilon*}(T)}{30}T^4,\,\,
p_*(T)\!\equiv\!\frac{\pi^2 g_{p*}(T)}{30}T^4,\,\, s_*(T)\!\equiv\!\frac{\epsilon_*(T)\!+\!p_*(T)}{T}\!
=\!\frac{2\pi^2 g_{s*}(T)}{45}T^3.
\end{equation}

Recalling that Eq (\ref{HiggsEdensQ}) gives us the value of the Higgs--field energy
density immediately after the big bang, using Eq (\ref{energydens}) with $g=1$ and $m=\mu_H$,
we obtain big bang temperature $T_B$ by solving for $T_B$ equation $\epsilon_*(T_B) =
\mu_H^4/16\lambda$. Then, using Eq (\ref{pressure}) and (\ref{entropydens}), we can calculate
fluid pressure $p_*(T_B)$,  Higgs--boson density $n_*(T_B)$ and entropy density $s_*(T_B)$.
Numerical computations carried out by a Matlab routine gave:
\vspace*{-2mm}
\begin{eqnarray}
\label{bigbangdata}
& & \epsilon_*(T_B) = \frac{\mu_H^4}{16\,\lambda} \simeq  1.186\times 10^8\,\hbox{GeV}^4
\quad\hbox{energy density at big bang}\,;\nonumber\\
& & T_B \simeq 141.0\,\hbox{GeV}\quad\hbox{big bang temperature}\,;\nonumber \\
\label{epsstar}
& & g_{\epsilon*}(T_B) = \frac{30}{\pi^2}\,\frac{\epsilon_*(T_B)}{T^4_B} \simeq 0.9121
\quad\hbox{effective degeneracy of }\,\epsilon_*(T_B)\,;\\
\label{higgspress}
& & p_*(T_B) \simeq 3.554\times 10^7\, \hbox{GeV}^4\quad\hbox{Higgs field pressure at big bang}\,;\\
\label{gpstar}
& & g_{p*}(T_B) = \frac{30}{\pi^2}\,\frac{p_*(T_B)}{T^4_B} \simeq 0.2733
\quad\hbox{effective degeneracy of }\,p_*(T_B)\,;\\
\label{cs2}
& &\frac{p_*(T_B)}{\epsilon_*(T_B)} = \frac{g_{p*}(T_B)}{g_{\epsilon*}(T_B)}
\simeq  0.2997\,\,(\hbox{relativistic limit = }1/3)\,; \\
\label{higgdens}
& & n_*(T_B) \simeq 2.655 \times 10^5\,\hbox{GeV}^3\quad \hbox{Higgs--boson density at big bang\,;}\\
& & s_*(T_B) = \frac{\epsilon_*(T_B) + p_*(T_B)}{T_B} \simeq 1.093\times 10^6\,
\hbox{GeV}^3\,\,\hbox{entropy density at big bang}\,; \nonumber\\
\label{gsstar}
& & g_{s*}(T_B) = \frac{45}{2\pi^2}\frac{s_*(T_B)}{ T_B^3} \simeq
0.8883\quad\hbox{effective degeneracy of }\,s_*(T_B)\,;\\
\label{gestarder}
& & \frac{d e_*(T_B)}{d T_B} = 3.513\times 10^6\,\hbox{GeV}^3\,;\quad
\frac{d g_{e*}(T_B)}{dT_B} = 1.142\times 10^{-3}\,\hbox{GeV}^{-1}\,;\\
\label{geptarder}
& & \frac{d p_*(T_B)}{d T_B} = 1.093\times 10^6\,\hbox{GeV}^3\,;\quad
\frac{d g_{p*}(T_B)}{dT_B} = 6.513\times 10^{-4}\,\hbox{GeV}^{-1}\,;\\
\label{gsstarder}
& & \frac{d s_*(T_B)}{d T_B} = 2.490\times 10^4\,\hbox{GeV}^2\,;\quad
\frac{d g_{s*}(T_B)}{dT_B} = 1.345\times 10^{-3}\,\hbox{GeV}^{-1}\,.
\end{eqnarray}
\centerline{\small Table 1. Magnitudes of most significant thermodynamic quantities at big bang.}

These may be compared with the cosmic--background data observed today:
\vspace{-4mm}
\begin{eqnarray}
\label{cosmicdata}
& & T_{BK}= 2.726 \, \mbox{$^{\mbox{\tiny o}}$K} = 2.350\times 10^{-13}\,
\hbox{GeV}\quad\hbox{(present cosmic--background temperature)}\,;  \nonumber \\
& & g_{*\epsilon}(T_{BK}) \simeq 3.738\,; \quad \epsilon_*(T_{BK}) = \frac{\pi^2}{30}\,g_{*\epsilon}(T_{BK})
\,T^4_{BK}\simeq 3.750\times 10^{-51}\,\hbox{GeV}^4\,;\nonumber\\
& & g_{*s}(T_{BK})\simeq 3.938\,;  \quad s_*(T_{BK})=\frac{2\pi^2}{45}\,g_{*s}(T_{BK})\,T_{BK}^3 \simeq 2.242\times 10^{-38}\,\hbox{GeV}^3\,.
\vspace{-4mm}
\end{eqnarray}
\centerline{\small Table 2. Magnitudes of most significant thermodynamic quantities today.}

Here, $g_{*\epsilon}(T_{BK})$, $s_*(T_{BK})$ are respectively the energy density and entropy density of
photons and neutrinos in the cosmic background, and $g_{*\epsilon}(T_{BK})$, $g_{*s}(T_{BK})$ their respective
degeneracy factors (contributions from other particles are negligible) \cite{KOLB} \cite{EGAN} \cite{MANGANO}.

From Eq (\ref{longcurrcons}), we can derive the following constraint for entropy densities
$s_*(T_1)$ and $s_*(T_2)$, respectively at time $\tau_1$ and $\tau_2$
\begin{equation}
\label{generalentropyratio}
\bigg[\frac{s_*(T_1)}{s_*(T_2)}\bigg]^{1/3}\!\!\! =\frac{c(\tau_2)\, s(\tau_2)}{c(\tau_1)\,s(\tau_1)}=
\bigg[\frac{g_{*s}(T_1)}{g_{*s}(T_2)}\bigg]^{1/3}\frac{T_1}{T_2}\,.
\end{equation}
In particular, for entropy densities $s_*(T_B)$ and $s_*(T_{BK})$, respectively at kinematic big--bang time $\tau_j$
and universe age $\tau_u$, we have
\begin{equation}
\label{entropyratio}
\bigg[\frac{s_*(T_{BK})}{s_*(T_B)}\bigg]^{1/3}\!\!\! =\frac{c(\tau_c)\, s(\tau_c)}{c(\tau_u)\,s(\tau_u)}=
\bigg[\frac{g_{*s}(T_{BK})}{g_{*s}(T_B)}\bigg]^{1/3}\frac{T_{BK}}{T_B}\simeq  2.738\times 10^{-15}.
\end{equation}

\subsection{Self--consistent determination of the cosmological constant}
\label{cosmconstandscale}
We are now in a position to calculate the most significant constants of inflationary cosmology by a self--consistent
method based on the crossing of two independent determinations of scale factor $s(\tau)$ at big bang kinematic
time $\tau_j\simeq \tau_c$,  so that the age of the universe in proper time units $\tilde\tau_U$ only is a free
parameter. Since this age is affected by a certain imprecision, so will the results of our computations.
However, among the various determinations of $\tilde\tau_U$ mentioned in \S~\ref{caveats}, we retain here,
in a first instance, the value of 15 Gyr $\simeq 4.35\times 10^{17}$s, because only this age is standard--model
independent. Using Eq (\ref{tildetau2tau}), we can find the corresponding universe age in kinematic--time
units $\tau_U$ as a function of critical time $\tau_c$ and $\tilde\tau_U$:
$$
\tau_U \simeq \tau_c +  \frac{\tilde\tau_U}{2} \Bigg(1+\sqrt{1+ \frac{2\tau_c}{\tilde\tau_U}}\,\,\Bigg).
$$

The first determination  of the scale factor at critical time $\tau_c$, which we call the
{\em entropic determination},  is provided by Eq (\ref{entropyratio}):
\begin{equation}
\label{sAj}
s_1(\tau_c) \simeq \frac{\sinh\big(2\tau_U/\tau_c\big)}{\sinh(2)}
\bigg(1 - \frac{\tau_c^2}{\tau_U^2}\bigg) \frac{T_{BK}}{T_B}
\bigg[\frac{g_{*s}(T_{BK})} {g_{*s}(T_B)}\bigg]^{1/3},
\end{equation}

The second determination of the same scale factor, which we call the {\em energetic determination},
is provided by Eq (\ref{exactR}):
\begin{equation}
\label{initialthetamunu}
s_2(\tau_c) \simeq \bigg(\frac{-R \lambda\,\sigma_0^2}{6\,\mu^4}\bigg)^{1/4}\!\!\!=
\bigg(\frac{\lambda\,\Lambda}{3\,\kappa\,\mu^4_H}\bigg)^{1/4}\!\!\!=
\bigg(\frac{2\lambda}{\kappa}\bigg)^{1/4}\!\!\frac{2}{\mu_H\sqrt{\tau_c}}\,,
\vspace{-1mm}
\end{equation}
where, as discussed in \S\,\ref{Curvedhyperbspacetime}, $\rho_{\hbox{\tiny vac}}=-4\,R/\kappa =\Lambda/\kappa$
and $\tau_c = 2\tau_\Lambda=2 \sqrt{3/\Lambda}$.

\newpage
We can then calculate both $\tau_c$ and $s(\tau_c)$, and all other quantities which depend on them, by solving
numerically equation $s_1(\tau_c)=s_2(\tau_c)$. The results of computations carried out by a Matlab routine,
for a presumed universe age $\tilde\tau_U$ of about 15 Gyr, are shown in Fig.\,13.
\begin{figure}[!h]
\vspace{-1mm}
\centering
\includegraphics[scale=0.8]{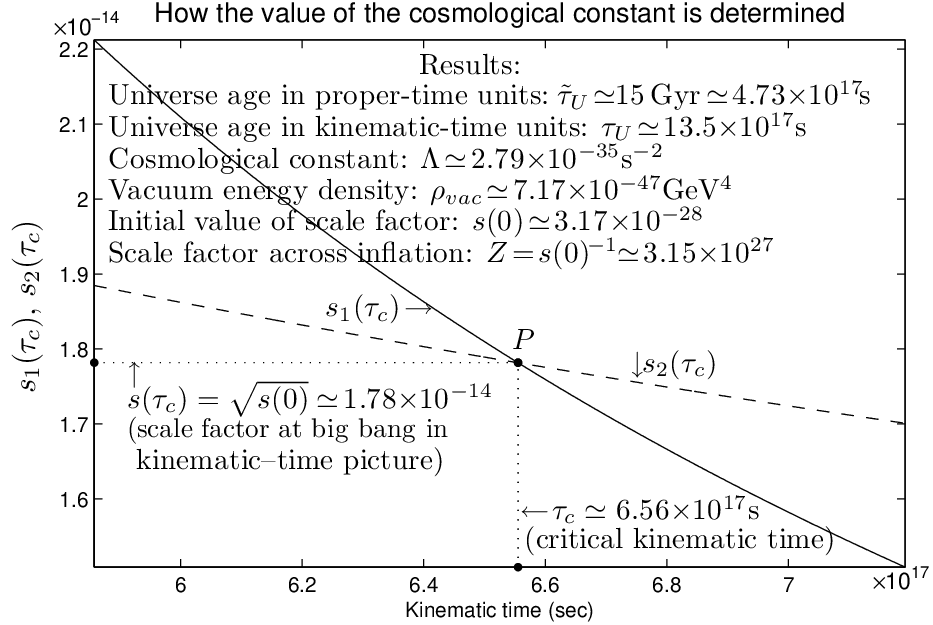}
\vspace{-4mm}
\caption{\small Big--bang kinematic time $\tau_c$, scale factor $s(\tau_c)$ and
consequently cosmological constant $\Lambda = 6/\tau_c^2$ are found by crossing curves
$s_1(\tau_c)$ and $s_2(\tau_c)$, which are respectively determined by entropic formula
for $s(\tau_c)$ and gravitational equation at $\tau =0$.}
\end{figure}

It is interesting to compare these results with those obtained by assuming different
universe ages. Here is a list:

\quad

\centerline{
\begin{tabular}{|c|c|c|c|c|}
\hline
$\tilde\tau_U$\quad & 13.8 Gyr & 14.8 Gyr & 15 Gyr & \!\!19.5 Gyr\\
\hline
$\tau_U$\quad& $1.22\!\times\!10^{18}$s & $1.31\!\times\!10^{18}$s & $1.35\!\times\!10^{18}$s & $1.87\!\times\!10^{18}$s\\
\hline
$\Lambda$\quad& $3.51\!\times\!10^{-35}$s$^{-2}$  & $2.90\!\times\!10^{-35}$s$^{-2}$ &$2.79\!\times\!10^{-35}$s$^{-2}$ &$1.35\!\times\!10^{-35}$s$^{-2}$\\
\hline
$\rho_{\mbox{\tiny vac}}$&  $9.00\!\times\!10^{-47}$GeV$^4$& $7.44\!\times\!10^{-47}$GeV$^4$ &$7.17\!\times\!10^{-47}$GeV$^4$&$3.46\!\times\!10^{-47}$GeV$^4$\\
\hline
$s(\tau_c)$ & $1.89\!\times\!10^{-14}$ & $1.80\!\times\!10^{-14}$&$1.78\!\times\!10^{-14}$ &$1.48\!\times\! 10^{-14}$\\
\hline
$\tau_c$\quad& $5.85\!\times\!10^{17}$s&$6.44\!\times\!10^{17}$s&$6.56\!\times\!10^{17}$s & $9.44\!\times\!10^{17}$s\\
\hline
$s(0)$& $3.56\!\times\!10^{-28}$ & $3.23\!\times\!10^{-28}$ & $3.17\!\times\!10^{-28}$ & $2.20\!\times\!10^{-28}$\\
\hline
$Z$\quad& $2.81\!\times\!10^{27}$  & $3.09\!\times\!10^{27}$  & $3.15\!\times\!10^{27}$ & $4.54\!\times\!10^{27}$ \\
\hline
\end{tabular}}

\subsection{The lower bound of cosmic microwave background anisotropies}
\label{CMBanysot}
The anisotropies of the temperature of cosmic microwave background (CMB) observed through spatial or terrestrial infrared--sensitive 
telescopes appear as a jumble of pale spots which cover the dark areas of the celestial sphere with spherical--harmonic distribution
ranging from $\,\ell =$ 2 to 2000 and power spectrum extending from $\simeq 30$ to $6\times 10^3 \mu$K$^2$
\cite{HINSHOW} \cite{WRIGHT} \cite{SHIROKOFF}. They are believed to be long--delayed effects of quantum fluctuations which 
occurred in the vacuum state of the universe at the moment of the big bang and were hugely amplified by the accelerated stage of inflation. 
But, as we shall prove below, within the framework of CGR, they can most simply be ascribed to the thermal fluctuation of the bulk of Higgs
bosons at the moment of its sudden creation, almost immediately ``clotted'' by incipient gravitational collapse. This was
possible despite the temperature of about 141 GeV because, in virtue of Weyl scale--factor $s(\tau)$, the
gravitational forces at the moment of the big--bang $\tau_j\simeq \tau_c$ were $Z= 1/s(\tau_c)^2\approx 10^{27}$
times larger than they are today.

The mechanism of gravitational collapse was investigated in 1902 by Jeans \cite{JEANS}, who showed that a
homogeneous sphere of a non--relativistic gravitating fluid becomes unstable as its radius exceeds a critical
value $R_J$, known as the radius of Jeans. A simple determination of $R_J$ can be provided by
requiring gravitational energy $U_G$ of the sphere plus its thermal energy $U_T$ to be negative. In the
Newtonian approximation, the gravitational potential is related to matter density by equation
$\nabla^2\Phi = 4\pi G\rho$, where $G\equiv\kappa/8\pi$ is the gravitational coupling constant
and $\rho$ the mass density of the fluid. Since the gravitational potential at the surface of a
sphere of radius $R$ and mass $M =4\pi\rho R^3/3$ is $\Phi(R) = GM/R$, and, on the other hand,
the contribution to $U_G$ of the spherical shell of radius $R$ and thickness $dR$ is
$dU_G= -\Phi(R)\,4\pi\rho R^2 dR$, by integration we find:
\begin{equation}
\label{U_G}
U_G = - \frac{16\,\pi^2 G\,\rho^2\, R^5}{15} = - \frac{3G\,\rho^2\, V^2}{5R}\,,\quad\mbox{where }\, V=\frac{4}{3}\,\pi R^3\,.
\end{equation}
With fluid temperature $T$ and entropy density $s(T)$, the thermal energy of the sphere is
\begin{equation}
\label{U_T}
U_T = c_V\,  T\, V \,,\quad\hbox{with }\, c_V = T\, \frac{d s}{dT}\,,
\end{equation}
where $c_V$ is the specific thermal capacity of the fluid at constant volume. We
therefore obtain
\begin{equation}
\label{R_Ggas}
R_J=\sqrt{\frac{5\,Tc_V}{4\pi G\,\rho^2}}\,.
\end{equation}

When we try to determine the Jeans radius for a homogeneous sphere of Higgs--boson gas
at big--bang temperature $T_B$, we encounter four main problems:
\begin{itemize}

\item[(1)] In passing from standard cosmology to CGR, we must describe the fluid in
the hyperbolic coordinates of accelerated--inflated Milne spacetime $\widetilde M^+$
(Fig.\,11).

\item[(2)] Since at the moment of their creation the Higgs bosons are nearly relativistic, mass density $\rho$
in equation $\nabla^2\Phi = 4\pi G\,\rho$ must be replaced by $\rho +3p -\rho_{\mbox{\tiny vac}}$.
This is due to the fact that, as reported on the ends of \S\,3.4 and \S\,3.6 of Part II, in the static Newtonian
approximation, gravitational potential $\Phi$ is related to the spatially perturbed component $R_{00}$ of the Ricci tensor by equation
$\nabla^2 \Phi \simeq R_{00}$, while $R_{00}$ is related to matter EM--tensor $\Theta^M_{\mu\nu}$, and its trace $\Theta^M$,
by equation $R_{00} = \Theta^M_{00}- \frac{1}{2}\Theta$, which therefore gives $\nabla^2 \Phi = 4\pi G(\rho+3 p -
\rho_{\mbox{\tiny vac}})$.

\item[(3)] Because of the underlying conformal invariance of CGR, we cannot neglect the powerful actions of
acceleration factor $c(\tau)/\tau = \sinh(\tau/\tau_\Lambda)$ and Weyl factor $s(\tau)$, which respectively
become $\sinh(\tau_c/\tau_\Lambda) = \sinh 2$ and $s(\tau_c)= \sqrt{s(0)}$ at big bang. This means that,
at big bang, the radius of Jeans must be multiplied by $\sinh 2$ and all constants of dimension $n$ must
be multiplied by $s(\tau_c)^n$.

\item[(4)] We must be able to explain how and why the same pattern of thermal fluctuations created at
big bang may then resurface after the recombination epoch.
\end{itemize}

The first difficulty can be circumvented by focusing on those regions
of accelerated Milne spacetime ${\cal M}^+$ which are close to the future--cone axis
(Fig.\,10 of \S~\ref{caveats}). Since, in passing from the kinematic--time picture of ${\cal M}^+$
to the proper--time picture of inflated--accelerated Milne spacetime $\widetilde{\cal M}^+$
(Fig.\,11 of \S~\ref{caveats}), these regions flatten considerably in the neighborhood of the
time axis, metric $\widetilde{\cal M}^+$ is well approximated by the RW metric of the
standard model. We can therefore safely replace proper time $\tilde\tau$ of
conical spacetime $\widetilde{\cal M}^+$  with proper time $t$ of
the cylindrical spacetime of standard cosmology. Correspondingly, we can safely replace
scale factor $\tilde a(\tilde\tau) = \tilde c(\tilde\tau)\,\tilde s(\tilde\tau)$ of the metric
described by Eq (\ref{tildemetrictensor}) with scale factor $a_{RW}(t)$ of the RW metric.

The second difficulty can be overcome by replacing $\rho$ with $\rho +3p$ in Eq (\ref{R_Ggas}), since
$\rho_{\mbox{\tiny vac}}$ is comparatively negligible. Then, in consideration of Eq (\ref{cs2}),
the energy density at big bang $\epsilon_*(T_B)$ must be replaced by $\bar\epsilon_*(T_B) = \epsilon_*(T_B)+3p_*(T_B)
\simeq 1.9 \,\epsilon_*(T_B)\simeq 2.25\times 10^8$GeV$^4$. Correspondingly, entropy density $s$ and
specific thermal capacity density $c_V = Tds/dT$, which appear in Eq (\ref{U_T}), must respectively be
replaced by effective entropy density $s_*(T_B)= 1.093\times 10^6$ GeV$^3$ and effective specific thermal
capacity
$$
c_*(T_B)= T_B\,\frac{d s_*(T_B)}{dT_B} \simeq 3.51\times 10^6\,\,\hbox{GeV}^3\,,\quad \hbox{then }\, \frac{U_G}{V_J}
= T_B\,c_*(T_B)\simeq 4.95\times 10^8\,\,\hbox{GeV}^4\,.
$$
Here, Eqs (\ref{epsstar}), (\ref{higgdens}), (\ref{gsstar}) and (\ref{gsstarder}) of Table 1 of
\S\,\ref{entropycourse} are exploited. For the sake completeness, we add to these the particle density of
Higgs--bosons $n_*(T_B) = 2.655\times 10^5$ and identify $V$ with Jeans' sphere volume $V_J$.

The third difficulty can be solved by multiplying $R_j$ by $\sinh2$ and replacing $G$ with $G/s(\tau_c)^2 = Z G$,
since $G$ has dimension $n =-2$. This means that, for a proper--time universe age of $\simeq 15$ Gyr, the gravitational
coupling parameter at big bang is $Z \simeq 3.15\times10^{27}$ times larger than in GR (see the list of constants on
the end of the previous section).

The fourth difficulty can be solved by carrying out spatial variations $\delta_{\vec\rho}$
of Eq (\ref{generalentropyratio}), which yields
$$
\delta_{\vec\rho}\ln \Bigg\{\frac{a(\tau_2)}{a(\tau_1)}\bigg[\frac{g_{*s}(T_1)}{g_{*s}(T_2)}\bigg]^{1/3}\Bigg\}=
\delta_{\vec\rho}\ln \frac{T(\tau_2,\vec\rho\,)}{T(\tau_1,\vec\rho\,)}=
\frac{\delta_{\vec\rho} T(\tau_2,\vec\rho\,)\,}{T(\tau_2,\vec\rho\,)\,}-\frac{\delta_{\vec\rho}
T(\tau_1,\vec\rho\,)\,}{T(\tau_1,\vec\rho\,)\,} =0.
$$
Here, the expression on the left vanishes because $a(\tau)$ and $ g_{*s}(\tau)$ are isotropic, whereas $T(\tau, \vec \rho\,)$ is not,
since it depends on the direction $\vec\rho$ of the polar geodesic stemming from the future--cone origin. This means
that the spatial pattern of temperature and entropy variations on the large scale remains the same in all the expanding hyperboloids.

In sum, by performing all the substitutions indicated above and carrying out the numerical computations by
a Matlab routine, we can easily verify that the critical radius of the Jeans sphere at kinematic time
$\tau_j\simeq \tau_c$ is
$$
R_J = \sinh 2\sqrt{\frac{5\,c_*(T_B)\,T_B}{4\pi Z\, G\,\bar \epsilon_*(T_B)\,\epsilon_*(T_B)}}
\simeq 67.7\,\hbox{GeV}^{-1}\,\simeq 13.3\,\hbox{fm}\,\,\hbox{(at big bang)} \,,
$$
from which we derive the following additional results
\begin{itemize}
\vspace{-1mm}
\item[] $V_J = (4/3)\,\pi\,R_J^3 \simeq  1.30\times 10^6$\,GeV$^{-3}$ (volume of Jeans sphere at big bang);
\vspace{-1mm}
\item[] $N_J = n_*(T_B)\,V_J \simeq 3.45\times 10^{11}$ (number of Higgs bosons in Jeans sphere);
\vspace{-1mm}
\item[] $\Delta N_j = 1/\sqrt{N_J} \simeq  1.70\times 10^{-6}$ (statistical standard deviation of $N_J$ at big bang).
\end{itemize}

Since the anisotropies of smaller size come from collapsing regions of minimum size, i.e., the Jeans spheres
described above, we infer that the minimum standard deviation of CMB temperature observable in the sky is related
to the standard deviation $\Delta N_j$ of the Higgs boson number at big bang by equation $\Delta T_{BK} =
T_{BK}\Delta N_j \simeq 4.64\,\mu$K. Thus, the minimum of the spectral power
of CMB anisotropies is $W_{BK} = \Delta\,T_{BK}^2  \simeq 2.15$\,$\mu$K$^2$.
In Fig.\,14, the predicted lower bound of $W_{BK}$ is shown for comparison with astronomic data.
\begin{figure}[!h]
\vspace{-8mm}
\begin{center}
\scalebox{0.26}{\includegraphics{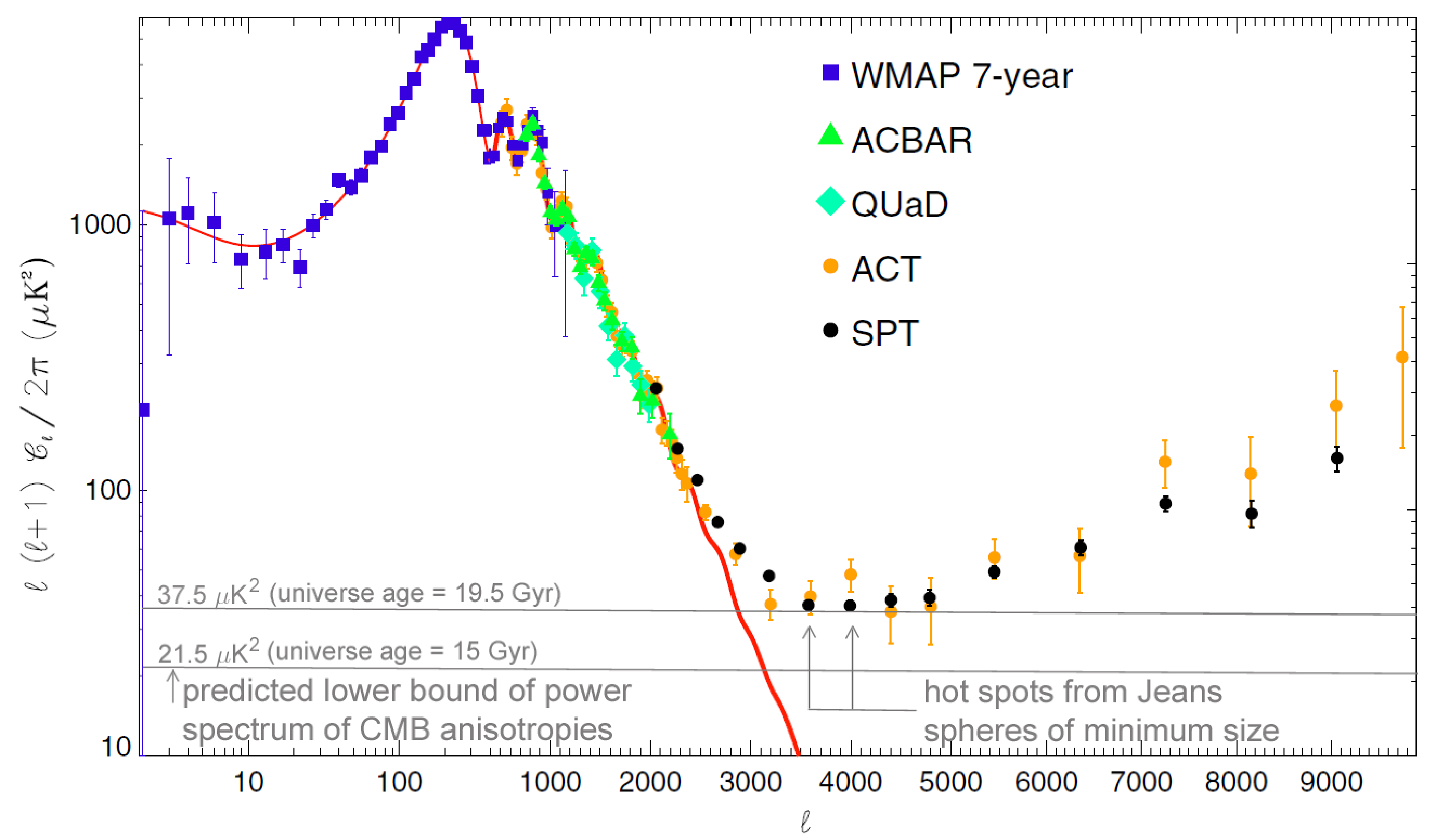}}
\end{center}
\vspace{-10mm}
\caption{\small Lower bounds of CMB anisotropies in $\mu$K$^2$ compared with
data from five astronomical missions: WMAP (Wilkinson Microwave Probe Telescope, 2001--2008);
ACBAR (Arcminute Cosmology Bolometer Array Receiver, 2002--2006); QUaD (Q\&U
Extragalactic Survey Telescope + Degree Angular Scale Interferometer, 2003);
ACT (Atacama Cosmology Telescope, 2014); STP (South Pole Telescope, 2007--2011).
SPT data with 3.5\% calibration error from figure 4 of paper by Shirokoff {\em et al.} (2011).
Fit would be perfect if universe age were $\simeq 19.5$ Gyr.}
\vspace{-2mm}
\end{figure}

These results suggest that the age of the universe is larger than that assumed in \S\,\ref{caveats},
but still in the range of $15.6\pm 4.6$ Gyr inferred by Cowan {\em et al.} (1999) from
the abundances of long--living radioactive elements in extremely old halo stars.
Assuming an age of 19.5 Gyr, we derive cosmological parameters $\Lambda \simeq 1.35\times 10^{-35}$s$^{-2}$,
$\rho_{\mbox{\tiny vac}} \simeq 3.46\times 10^{-47}$GeV$^4$, $\tau_c \simeq 8.44\times 10^{-14}$s,
$Z \simeq 4.54\times 10^{27}$, $s(\tau_c) = Z^{-1/2} \simeq 1.48\times 10^{-14}$
(see end of \S~\ref{cosmconstandscale}), and the lower bound of CMB spectral power $W_{BK}
\simeq 37.5$\,$\mu$K$^2$; all of which seem in fact to be closer than previous ones to the values
usually reported in the current literature.

\centerline{--------------------------------------}

\centerline{---------------}


\begin{thebibliography}{99}
\bibliographystyle{plain}
\addcontentsline{toc}{section}{References}

\bibitem{PART1} Nobili, R.: THE CONFORMAL UNIVERSE I: Theoretical Basis of Conformal General
Relativity. {\small \em FINAL VERSION}. arXiv:1201.2314v4 [hep-th] (2016).

\bibitem{PART2} Nobili, R.: THE CONFORMAL UNIVERSE II: Conformal Symmetry, its Spontaneous Breakdown
and Higgs Fields in Conformally Flat Spacetime. {\small \em FINAL VERSION}. arXiv:1201.3343v5 [hep-th] (2016).

\bibitem{ILHAN} Ilhan, I.B, and Kovner, A.: Some Comments on Ghosts and Unitarity: The Pais-Uhlenbeck Oscillator Revisited.
{\em Phys. Rev. D} 88:044045-1--044045-12 (2013); DOI: 10.1103/PhysRevD.88.044045.

\bibitem{CARTAN} Cartan, E.:  Les espaces \`a connexion conforme.  {\em Ann. Soc. Pol. Math.} 2,  171--221  (1923);
Sur les variet\'es \`a connexion affine et la th\'eorie de la relativit\'e g\'en\'eralis\'ee. Premi\`ere partie
{\em Ann. Ec. Norm.} 41, 1--25 (1924); Deuxi\`eme partie  {\em Ann. Ec. Norm.} 42, 17--88 (1925).

\bibitem{EISENHART} Eisenhart, L.P.:  {\em Riemannian Geometry.} pp. 82--92, Princeton
University Press (1949).

\bibitem{PEACOCK} Peacock, J.A. {\em Cosmological Physics.} Cambridge University Press, UK (1999).

\bibitem{MUKHANOV} Mukhanov, V. {\em Physical Foundation of Cosmology.} Cambridge University Press, UK (2005).

\bibitem{COWAN} Cowan, J.J {\em et al.}: r-Process abundances and chronometers in metal---poor stars.
{\em The Astrophysical Journal}, 521:194--205 (1999).

\bibitem{FUBINI} Fubini, S.: A New Approach to Conformal Invariant Field
Theories.  {\em Il Nuovo Cimento}. 34A, 521--554 (1976).

\bibitem{WALKER} Walker, A.G. (1937) On Milne's Theory of World Structure. {\em
Proc. London Math. Soc}. 42:90--127.

\bibitem{ROBERTSON} Roberson, H.P.: Kinematics and World--Structure. Part I
{\em The Astrophysical Journal} 82:284--301 (1935); Part II;
{\em The Astrophysical Journal} 83:187--201 (1936); Part III,
{\em The Astrophysical Journal} 83:257--271 (1936).

\bibitem{BEHAR} Behar, S. and Carmeli, M.: (2000) Cosmological Relativity: A New Theory of Cosmology.
{\em Intern. J. Theor. Phys.} 39:1375--1396; (astro-ph/0008352);
Carmeli, M. and Kuzmenko. T.: (2001) Value of the Cosmological Constant:
Theory versus Experiment; (arXiv:astro-ph/0102033v2).

\bibitem{FRIEDMANN} Friedmann, A.A.: (1922) On the Curvature of Space
(English translation); {\em General Relativity and Gravitation}, 31:1991--2000 (1999).

\bibitem{LEMAITRE} Lema\^itre, Abb\'e G.: A Homogeneous Universe of Constant Mass and
Increasing Radius accounting for the Radial Velocity of Extra--galactic Nebulae.
Translated from  {\em Annales de la Soci\'et\'e scientifique de Bruxelles}. Tome XLVII,
s\'erie A, premi\`ere partie, pp. 483--490 (1931).

\bibitem{UMEZAWA2} Umezawa, H. {\em Advanced Field Theory. Micro, Macro, and Thermal Physics} (\S\,2.3).
American Institute of Physics, New York (1993).

\bibitem{KIBBLE} Kibble, T.W.B.: Coherent Soft-Photon States and Infrared Divergences. I. Classical Currents,
{\em Journal of Math. Phys.} 9:315-324 (1968);  Coherent Soft-Photon States and Infrared Divergences.
II. Mass-Shell Singularities of Green's Functions. {\em Physical Rev.} 173:1527-1535 (1968); Coherent
Soft-Photon States and Infrared Divergences. III. Asymptotic States and Reduction Formulas.
{\em Physical Rev.} 174:1882-1901 (1968);  Coherent Soft-Photon States and Infrared
Divergences. IV. The Scattering Operator. {\em Physical Rev.} 175:1624-1640 (1968).

\bibitem{BROUT} Brout, R., Englert, F. and Gunzig, E.: The Creation of the
Universe as a Quantum Phenomenon. {\em Annals od Physics}. 115, 78--106 (1978).

\bibitem{PLANCK} Adam, R. {\em et al.}:  Planck 2015 results. I.
Overview of products and scientific results. (Table 9) arXiv:1502.01582v2 [astro-ph.CO] (2015).

\bibitem{RIESS} Riess, A.G,  {\em et al.}: Observational evidence from supernovae for an accelerating
universe and a cosmological constant. {\em The Astronomical Journal}, 116:1009--1038 (1998)

\bibitem{PERLMUTTER} Perlmutter, S. {\em et al.}: Measurements of $\Omega$ and $\Lambda$ from
42 high--redshift supernovae. {\em The Aastrophysical Journal}, 517:565--586 (1999).

\bibitem{JONA} Jona--Lasinio, G.: Relativistic Field Theories with Symmetry--Breaking Solutions.
{\em Il Nuovo Cimento} 34:1790--1795 (1964).

\bibitem{COLEMAN1} Coleman, S. and Weinberg, E.: Radiative Corrections as the Origin of Spontaneous
Symmetry Breaking. {\em Phys. Rev. D} 7:1888--1910 (1973); {\em Aspects of Symmetry. Selected Erice Lectures.}
Cambridge University Press (1985).

\bibitem{JACKIW} Jackiw, R.: Functional Evaluation of the Effective Potential. {\em Phys. Rev. D} 9:1686--1701 (1974).

\bibitem{WEINBERG1} Weinberg, S.:  The cosmological constant problem. {\em Rev. of Modern
Phys.} 81, 1--23 (1989).

\bibitem{MARTIN} Martin, J.: Everything You Always Wanted To Know About The Cosmological Constant Problem
(But Were Afraid To Ask); arXiv:1205.3365 [astro-ph.CO] (2012).

\bibitem{VELTMAN} Veltman, M.: The infrared--ultraviolet connection. {\em Act. Phys. Pol.} B12:437--457 (1981).

\bibitem{STELLE} Stelle, K.S.: Renormalization of higher--derivative quantum
gravity. {\em Phys. Rev. D}. 16, 953--969 (1977).

\bibitem{TOMBOULIS1} Tomboulis, E.: Renormalizability and Asymptotic Freedom in
Quantum Gravity. {\em Phys. Lett.}, 97B, 77--80 (1980).

\bibitem{ARAKI} Araki, A. and Woods, E.J.: Representations of the Canonical Commutation
Relations Describing a Nonrelativistic Infinite Free Bose Gas. {\em J. Math. Phys.}
4:637--662 (1963).

\bibitem{KUBO} Kubo, R.: The fluctuation--dissipation theorem. {\em Reports on Progress in Physics}, 29:255-284 (1966).

\bibitem{HAAG} Haag, R.: {\em Local Quantum Physics: Fields, Particles, Algebras}.
pp 13--17, Springer--Verlag, Berlin (1992).

\bibitem{VNEUMANN} v. Neumann, J.: On infinite direct products. {\em Compositio Mathematica}. 6:1--77. P.Nortdhoff,
Gr\"oningen (1939).

\bibitem{BRATTELI} Bratteli, A. and Robinson, D.W.: {\em Operator Algebras and Quantum Statistical Mechanics 1, 2}.
Springer (2002).

\bibitem{UMEZAWA1} Umezawa, H., Matsumoto, H. and Tachiki, M. {\em Thermo Field Dynamics and Condensed
States}. North--Holland Pub. Comp. (1982).

\bibitem{BOGOLIUBOV} Bogoliubov, N.N.: On the theory of superfluidity, {\em Journal of Physics} 11: 23–32
(1947).

\bibitem{MANN} Mann,A. Revzen,M.,  Umezawa, H and Yamanaka, Y: Relation between quantum and thermal fluctuations.
{\em Phys.Lett.A} 140:475–-478 (1989).

\bibitem{TAKAHASHI} Takahashi, Y. and Umezawa, H., A General Theory of Expanding
Systems. I. Formulation. {\em Il Nuovo Cimento}.  6:1324--1334 (1957).

\bibitem{WEINBERG4} Weinberg, S.: {\em Cosmology}. \S\,3.1. Oxford University Press (2008).

\bibitem{WALD} Wald, R.M. {\em General Relativity}. Ch. 5,  The University of Chicago Press (1984).

\bibitem{KOLB} Kolb, E.W. and Turner, M.S.: {\em The Early Universe}, page 76.  Addison--Wesley (1990)

\bibitem{EGAN} Egan, C.A. and Lineweaver, C.H.: A Larger Estimate of the Entropy of the Universe.
{\em The Astrophysical Journal}, 710:1825–1834 (2010)

\bibitem{MANGANO} Mangano, G., Mielea, G., Pastor, S. and Pelosoc, M.: A precision calculation of the
effective number of cosmological neutrinos. {\em Phys. Lett. B} 534:8–16 (2002).

\bibitem{HINSHOW} Hinshow, G. {\em et al.}: Three--Year Wilkinson Microwave Amisotropy Probe (WMAP) Observations:
Temperature Analysis. {\em The Astrophysical Journal Supplement Series}, 170:288--334 (2007).

\bibitem{WRIGHT} Wright, E.L.: Acoustic Waves in the Early Universe. {\em Journal of Physics: Conference Series},
118:1--8 (2008); DOI:10.1088/1742-6596/118/1/012007.

\bibitem{SHIROKOFF} Shirokoff {\em et al.}: Improved Constraint on Cosmic Microwave Background Secondary Anisotropies
from the Complete 2008 South Pole Telescope Data. {\em The Astrophysical Journal}, 736:61--82 (2011).

\bibitem{JEANS} Jeans, J.H.: The Stability of a Spherical Nebula. {\em Philosophical Transactions of the Royal Society}
A 199:1–-53 (1902).
\end{thebibliography}
\end{document}